\documentclass[aps,pra,floatfix,superscriptaddress,twocolumn]{revtex4-2}

\usepackage{amsmath,amssymb,graphicx,color,braket,xcolor}
\usepackage[ocgcolorlinks,colorlinks=true,linkcolor=blue,citecolor=red,linktocpage=true]{hyperref}
\usepackage{bbm}

\usepackage[normalem]{ulem}

\usepackage{physics}
\usepackage{comment}

\begin{document}

\title{Quasiprobabilities in quantum thermodynamics and many-body systems}

\author{Stefano Gherardini}
\email{stefano.gherardini@ino.cnr.it}
\affiliation{Istituto Nazionale di Ottica del Consiglio Nazionale delle Ricerche (CNR-INO), Largo Enrico Fermi 6, I-50125 Firenze, Italy.}
\affiliation{European Laboratory for non-linear Spectroscopy, Università di Firenze, I-50019 Sesto Fiorentino, Italy.}

\author{Gabriele De Chiara}
\email{g.dechiara@qub.ac.uk}
\affiliation{Centre for Quantum Materials and Technology, School of Mathematics and Physics, Queen’s University Belfast, Belfast BT7 1NN, United Kingdom.}

\date{\today}

\begin{abstract}
In this tutorial, we present the definition, interpretation and properties of some of the main quasiprobabilities that can describe the statistics of measurement outcomes evaluated at two or more times. Such statistics incorporate the incompatibility of the measurement observables and the state of the measured quantum system. We particularly focus on Kirkwood-Dirac quasiprobabilities and related distributions.
We also discuss techniques to experimentally access a quasiprobability distribution, ranging from the weak two-point measurement scheme, to a Ramsey-like interferometric scheme and procedures assisted by an external detector. 
Once defined the fundamental concepts following the standpoint of joint measurability in quantum mechanics, we illustrate the use of quasiprobabilities in quantum thermodynamics to describe the quantum statistics of work and heat, and to explain anomalies in the energy exchanges entailed by a given thermodynamic transformation. On the one hand, in work protocols, we show how absorbed energy can be converted to extractable work and vice versa due to Hamiltonian incompatibility at distinct times. On the other hand, in exchange processes between two quantum systems initially at different temperatures, we explain how quantum correlations in their initial state may induce cold-to-hot energy exchanges, which are unnatural between any pair of equilibrium nondriven systems. We conclude the tutorial by giving simple examples where quasiprobabilities are applied to many-body systems: scrambling of quantum information, sensitivity to local perturbations, and quantum work statistics in the quenched dynamics of models that can be mapped onto systems of free fermions, for instance the Ising model with a transverse field. Throughout the tutorial, we meticulously present derivations of essential concepts alongside straightforward examples, aiming to enhance comprehension and facilitate learning.
\end{abstract}

\maketitle

\tableofcontents

\section{Introduction}

In this paper we provide a tutorial that delves into the use of quasiprobabilities and their associated distributions within the realm of quantum science and with specific applications in quantum thermodynamics and many-body quantum systems.

As evident from numerous studies~\cite{PhysRevLett.101.020401,ferrie2011quasi,RyuPRA2013,HalpernPRA2018,lostaglio2018quantum,alonso2019out,Lostaglio2020certifying,levy2020quasiprobability,ArvidssonShukurJPA2021,PhysRevLett.127.190404,LostaglioQuantum2023,hernandez2022experimental,DeBievreArXiv2022,zhang2022quasiprobability,Cerisola2023aWigner,FrancicaPRE2023,SantiniPRB2023,SagarQSL2024,ArvidssonShukurReview2024}, quasiprobabilities have garnered significant interest within the quantum community.
In fact, they are a proper tool to describe the statistics of the outcomes resulting from consecutive events in several areas of quantum mechanics and associated technologies, from foundations to quantum devices. 
The most appealing feature of a quasiprobability is its capability to embody the incompatibility of quantum observables, due to their nonzero commutator, that are evaluated at different times of a given experiment. This incompatibility is singled out by the fact that the distribution of the measurement outcomes admits nonclassical traits. The latter ones, expressed in the form of `negative' and `nonreal probabilities', reflect the effects entailed by the Heisenberg uncertainty principle that indeed concerns the impossibility of concurrently measuring two complementary and incompatible properties of a quantum system in contiguous times.

In the following, we are going to introduce the concept of quasiprobability from theoretical arguments, and we then outline their main properties, especially those concerning the loss of positivity. After that, we will address measurement procedures that allow observation of the `negative' and `nonreal probabilities' of pairs of measurement outcomes, by showing that such genuinely quantum features have a clear physical interpretation inherently related to quantum coherence and correlations~\cite{BudiyonoPRA2023,BudiyonoJPA2023}. We are particularly interested in pedagogically explaining those physical interpretations that can be linked to thermodynamic quantities under out-of-equilibrium conditions (work, heat and their distributions)~\cite{Solinas2016probing,lostaglio2018quantum,levy2020quasiprobability,beyer2021joint,SolinasPRA2022,maffei2022anomalous,hernandez2022experimental,FrancicaPRE2022,FrancicaPRE2022Most,Cerisola2023aWigner}, and in the rich arena of many-body quantum systems~\cite{HalpernPRA2018,dressel2018strengthening,alonso2019out,LostaglioQuantum2023,zhang2022quasiprobability,FrancicaPRE2023,SantiniPRB2023}. In this regard, interpretations of the well-known Loschmidt echo and out-of-time-ordered correlators (OTOCs) for many-body systems in terms of quasiprobabilities are discussed and illustrated with step-by-step worked examples. Finally, the tutorial is concluded with a discussion containing some perspectives on possible theoretical studies and experimental observations of quasiprobabilities in current quantum platforms~\cite{lupu2021negative,maffei2022anomalous,hernandez2022experimental,ArvidssonShukurReview2024}.

This tutorial is targeted at graduate students and researchers, both theoretical and experimental, with a basic knowledge of quantum mechanics. Specifically, we aim to provide both the formal definitions that lay the foundations for quasiprobabilities in quantum science, and simple analytical examples helping to understand the fundamental concepts and applications of the methodology. We would also give some hints on new directions on the use of quasiprobabilities that have not yet been explored, especially in quantum thermodynamics and many-body quantum systems.

\section{Quasiprobabilities}
\label{section_quasiprobs}

In this section, we introduce the concept of \emph{quasiprobabilities} in quantum mechanics, following the seminal papers by Kirkwood~\cite{kirkwood1933quantum} and Dirac~\cite{dirac1945analogy} in the 1930s and 1940s, respectively. 
There have been several approaches that brought to light the notion of a `negative probability' and crucially even probabilities represented by a complex number. In this tutorial, we choose to approach this topic from the fundamental standpoint of \emph{joint measurability} in quantum mechanics.

Under conditions we are going to detail, Kirkwood-Dirac quasiprobabilities (KDQ) can take both negative and imaginary values. `Negative probability' and even `probabilities represented by a complex number' can be explained from the fundamental standpoint of \emph{joint measurability} in quantum mechanics. 

For this purpose, let us set the theoretical framework. First, consider a quantum state preparation that generates a generic density operator $\rho$ that, by definition, is a Hermitian, semi-definite operator with trace $1$. Then, we define two quantum observables, associated to two Hermitian operators, $\mathcal{O}_{1}(t_1)$ and $\mathcal{O}_{2}(t_2)$, that we measure at two distinct times $t_1$ and $t_2$ with $t_1 < t_2$. The observables can be generally expressed using their spectral decomposition as
\begin{eqnarray}
   \mathcal{O}_{i}(t_i)&=&\sum_{s_i}o_{s_{i}}(t_i)\Pi_{s_i}(t_i), \label{eq:O_ti}
\end{eqnarray}
in terms of their eigenvalues $o_{s_{i}}(t_i)$ and the associated projectors $\Pi_{s_i}(t_i)$ onto the corresponding eigenspace. 

In the general case, $\rho$, $\mathcal{O}_{1}(t_1)$ and $\mathcal{O}_{2}(t_2)$ do not commute with each other. Moreover, in the time interval $[t_1,t_2]$---after the first measurement of $\mathcal{O}_{1}(t_1)$ and before the second measurement of $\mathcal{O}_{2}(t_2)$---the state of the quantum system can be subject to a generic quantum process described by a completely positive trace preserving (CPTP) quantum map $\Phi:\rho(t_1)\mapsto\Phi[\rho(t_1)]=\rho(t_2)$, which operates on and returns density operators. For simplicity, we omit the explicit time dependence of $\Phi$ to avoid making the notation too heavy.
We introduce its Kraus representation such that $\Phi[\rho]=\sum_{\alpha}K_{\alpha}\rho K_{\alpha}^{\dagger}$ and $\sum_{\alpha}K_{\alpha}^\dagger K_{\alpha}=\mathbb{I}$, where $\mathbb{I}$ is the identity operator. The reader interested in the main properties of quantum maps can refer to Refs.~\cite{breuer2002theory,CarusoRMP2014}.

In this section, we discuss how to characterize the statistics of measurements outcomes from the \emph{two-time evaluation} of $\mathcal{O}_{1}(t_1)$ and $\mathcal{O}_{2}(t_2)$, by also taking into account the noncommutativity of the involved operators, i.e., the initial density operator $\rho$ and the two observables. We explicitly wrote ``two-time evaluation'' in order to clearly distinguish it from the wording \emph{sequential measurements}. In fact, as we will explain in a while, a procedure based on sequential measurements is necessarily \emph{invasive}. As a result, (a) the measured system is perturbed; (b) initial quantum coherence in $\rho$ (with respect to the basis decomposing $\mathcal{O}_{1}(t_1)$) is destroyed; (c) the statistics of the outcome pairs $(o_{s_1},o_{s_2})$ resulting from sequentially measuring $\mathcal{O}_{1}(t_1)$ and $\mathcal{O}_{2}(t_2)$ changes.

Because of these reasons, we are going to consider an approach that, even in the general noncommutative case, can return a statistics of $(o_{s_1},o_{s_2})$ that is exempt from the invasiveness of the measurement apparatus, and is not affected by the first measurement of a sequential procedure, at least in some statistical moments~\cite{HofmannNJP2014}. This aspect should not be surprising since it is known that~\cite{halliwell2016leggett} when noncommuting observables are taken into account, there is no unique formula to describe the joint probabilities of $(o_{s_1},o_{s_2})$ that has both a correspondence with the classical theory of probability and at the same time is returned by a noninvasive measurement routine; see Sec.~\ref{sec:no-go_theorem} for more details on this aspect.

Now let us discuss in greater depth the invasiveness under the joint measurability problem, by introducing the celebrated \emph{two-point measurement} (TPM) scheme~\cite{CampisiRMP2011}.

\subsection{Sequential projective measurements and the TPM scheme}\label{sec:TPM_scheme}

The TPM scheme is the procedure to characterize the statistics of $(o_{s_1},o_{s_2})$ by means of sequential measurements, which has a correspondence with the classical theory of probability~\cite{JarzynskiPRX2015}. Albeit the process underlying the measurement outcomes has quantum traits, the results from such a measurement procedure can also be described by a probability distribution that obeys Kolmogorov's axioms of probability. The Kolmogorov's axioms are as follows: (i) the probability of getting a measurement outcome is a nonnegative real number; (ii) the probability to measure at least one of the outcomes is $1$; (iii) the probability to measure any countable sequence of mutually exclusive measurement outcomes is equal to the sum of the probabilities for each outcome.

Operationally, the two quantum observables $\mathcal{O}_{1}(t_1)$ and $\mathcal{O}_{2}(t_2)$ are measured at times $t_1$ and $t_2$, respectively, and a pair of outcomes $(o_{s_1},o_{s_2})$ is obtained. The probability distribution associated to any outcome pair is determined by repeating the sequential measurement procedure several times. Formally, the joint probability to get $(o_{s_1},o_{s_2})$, according to the TPM scheme, is
\begin{equation}\label{eq:joint_TPM}
    p(s_1,s_2) = {\rm Tr}\left[ \Pi_{s_2}^{H}(t_2)\Pi_{s_1}(t_1)\rho\,\Pi_{s_1}(t_1) \right],
\end{equation}
where $\rho$ is the initial quantum state (prepared at time $t_1$), and $\Pi_{s_2}^{H}(t_2)=\Phi^{\dagger}[\Pi_{s_2}(t_2)]=\sum_{\alpha}K_{\alpha}^{\dagger} \Pi_{s_2}(t_2) K_{\alpha}$ is one of the projectors of $\mathcal{O}_2(t_2)$ evolved in the Heisenberg picture. For unitary dynamics, $\Pi_{s_2}^{H}(t_2) = U^{\dagger}\Pi_{s_2}(t_2) U$, where we have introduced the unitary evolution operator of the system $U \equiv \mathbb{T}\exp(-(i/\hbar)\int_{t_1}^{t_2}\mathcal{H}(t)dt)$ with $\hbar$ denoting the reduced Planck's constant, set to 1 for simplicity from now on, $\mathbb{T}$ the time-ordering operator and $\mathcal{H}(t)$ the Hamiltonian of the system at time $t$.

A procedure based on sequential projective measurements is \emph{invasive} as it violates the \emph{no-signaling in time} condition~\cite{KoflerPRA2013}. To better single out this aspect, we introduce the generic joint probability $d(s_1,s_2)$ at two times, depending on the outcome pairs $(o_{s_1},o_{s_2})$. We do not explicitly provide a specific expression for $d(s_1,s_2)$, as we aim to study the mathematical properties that any joint probability needs to satisfy assuming a (non)invasive measurement procedure. Thus, if applied to our case-study, the no-signaling in time condition states that the statistics of $(o_{s_1},o_{s_2})$ that is returned by a noninvasive measurement apparatus must fulfill the condition 
\begin{equation}\label{eq:no_signaling_in_time}
   \sum_{s_1}d(s_1,s_2) = p_{s_2}(t_2) = {\rm Tr}\left[\Pi_{s_2}^{H}(t_2)\rho\right].
\end{equation}
%
%
In other terms, the requirement for the noninvasiveness is that, marginalizing the distribution over the outcomes $s_1$ of the first observable $\mathcal{O}_{1}(t_1)$ at time $t_1$, we recover the \emph{unperturbed} single-time probability $p_{s_2}(t_2)$ associated to the outcomes of the second observable $\mathcal{O}_2(t_2)$ at $t_2$.

In this respect, noninvasiveness can be considered as a synonymous of \emph{unperturbed marginals}. The validity of Eq.~\eqref{eq:no_signaling_in_time} is a necessary condition for \emph{macrorealism}~\cite{LeggettPRL1985} and, interestingly, Eq.~\eqref{eq:no_signaling_in_time} can be violated even in situations where no violation of Leggett-Garg inequalities is allowed~\cite{KoflerPRA2013}. The violation of the no-signaling in time condition marks the main consequence of the joint measurability problem due to the incompatibility of the involved quantum operators $\rho$, $\mathcal{O}_{1}(t_1)$ and $\mathcal{O}_{2}(t_2)$.

The question that now arises is: ``What information is erased by using the TPM scheme in the attempt of attaining the statistics of $(o_{s_1},o_{s_2})$?'' or equivalently ``How invasive is a procedure based on sequential measurements?'' We are going to show that the TPM scheme is noninvasive \emph{if and only if} $[\rho,\Pi_{s_1}(t_1)]=0$ or $[\Pi_{s_1}(t_1),\Pi_{s_2}^{H}(t_2)]=0$. 
Otherwise,  Eq.~\eqref{eq:no_signaling_in_time} would be violated. In order to determine what information is erased by the TPM scheme, let us compare the final-time probability $p_{s_2}(t_2) = {\rm Tr}[\Pi_{s_2}^{H}(t_2)\rho]$ and $\sum_{s_1}p(s_1,s_2)$. The latter equals
\begin{eqnarray}
    &&\sum_{s_1}p(s_1,s_2) = {\rm Tr}\left[ \Pi_{s_2}^{H}(t_2)\sum_{s_1}\Pi_{s_1}(t_1)\rho\,\Pi_{s_1}(t_1) \right]=\nonumber \\
    &&={\rm Tr}\left[ \Pi_{s_2}^{H}(t_2)\mathcal{D}_{1}[\rho] \right],
    \label{eq:margs1}
\end{eqnarray}
where the superoperator 
\begin{equation}\label{eq:dephasing_operator}
    \mathcal{D}_{1}[\rho] \equiv \sum_{s_1} \Pi_{s_1}(t_1)\rho\,\Pi_{s_1}(t_1) = \sum_{s_1}p_{s_1}(t_1)\Pi_{s_1}(t_1)
\end{equation}
denotes the \emph{dephasing channel}, which is defined over the eigenbasis of the quantum observable $\mathcal{O}_{1}(t_1)$ and is applied to the initial density operator $\rho$. In Eq.~\eqref{eq:dephasing_operator}, 
$p_{s_1}(t_1) = {\rm Tr}[ \Pi_{s_1}(t_1)\rho ]$. Hence, if we compare $p_{s_2}(t_2)$ and $\sum_{s_1}p(s_1,s_2)$, we can see that the first measurement of the TPM scheme erases the \emph{quantum coherence} contained in $\rho$ once projected onto the eigenbasis of $\mathcal{O}_{1}(t_1)$. 

It is worth noting that the initial density operator can always be linearly decomposed in the basis of $\mathcal{O}_{1}(t_1)$ as 
\begin{equation}\label{eq:linear_decomp_rho}
    \rho = \mathcal{D}_{1}[\rho] + \chi,
\end{equation}
where 
\begin{equation}\label{eq:initial_quantum_coherence}
    \chi \equiv \sum_{s_1 \neq s_1'}\rho_{s_1,s_1'}|s_1\rangle\!\langle s_1'|
\end{equation}
is the operator containing the off-diagonal elements of $\rho$, with ${\rm Tr}[\chi]=0$.

As an example, let us consider a qubit as the quantum system and $\mathcal{O}_{1}(t_1) =\sigma^z$ as the quantum observable at time $t_1$. We have introduced the Pauli matrices $\{\sigma^x, \sigma^y,\sigma^z\}$~\footnote{
Throughout the tutorial, we recall that any Pauli matrix $\sigma$ along the $x$-, $y$- and $z$-axis (i.e., $\sigma^x$, $\sigma^y$ or $\sigma^z$) is a $2\times 2$ complex operator that is Hermitian ($\sigma^{\dagger}=\sigma$), involutory ($\sigma^{2}=\mathbb{I}$) and unitary ($\sigma^{-1}=\sigma^{\dagger}$ entailing that $\sigma^{\dagger}\sigma=\mathbb{I}$).}.
Then, 
\begin{equation}
    \chi = \rho_{0,1}|0\rangle\!\langle 1| + \rho_{1,0}|1\rangle\!\langle 0|,
    \label{eq:chispin12}
\end{equation}
where $|0\rangle$, $|1\rangle$ are the two eigenstates of $\sigma^z$, $\rho_{i,j}=\bra i \rho \ket j$, and, due to the Hermiticity of $\rho$, $\rho_{1,0}=\rho_{0,1}^{*}$. 

\subsection{No-go theorem for sequential outcomes statistics}
\label{sec:no-go_theorem}

Previously, we have outlined that a procedure based on sequential measurements fails in recovering the marginal distribution $p_{s_2}(t_2)$ from the joint TPM distribution $p(s_1,s_2)$ of the pairs $(o_{s_1},o_{s_2})$ with respect to the measurement outcomes $o_{s_{1}}$ of $\mathcal{O}_{1}(t_1)$, see Eq.~\eqref{eq:margs1} and the related discussion. This directly violates the no-signaling in time condition Eq.~\eqref{eq:no_signaling_in_time}, and establishes that the measurement procedure is invasive. The origin of such violation lies in the fact that at least one of the commutators $[\rho,\Pi_{s_1}(t_1)]$ and $[\Pi_{s_1}(t_1),\Pi_{s_2}^{H}(t_2)]$ is different from zero.

This conclusion is related to a deeper statement summarised by the \emph{no-go theorem} reported in Ref.~\cite{LostaglioQuantum2023}, which is 
less restrictive than the formulation firstly proven in Ref.~\cite{PerarnauLlobetPRL2017}. The no-go theorem states that the following three properties cannot be valid simultaneously for any initial density operator $\rho$ \emph{if and only if} $[\Pi_{s_1}(t_1),\Pi_{s_2}^{H}(t_2)] \neq 0$ for some pair $(o_{s_1},o_{s_2})$:
\begin{enumerate}
    \item[i)]
    The probability distribution of the pairs $(o_{s_1},o_{s_2})$, defined by the generic joint probabilities $d(s_1,s_2)$ in the time interval $[t_1,t_2]$, obeys the Kolmogorov's axioms of the classical theory of probability.
    \item[ii)]
    The joint probabilities $d(s_1,s_2)$ lead to \emph{unperturbed marginals}:
    \begin{eqnarray}
        \sum_{s_1}d(s_1,s_2) &=& p_{s_2}(t_2), \\
        \sum_{s_2}d(s_1,s_2) &=& p_{s_1}(t_1).
    \end{eqnarray}
    \item[iii)]
    The joint probabilities $d(s_1,s_2)$ are \emph{linear} functions of the initial density operator $\rho$. Formally, this means that, given a linear combination $\rho=\sum_{k}a_{k}\rho_{k}$, then $d(s_1,s_2,\rho) = \sum_{k}a_{k}d(s_1,s_2,\rho_k)$.
\end{enumerate}

The three properties i)-iii) are all simultaneously satisfied  under the assumption of the commutative condition $[\Pi_{s_1}(t_1),\Pi_{s_2}^{H}(t_2)] = 0$, and the probability distribution that fully characterizes the statistics of the pairs $(o_{s_1},o_{s_2})$ is the one returned by the TPM scheme.

In the following and throughout the tutorial, we will give up the property i). As a consequence we can no longer employ sequential projective measurements to characterize the statistics of $(o_{s_1},o_{s_2})$. Avoiding the direct application of the TPM scheme on the quantum system under scrutiny may completely eliminate the invasiveness of the measurement procedure and allow us to recover unperturbed marginal distributions [property ii)]. Such a requirement for the generic joint probabilities $d(s_1,s_2)$ is well justified if we want that our knowledge on the fluctuations of the pairs $(o_{s_1},o_{s_2})$ is not decreased by the quantum \emph{measurement back-action}. We therefore require that the no-signaling in time condition is fulfilled. Furthermore, we demand that the probability distribution of $(o_{s_1},o_{s_2})$ exhibits \emph{linearity}, in conformity with the property iii). In this way, for any variation of $\rho$, one does not need to repeat from scratch the experimental procedure (which, as noted earlier, should not be sequential) to determine $d(s_1,s_2)$. 

As a result, by linearly decomposing $\rho$ as in Eq.~\eqref{eq:linear_decomp_rho}, in terms of its diagonal and off-diagonal parts with respect to the eigenbasis of $\mathcal{O}_{1}(t_1)$, we recover the results of the TPM scheme whenever $\chi={\bf 0}$, with ${\bf 0}$ denoting the matrix with all zeros. 
In addition, another consequence of the linearity property is that the procedure for measuring $d(s_1,s_2)$ can be independent on the initial density operator $\rho$, as it is customary in the classical case.

\subsection{Beyond the two-point measurement scheme: quasiprobability approach}\label{sec:beyond_TPM}

Under the noncommutativity hypothesis $[\Pi_{s_1}(t_1),\Pi_{s_2}^{H}(t_2)] \neq 0$ for some pair $(o_{s_1},o_{s_2})$, dropping the property i) of the no-go theorem mentioned in Sec.~\ref{sec:no-go_theorem} allows for the introduction of a \emph{quasiprobability} distribution (QD), whose terms can be nonpositive (i.e., negative real numbers or even complex numbers), albeit still summing to $1$. In general, there is \emph{not} a \emph{unique} QD due to \emph{ordering ambiguities} in how the QD is defined (see Refs.~\cite{halliwell2016leggett,ArvidssonShukurJPA2021,PhysRevLett.127.190404,DeBievreArXiv2022,LostaglioQuantum2023}). As a consequence, there are, in principle, infinite QD that are \emph{linear} in the initial state $\rho$ and lead to unperturbed marginals, at both the initial and final times $t_1$ and $t_2$.

Let us introduce quasiprobabilities. We start from the expression for the generic joint probabilities $d(s_1,s_2)$ and assign a linear operator $M(s_1,s_2)$ to each pair $(o_{s_1},o_{s_2})$ of measurement outcomes. Without loss of generality, we can write:
\begin{equation}\label{eq:generic_d_s1_s2}
    d(s_1,s_2) = {\rm Tr}\left[ M(s_1,s_2)\rho \right].
\end{equation}
From classical probability theory, in the case of the TPM scheme [see Sec.~\ref{sec:TPM_scheme}], we find that
\begin{equation}\label{eq:M_TPM}
M(s_1,s_2) = M_{\rm TPM}(s_1,s_2) \equiv \Pi_{s_1}(t_1)\Pi_{s_2}^{H}(t_2)\Pi_{s_1}(t_1),   
\end{equation}
which returns Eq.~\eqref{eq:joint_TPM}. 
Multiplying the observable $\Pi_{s_2}^{H}(t_2)$ by the projector $\Pi_{s_1}(t_1)$ on both the left and right sides is equivalent to performing a projective measurement entailing the collapse of the measured quantity. In order to overcome this, the minimal change is to remove one projector $\Pi_{s_1}(t_1)$. We can set either
\begin{equation}
M(s_1,s_2) = M_{\rm KDQ\,1}(s_1,s_2) \equiv \Pi_{s_2}^{H}(t_2)\Pi_{s_1}(t_1),    
\end{equation}
or 
\begin{equation}
M(s_1,s_2) = M_{\rm KDQ\,2}(s_1,s_2) \equiv \Pi_{s_1}(t_1)\Pi_{s_2}^{H}(t_2).  
\end{equation}
Substituting the linear operators $M_{\rm KDQ\,1}(s_1,s_2)$ or $M_{\rm KDQ\,2}(s_1,s_2)$ in Eq.~\eqref{eq:generic_d_s1_s2} gives two perfectly valid KDQ.

The difference of applying $M_{\rm KDQ\,1}$ or $M_{\rm KDQ\,2}$ on $\rho$ is that they operate on off-diagonal terms of $\rho$ with exchanged indexes [$(o_{s_1},o_{s_2}) \longleftrightarrow (o_{s_2},o_{s_1})$]. In fact, one can compute that 
\begin{eqnarray*}
    {\rm Tr}\left[ \Pi_{s_2}^{H}\Pi_{s_1}\rho \right] &=& {\rm Re}\,{\rm Tr}\left[ \Pi_{s_2}^{H}\Pi_{s_1}\rho \right] + i\,{\rm Im}\,{\rm Tr}\left[ \Pi_{s_2}^{H}\Pi_{s_1}\rho \right]\\
    {\rm Tr}\left[ \Pi_{s_1}\Pi_{s_2}^{H}\rho \right] &=& {\rm Re}\,{\rm Tr}\left[ \Pi_{s_2}^{H}\Pi_{s_1}\rho \right] - i\,{\rm Im}\,{\rm Tr}\left[ \Pi_{s_2}^{H}\Pi_{s_1}\rho \right],
\end{eqnarray*}
whence
\begin{equation}
    {\rm Tr}\left[ \rho\,\Pi_{s_1}\Pi_{s_2}^{H} \right] = {\rm Tr}\left[\left( \Pi_{s_2}^{H}\Pi_{s_1}\,\rho \right)^{\dagger}\right] = {\rm Tr}\left[ \Pi_{s_2}^{H}\Pi_{s_1}\,\rho \right]^{*},    
\end{equation}
thus meaning that the KDQ ${\rm Tr}[ M_{\rm KDQ\,1}(s_1,s_2)\rho ]$ and ${\rm Tr}[ M_{\rm KDQ\,2}(s_1,s_2)\rho ]$ differ by their imaginary parts that are opposite in sign.

Both KDQ reduce to $p(s_1,s_2) = {\rm Tr}[ M_{\rm TPM}(s_1,s_2)\rho ]$, as in Eq.~\eqref{eq:joint_TPM}, if $[\rho,\Pi_{s_1}(t_1)] = 0$ or $[\Pi_{s_1}(t_1),\Pi_{s_2}^{H}(t_2)] = 0$. In this tutorial, without loss of generality, we will make use of the KDQ defined by $M_{\rm KDQ\,1}(s_1,s_2)$ that, from now on, we denote as
\begin{eqnarray}\label{eq:def_KDQ}
    q(s_1,s_2) &\equiv& {\rm Tr}\left[ M_{\rm KDQ\,1}(s_1,s_2)\rho \right] = \nonumber \\
    &=& {\rm Tr}\left[ \Pi_{s_2}^{H}(t_2)\Pi_{s_1}(t_1)\rho \right].
\end{eqnarray} 
The sign ambiguity in ${\rm Tr}[ M_{\rm KDQ\,1}(s_1,s_2)\rho ]$ and ${\rm Tr}[ M_{\rm KDQ\,2}(s_1,s_2)\rho ]$ can be overcome by taking the uniformly weighted sum of $M_{\rm KDQ\,1}(s_1,s_2)$ and $M_{\rm KDQ\,2}(s_1,s_2)$, i.e.,
\begin{eqnarray}
M_{\rm MHQ}(s_1,s_2) &\equiv& \frac{1}{2}\Big( M_{\rm KDQ\,1}(s_1,s_2) + M_{\rm KDQ\,2}(s_1,s_2) \Big) \nonumber \\
&=& \frac{1}{2}\left\{ \Pi_{s_2}^{H}(t_2), \Pi_{s_1}(t_1) \right\},
\end{eqnarray}
where $\{A,B\} \equiv AB + BA$ denotes the anticommutator of the generic operators $A, B$. In this way, we end up with the quasiprobability
\begin{eqnarray}
    q_{\rm MHQ}(s_1,s_2) &\equiv& {\rm Tr}\left[ M_{\rm MHQ}(s_1,s_2)\rho \right] = \nonumber \\
    &=& \frac{1}{2}{\rm Tr}\left[ \left\{ \Pi_{s_2}^{H}(t_2), \Pi_{s_1}(t_1) \right\}\rho \right] = \label{eq:q_MHQ_1} \\
    &=&{\rm Re}\,{\rm Tr}\left[ \Pi_{s_2}^{H}(t_2)\Pi_{s_1}(t_1)\rho \right] =\nonumber \\
    &=&{\rm Re}\left[ q(s_1,s_2) \right], \label{eq:q_MHQ_2}
\end{eqnarray}
commonly known as the Margenau-Hill quasiprobability (MHQ)~\cite{margenau1961correlation}.

Other quasiprobabilities have been considered in the literature; for example, for systems weakly interacting with a detector in order to avoid the invasiveness of the first measurement of the TPM scheme~\cite{RoncagliaPRL2014,DeChiaraNJP15,Solinas2016probing,SolinasPRA2017,Cerisola2017,SolinasPRA2021,SolinasPRA2022}. Given their affinity with MHQ and KDQ, we will discuss them in detail in Sec.~\ref{subsec:meas_QP}. Moreover, in Sec.~\ref{subsec:alternative_forms}, we will also give a short overview of alternative formulations of quasiprobabilities considered in the literature.

The quasiprobabilities defined in Eq.~\eqref{eq:def_KDQ} and Eq.~\eqref{eq:q_MHQ_2} fulfil properties ii) and iii) of the no-go theorem in Sec.~\ref{sec:no-go_theorem}, meaning that the no-signaling in time condition is fulfilled and the measurement procedure that allows to get a QD is independent of the initial state $\rho$. Moreover, the TPM statistics is recovered in the case in which $\rho$, $\mathcal{O}_{1}(t_1)$, $\mathcal{O}_{2}(t_2)$ all commute with each other. 

These characteristics are important requisites to build a consistent thermodynamic theory via quasiprobabilities. 
In particular, the property iii) of the no-go theorem, regarding the linearity on the initial state, is meaningful when drawing a correspondence with classical thermodynamics. In fact, any nonlinear dependence of thermodynamic quantities---work, heat and entropy---on the initial state seems to be in contradiction with their standard definition in classical thermodynamics that does not change depending on the way the phase-space distribution taken as the input ensemble is split. 

\subsubsection{nonpositivity}\label{subsec:non_positivity}

KDQ naturally encode temporal correlations between the measurement outcomes of the quantum observables $\mathcal{O}_{1}(t_1)$ and $\mathcal{O}_{2}(t_2)$. As explained in Sec.~\ref{sec:no-go_theorem}, in relation to the no-go theorem, the quasiprobabilities $q(s_1,s_2)$ can be \emph{nonpositive}, although they are still subject to the normalisation constraint
\begin{equation}\label{eq:probability_conservation}
    \sum_{s_1,s_2}q(s_1,s_2) = 1.
\end{equation}

In the case of KDQ, nonpositivity can mean the following two facts: 
\begin{itemize}
\item[I)] 
the real part of $q(s_1,s_2)$ is negative;
\item[II)] 
$q(s_1,s_2)$ is a complex number with a nonzero imaginary part.
\end{itemize}

From a mathematical point of view, the onset of nonpositivity in KDQ is a consequence of the fact that the product of two quantum observables (say, $A$ and $B$) can always be
decomposed as the linear combination of self-adjoint operators as $AB = \{A,B\}/2 + i[A,B]/(2i)$. Thus, for the product $\Pi_{s_2}^{H}(t_2)\Pi_{s_1}(t_1)$, we have 
\begin{equation}\label{eq:product_projectors}
    \Pi_{s_2}^{H}(t_2)\Pi_{s_1}(t_1) = \frac{ \left\{\Pi_{s_2}^{H},\Pi_{s_1}\right\} }{ 2 } + i\,\frac{ \left[\Pi_{s_2}^{H},\Pi_{s_1}\right] }{ 2i }\,.
\end{equation}
The first term on the right-hand-side of Eq.~\eqref{eq:product_projectors} gives rise to the MHQ---the real part of the KDQ---when evaluated (i.e., averaged) with respect to the initial density operator $\rho$. Then, looking at the second term of Eq.~\eqref{eq:product_projectors}, it is evident that a necessary condition for the KDQ to have an imaginary part is the noncommutativity of $\Pi_{s_1}(t_1)$ and $\Pi_{s_2}^{H}(t_2)$. In this regard, we stress that if $[\Pi_{s_1}(t_1),\Pi_{s_2}^{H}(t_2)]=0$, then $\Pi_{s_2}^{H}(t_2)\Pi_{s_1}(t_1) = \Pi_{s_1}(t_1)\Pi_{s_2}^{H}(t_2)\Pi_{s_1}(t_1)$, and the KDQ coincides with the TPM probabilities. 
In order to ensure the validity of Eq.~\eqref{eq:probability_conservation}, the imaginary parts of KDQ must cancel each other out.

In Sec.~\ref{sec:simple_case_study}, we will show a simple case study that directly connects the imaginary parts of the KDQ with the presence of imaginary coherence terms in the initial density operator $\rho$, with respect to the eigenbasis of $\mathcal{O}_{1}(t_1)$. Hence, if we ignored the imaginary parts of $q(s_1,s_2)$, we would exclude some information stemming from quantum coherence and correlations that may emerge in the quantum statistics of $(o_{s_1},o_{s_2})$. 

The occurrence of nonpositivity can be considered as a nonclassical feature in the statistics of the measurement pairs, underlining the presence of genuinely quantum features due to the interplay of quantum dynamics and measurement. From here on, we will refer to this with the term \emph{nonclassicality}. The formal conditions and experimental routines allowing to identify nonclassicality take also the name of \emph{quantum contextuality}~\cite{PhysRevLett.101.020401}. In a scenario involving a quasiprobability distribution describing the occurrence of a given pair of measurement outcomes at two times, an experimental protocol is defined contextual if it is able to yield nonpositive values, provided incompatibility of noncommuting operators plays a role. No classical stochastic process can explain such a behaviour. 
\paragraph{\bf noncommutativity is a necessary condition.} For the MHQ $q_{\rm MHQ}$, see Eq.~\eqref{eq:q_MHQ_2}, it has been recently shown that the pairwise noncommutativity of the initial density operator and the quantum measurement observables is only a \emph{necessary but not sufficient condition} for nonpositivity (i.e., negativity of the MHQ). This means that there are counterexamples where $[\rho,\Pi_{s_1}(t_1)] \neq 0$ and/or $[\Pi_{s_1}(t_1),\Pi_{s_2}^{H}(t_2)] \neq 0$, but still ${\rm Re}\left[ q(s_1,s_2)\right] \geq 0$~\cite{GoldsteinPRL1995}. A detailed analysis of this aspect can be found in Ref.~\cite{ArvidssonShukurJPA2021}, where it has been formulated with the wording \emph{negativity is stronger than noncommutativity}.
\paragraph{\bf Direct link with weak values.} The nonpositivity of KDQ can find a physical interpretation from their direct connection with \emph{weak values}~\cite{aharonov1988how,HofmannPRL2012,Dressel2014colloquium} that, indeed, are \emph{conditional} KDQ~\cite{LostaglioQuantum2023}. To see this, we set $\Pi_{s_2}^{H}(t_2)=|\widetilde{s}_2\rangle\!\langle \widetilde{s}_2|$, where $\ket{\widetilde{s}_2}=U\ket{s_2}$, under the hypothesis that the dynamics is unitary, and $\Pi_{s_2}^{H}(t_2)$ is a rank-1 projector. Moreover, we take $\rho=\ket\psi\bra\psi$ as a pure state. Hence, from Eq.~\eqref{eq:def_KDQ} one has that
\begin{equation}
    \frac{ q(s_1,s_2) }{ p_{s_2}(t_2) } = \frac{ \langle\psi|\widetilde{s}_2 \rangle\!\langle \widetilde{s}_2|\Pi_{s_1}(t_1)|\psi\rangle }{ |\langle\psi|\widetilde{s}_2 \rangle|^2 } = \frac{ \langle \widetilde{s}_2|\Pi_{s_1}(t_1)|\psi\rangle }{ \langle\widetilde{s}_2|\psi\rangle }\,,
\end{equation}
where 
\begin{equation}
    \frac{ \langle \widetilde{s}_2|\Pi_{s_1}(t_1)|\psi\rangle }{ \langle\widetilde{s}_2|\psi\rangle } \equiv \langle\Pi_{s_1}(t_1)\rangle_{\rm WV}
\end{equation}
is the original definition of the weak value (WV) of the projector $\Pi_{s_1}(t_1)$ with initial state $|\psi\rangle$ and postselection $|\widetilde{s}_2\rangle$. In this way, the weak value $\langle\mathcal{O}_1(t_1)\rangle_{\rm WV}$ of the observable $\mathcal{O}_1(t_1)$ is obtained by averaging the outcomes $o_{s_1}(t_1)$ over the conditional KDQ $\langle\Pi_{s_1}(t_1)\rangle_{\rm WV} = q(s_1,s_2)/p_{s_2}(t_2)$. Formally, we have that
\begin{equation}
    \langle\mathcal{O}_1(t_1)\rangle_{\rm WV} \equiv \frac{ \langle\widetilde{s}_2|\mathcal{O}_1(t_1)|\psi\rangle }{ \langle\widetilde{s}_2|\psi\rangle } = \sum_{s_1}o_{s_1}(t_1)\langle\Pi_{s_1}(t_1)\rangle_{\rm WV}\,.
\end{equation}
We recall that the weak values can be obtained via a \emph{weak measurement} that is performed on both a properly chosen preselected quantum state and a post-selected one. Weak values are called \emph{anomalous} when $\langle\mathcal{O}_1(t_1)\rangle_{\rm WV}$ lies outside the spectrum of $\mathcal{O}_1(t_1)$ (see Refs.~\cite{Johansen04,PuseyPRL2014,KunjwalPRA2019,AbbottQuantum2019}). In order to guarantee such anomaly, it is required that the (possibly mixed) pre and postselection states have quantum coherence with respect to the eigenbasis of $\mathcal{O}_1(t_1)$ and that the corresponding KDQ exhibits negativity~\cite{WagnerPRA2023}. Moreover, the generalization of weak values to mixed density operators (instead of pure quantum states), can be a complex number, and this is evidently in a one-to-one correspondence with complex KDQs~\cite{hofmann2011role,DresselPRA2012,Johansen04,KunjwalPRA2019}. Overall, the occurrence of an anomalous weak value is identified by the nonpositivity of a KDQ, implying quantum contextuality~\cite{PuseyPRL2014,KunjwalPRA2019}. 
\paragraph{\bf nonpositivity functional.}We conclude this subsection by introducing the \emph{nonpositivity functional}~\cite{AlonsoPRL2019,ArvidssonShukurJPA2021,hernandez2022experimental,LostaglioQuantum2023} 
\begin{equation}\label{eq:def_aleph}
    \aleph \equiv - 1 + \sum_{s_1,s_2}\big|q(s_1,s_2)\big|
\end{equation}
that quantifies the `amount' of nonclassicality in the statistics of the outcome pairs $(o_{s_1},o_{s_2})$. It is worth noting that both the real and imaginary parts of the KDQ contribute to the nonclassicality, whereby if present one has that 
\begin{equation}
    \sum_{s_1,s_2}\big|q(s_1,s_2)\big| > 1 \, \Rightarrow \aleph > 0.
\end{equation}
As noticed in Ref.~\cite{ArvidssonShukurJPA2021}, one could quantify the negativity and nonreality of a KDQ distribution by using the functionals
\begin{eqnarray}
    \aleph_{\rm Re} &\equiv& -1 + \sum_{s_1,s_2}\Big|{\rm Re}\left[ q(s_1,s_2) \right] \Big| \\
    \aleph_{\rm Im} &\equiv& \sum_{s_1,s_2}\Big|{\rm Im}\left[ q(s_1,s_2) \right]\Big|
\end{eqnarray}
that act separately on the real and imaginary parts of KDQ, respectively. The condition $\aleph=0$ occurs when all the KDQ are positive real numbers~\footnote{The positivity meant here goes beyond the notion of `linear positivity' introduced by Goldstein and Page~\cite{GoldsteinPRL1995},
whereby noncommutativity conditions can coexist with ${\rm Re}\left[ q(s_1,s_2)\right] \geq 0$ to some extent such that the occurrence of $(o_{s_1},o_{s_2})$ is described by positive probabilities. This is because for the KDQ also imaginary parts are taken into account, and not only the negativity of the real part of $q(s_1,s_2)$.}.

\subsubsection{Comparing KDQ and TPM probabilities}\label{sec:simple_case_study}

In this subsection we are going to compare the KDQ $q(s_1,s_2)$ with the joint probabilities $p(s_1,s_2)$ returned by applying the TPM scheme. In this regard, notice that
\begin{equation}\label{eq:difference_among_KDQ_TPM}
    q(s_1,s_2) - p(s_1,s_2) = {\rm Tr}\left[\Pi_{s_2}^{H}(t_2)\Pi_{s_1}(t_1)\rho\,\Pi_{s_1}^{\perp}(t_1)\right]\,,
\end{equation}
where 
\begin{equation}\label{eq:P_s1_orthogonal}
\Pi_{s_1}^{\perp}(t_1) \equiv \mathbb{I} - \Pi_{s_1}(t_1)    
\end{equation}
is the projector orthogonal to $\Pi_{s_1}(t_1)$. Interestingly, as discussed in Ref.~\cite{halliwell2016leggett}, Eq.~\eqref{eq:difference_among_KDQ_TPM} quantifies the \emph{interference patterns} between the two different sequential pairs of projectors, also known in the literature as \emph{quantum histories}, $(\Pi_{s_1}(t_1),\Pi_{s_2}^{H}(t_2))$ and $(\Pi_{s_1}^{\perp}(t_1),\Pi_{s_2}^{H}(t_2))$. Moreover, the right-hand-side of Eq.~\eqref{eq:difference_among_KDQ_TPM} is also recovered from the so-called \emph{nondemolition quasiprobability} (NDQP)~\cite{SolinasPRE2015,Solinas2016probing,SolinasPRA2022}
\begin{equation}\label{eq:def_NDQP}
\mathfrak{q}(s_1,s_1',s_2) \equiv {\rm Tr}\left[\Pi_{s_2}^{H}(t_2)\Pi_{s_1}(t_1)\rho\,\Pi_{s_1'}(t_1)\right]
\end{equation}
with $s_1 \neq s_1'$. The NDQP is evidently defined over three indexes: two, $s_1$ and $s_1'$ (different each other), refer to two possible measurement outcomes of the quantum observable $\mathcal{O}_{1}(t_1)$ at time $t_1$, while $s_2$ refers to $\mathcal{O}_{2}(t_2)$ as it holds for $q(s_1,s_2)$. Thus, by marginalizing over $s_1' \neq s_1$, one directly obtains the difference between the KDQ and TPM (joint) probabilities:
\begin{equation}
    q(s_1,s_2) - p(s_1,s_2) = \sum_{s_1'\neq s_1}\mathfrak{q}(s_1,s_1',s_2)\,.
\end{equation}
It can be easily observed that if ${\rm Re}\left[ q(s_1,s_2)\right]<0$, then necessarily $\sum_{s_1'\neq s_1}{\rm Re}\left[ \mathfrak{q}(s_1,s_1',s_2) \right]<0$; moreover, when the KDQ $q(s_1,s_2)$ is a complex number, also $\sum_{s_1'\neq s_1}\mathfrak{q}(s_1,s_1',s_2)$ is a complex number with the same imaginary part of $q(s_1,s_2)$.

Let us now exemplify these concepts with a simple case study. We consider a spin-$1/2$ particle, first initialized in the generic density operator $\rho$. Then, the spin of the particle is consecutively measured along two orthogonal axes, the $z$ and $x$ axis, respectively, i.e. $\mathcal{O}_{1}(t_1)=\sigma^z$ and $\mathcal{O}_{2}(t_2)=\sigma^x$. Moreover, we assume $U=\mathbb{I}$, so that the system does not evolve between $t_1$ and $t_2$. Note that, under specific conditions~\cite{MelloAIP2015}, this setup is representative of the physics underlying the well-known \emph{Stern-Gerlach} experiments. 

At the end of this quantum process, the state of the system collapses onto one of the eigenstates of $\sigma^x$, namely $|-\rangle\!\langle -|$ or $|+\rangle\!\langle +|$, with $|\pm \rangle \equiv (|0\rangle \pm |1\rangle)/\sqrt{2}$. Let us denote the eigenvalues of the observables $\sigma^z = \ketbra{0}{0}-\ketbra{1}{1}$ and $\sigma^x = \ketbra{+}{+}-\ketbra{-}{-}$ with $z_0=1$, $z_1=-1$ and $x_{+}=1$, $x_{-}=-1$, respectively. The quantum process is inherently probabilistic, due to the stochastic nature of any quantum measurement. We thus need to calculate the probabilities of obtaining the pairs of measurement outcomes $(z_k(t_1),x_j(t_2))$ measured at times $t_1$ and $t_2$ with $k\in\{0,1\}$ and $j\in\{-,+\}$, for the initial density operator $\rho$. 

As previously anticipated, if $[\rho,\mathcal{O}_{1}(t_1)] \neq 0$, then the application of the TPM scheme (i.e., sequential projective measurements) does no longer suffice. This fact is confirmed by the direct computation of the differences $q(z_k,x_j) - p(z_k,x_j)$:
\begin{eqnarray}
    q(-1,-1) - p(-1,-1) &=& - \frac{ \rho_{0,1}^{*} }{2},\label{eq:comparison_1}\\
    q(-1,+1) - p(-1,+1) &=& \frac{ \rho_{0,1}^{*} }{2},\\
    q(+1,-1) - p(+1,-1) &=& - \frac{ \rho_{0,1} }{2},\\
    q(+1,+1) - p(+1,+1) &=& \frac{ \rho_{0,1} }{2},\label{eq:comparison_4}
\end{eqnarray}
where $\sum_{k,j}p(z_k,x_j)=\sum_{k,j}q(z_k,x_j)=1$ by construction. 

From Eqs.~\eqref{eq:comparison_1}-\eqref{eq:comparison_4}, it is also apparent that at least two of the differences $q(z_k,x_j) - p(z_k,x_j)$, among all four, exhibit negative real parts whenever the initial state $\rho$ does not commute with the quantum observable $\sigma^z$ at time $t_1$. Notably, such a negativity is preserved from applying a second measurement of the observable $\sigma^x$ immediately after the first. 

Of course, in the case $\rho_{0,1}=0$, the KDQ $q(z_k,x_j)$ reduces to the TPM joint probabilities $p(z_k,x_j)$, and the no-signaling in time condition is fulfilled. In contrast, in the case $\rho_{0,1}\neq 0$, the first measurement of $\sigma^z$ required by the TPM scheme turns out to be invasive for the joint statistics of the measurement outcomes $(z_k(t_1),x_j(t_2))$. 

We now provide the average of the difference of outcomes $\Delta o = x(t_2) - z(t_1)$ (thus, $\Delta o_{j,k} = x_j(t_2) - z_k(t_1)$) that is evaluated with respect to the KDQ $q(z_k,x_j)$. We have
\begin{eqnarray}\label{eq:Delta0_example}
    \langle\Delta o\rangle &\equiv& \sum_{j,k}\left( x_j(t_2) - z_k(t_1) \right)q(z_k,x_j)=\nonumber\\
    &=& 2\left( q(-1,+1) - q(+1,-1)\right)=\nonumber\\
    &=& 1 - 2\rho_{0,0} + 2{\rm Re}\left[ \rho_{0,1} \right].
\end{eqnarray}

By setting $\rho=\mathbb{I}/2$, it holds that $\langle\Delta o\rangle=0$ that stems from having all the KDQ equal to $1/4$. This finding is in accordance with the classical theory of probability applied to our case study. In fact, if the initial density operator of the spin-$1/2$ is mixed with both elements equal to $1/2$ (i.e., the spin of the particle is initially up or down with equal probability $1/2$), then the sequence of measurement outcomes $\pm 1$ obtained from applying two mutually uncorrelated operations (i.e., the sequential projective measurement of $\sigma^z$ and $\sigma^x$) is naturally equiprobable. As a result, on average the difference of the measurement outcomes $\Delta o$ is zero. 

Let us now add quantum coherence to the initial state $\rho$ with respect to the eigenbasis of $\sigma^z$, by taking 
\begin{equation}\label{eq:qubitstateexample}
 \rho = \frac{ \mathbb{I} }{2} + \chi,   
\end{equation}
with $\chi$ defined in Eq.~\eqref{eq:chispin12}. Hence, from Eq.~\eqref{eq:Delta0_example}, we obtain $\langle\Delta o\rangle=2{\rm Re}\left[ \rho_{0,1} \right]$, meaning that a correction to the ``classical'' result $\langle\Delta o\rangle=0$ has to be included. In this case study, such a correction is directly proportional to the quantum coherence of $\rho$.

From the previous discussion, it is evident that the nonclassicality due to the incompatibility of non commuting operators is very fragile and can be erased by external noise and dissipation in an open quantum system. To show this, let us assume that the initial state of the qubit at time $t_1$ is $\rho$ and that before the second measurement the qubit undergoes a pure-dephasing (PD) channel. The latter is described by the following map (superoperator acting on $\rho$):
\begin{equation}
    \Phi_{\rm PD}(\rho) = (1-p)\rho+p \sigma^z \rho \sigma^z,
\end{equation}
which realizes a phase flip occurring with probability $p$.
If we repeat the calculations leading to Eqs.~\eqref{eq:comparison_1}-\eqref{eq:comparison_4}, then the quantum coherence in the resulting new equations is reduced by a factor $1-2p$ as an effect of the action of the noisy channel. In particular, one gets:
\begin{eqnarray}
    q(-1,-1) &=& \frac{ \rho_{1,1} -(1-2p)\rho_{0,1}^{*} }2\,,\\
    q(-1,+1) &=& \frac{ \rho_{1,1} +(1-2p)\rho_{0,1}^{*} }{2}\,,\\
    q(+1,-1) &=& \frac{ \rho_{0,0} -(1-2p)\rho_{0,1} }{2}\,,\\
    q(+1,+1) &=& \frac{\rho_{0,0}  +(1-2p)\rho_{0,1}  }{2}\,.
\end{eqnarray}

\subsubsection{Distribution and characteristic function of KDQ}\label{sec:distr_and_char-func_KDQ}

As mentioned in the previous sections, the KDQ $q(s_1,s_2)$ describes the joint probability of the outcomes' pairs $(o_{s_1},o_{s_2})$ from measuring the quantum observables $\mathcal{O}_{1}(t_1)$ and $\mathcal{O}_{2}(t_2)$ at times $t_1$ and $t_2$, initial and final times of the quantum process in analysis, with $t_1 < t_2$. 
The individual outcomes $s_1$ and $s_2$ correspond to the eigenvalues of the observables in Eq.~\eqref{eq:O_ti}.

Let us introduce the generic difference of outcomes
\begin{equation}\label{eq:def_Delta_o}
    \Delta o \equiv o(t_2) - o(t_1)
\end{equation}
such that $\Delta o_{s_1,s_2} = o_{s_2}(t_2) - o_{s_1}(t_1)$. The number of values that $\Delta o$ can take depends on the combinations of all possible measurement outcomes at $t_1, t_2$. Therefore, the KDQ distribution of $\Delta o$ is defined by
\begin{equation}
    P[\Delta o] = \sum_{s_1,s_2}q(s_1,s_2)\,\delta\left(\Delta o - \Delta o_{s_1,s_2}\right),
\end{equation}
where $\delta(\cdot)$ is the Dirac $\delta$ function. We remark that the KDQ distribution $P[\Delta o]$ is \emph{not} unique due to ordering ambiguities entailed by the noncommutativity of $\rho$, $\mathcal{O}_1(t_1)$ and $\mathcal{O}_2(t_2)$, as discussed in Sec.~\ref{sec:beyond_TPM}. We also note that the distribution of $\Delta o$ provided by the TPM scheme is 
\begin{equation}\label{eq:prob_distribution_TPM}
    P_{\rm TPM}[\Delta o] = \sum_{s_1,s_2}p(s_1,s_2)\,\delta\left(\Delta o - \Delta o_{s_1,s_2}\right),
\end{equation}
where, as before, $p(s_1,s_2)$ denotes the TPM joint probabilities.

All the information about the statistics of the outcome pairs $(o_{s_1},o_{s_2})$ is also encoded in the characteristic function of $P[\Delta o]$ defined as its Fourier transform:
\begin{eqnarray}
    \mathcal{G}(u) &=& \int_{-\infty}^\infty P[\Delta o]e^{iu\Delta o}d\Delta o =\nonumber \\
    &=& \sum_{s_1,s_2}q(s_1,s_2)e^{iu\Delta o_{s_1,s_2}} =\nonumber \\
    &=& {\rm Tr}\left[ e^{-iu\mathcal{O}_1(t_1)}\rho\,\Phi^{\dagger}\left[e^{iu\mathcal{O}_2(t_2)}\right]\right].
    \label{eq:characteristic}
\end{eqnarray}

While in principle for a Fourier transform the variable $u$ is real, it may be useful to extend Eq.~\eqref{eq:characteristic} with $u$ as a complex number, as we will see in Sec.~\ref{subsec:quantum_work}. Interestingly, both the KDQ $q(s_1,s_2)$ and the characteristic function $\mathcal{G}(u)$ are quantum correlation functions, namely they can be obtained as the expectation value of the product of two operators (not  necessarily Hermitian, but defined at two times) on the initial density operator $\rho$. In general, the distribution $P[\Delta o]$ depends on the time duration of the quantum system dynamics. Of course, the time dependence of $P[\Delta o]$ is mirrored in a time-dependent characteristic function $\mathcal{G}(u)$. Both for $P[\Delta o]$ and $\mathcal{G}(u)$, the time-dependence is omitted, unless specified, to enhance the clarity of the presentation.

The characteristic function can be used to detect complex values in the KDQs. In fact, a violation of the equality $\mathcal{G}(-u)=\mathcal{G}(u)^*$ implies that the positive-semi-definite condition for $q(s_1,s_2)$ does no longer hold~\cite{LostaglioQuantum2023}. The equality $\mathcal{G}(-u)=\mathcal{G}(u)^*$ is violated only when $\Im[q(s_1,s_2)] \neq 0$, and such a violation serves as a witness of complex values in the KDQs, which are identified by the nonpositivity functional.

\subsection{Alternative formulations}
\label{subsec:alternative_forms}

We conclude this section of the tutorial, dedicated to the introduction of quasiprobabilities, by mentioning  other possible formulations, alternative to the ones already discussed before, able to identify the presence of nonclassical temporal correlations.

The first formulation to consider (even for historical reasons) are {\it phase-space distributions}. In quantum mechanics, particularly in quantum optics, phase-space distributions represent the state of a quantum system, e.g., a light mode or a spin ensemble, using quasiprobability distributions like the Wigner function~\cite{WignerPR1932}. These distributions combine position and momentum, providing a comprehensive framework to analyze quantum states, coherence, and entanglement in optical systems~\cite{ShePR1966,LeePR1995,kenfack2004negativity,Vogel2006,Schleich,KrummPRA2017}. It is important to remark that phase-space distributions describe the state of a quantum system rather than a quantum process involving two measurements at distinct time, as we have described in this tutorial. Negativity in phase-space distributions signals quantum coherence in the state and has been used as a quantifier of nonclassical states.

It is also worth mentioning the Keldysh-ordered {\it full counting statistics} (FCS)~\cite{LevitovJMP1996,NazarovEPJB2003,ClerkPRA2011,HoferPRL2016}. It has been originally introduced to study the fluctuations of a time-integrated quantum observable, like current fluctuations in quantum electronic conductors averaged over a given time interval. Formally, if $\mathcal{O}^{H}(t)$ is an observable in the Heisenberg picture at time $t$, the FCS corresponds to the probability distribution $P_{\tau}[I]$ of the time integral $I \equiv \int_{0}^{\tau} \mathcal{O}^{H}(t) \, dt$ of $\mathcal{O}^{H}(t)$. Notice that the probability distribution of $I$ depends on the time interval $\tau$. In a similar fashion of a KDQ or NDQP distribution, $P_{\tau}[I]$ can be obtained by means of the inverse Fourier transform of the moment generating function for $I$, namely
\begin{eqnarray}
    \mathcal{G}_{I}(\lambda) &\equiv& \int_{-\infty}^\infty P_{\tau}[I] e^{-i\lambda I}dI =\nonumber \\
    &=& {\rm Tr}\left[ e^{-i\mathcal{H}_{\lambda}\tau} \rho \, e^{i\mathcal{H}_{-\lambda}\tau} \right],
\end{eqnarray}
where $\mathcal{H}_{\lambda} \equiv \mathcal{H} + \lambda I/2$ with $\mathcal{H}$ denoting the Hamiltonian of the quantum system that is initialized in $\rho$. The quasiprobability distribution $P_{\tau}[I]$ is not always positive, signaling nonclassicality~\cite{HoferPRL2016}. In fact, from the unravelling of the FCS in the spirit of Feynman's path-integral approach, the negativity in $P_{\tau}[I]$ results from the {\it interference} of the products of probability amplitudes that comprise the distribution $P_{\tau}[I]$. In this context, each product of probability amplitudes forms a discrete trajectory that is weighted by the elements of the initial density operator corresponding to the initial position of each trajectory. If this kind of interference between pairs of trajectories is absent, then $P_{\tau}[I]$ is positive definite.

The unravelling of the probability distribution $P_{\tau}[I]$~\cite{HoferPRL2016} has marked similarities with the derivation of the nondemolition quasiprobability distribution~\cite{Solinas2016probing,SolinasPRE2015} in Sec.~\ref{sec:simple_case_study} that, we recall, identifies the interference between pairs of sequential measurement projectors. Moreover, the unravelling of the FCS is also at the basis of the so-called \emph{Keldysh quasiprobabilities}~\cite{Hofer2017quasiprobability,PottsPRL2019}. In fact, the Keldysh formalism applied for investigating fluctuations of noncommuting operators extends both the phase-space distributions in quantum optics and the FCS of a time-integrated quantum observable. This is evident from the fact that the Keldysh quasiprobability distribution coincides with the Wigner function when position and momentum operators are considered, and it reduces to the FCS when we are interested in an observable integrated over time. Interestingly, what allows for this extension is to account for the back-action exerted by a detector measuring the two-commuting observables. Also this feature is shared by nondemolition quasiprobabilities that can be attained using a detector-assisted measurement scheme, as we will detail in Sec.~\ref{sec:det_assisted_meas}. We conclude by noting that, similarly to the KDQ, the negativity of a Keldysh quasiprobability distribution requires the noncommutativity of the initial density operator with the first measurement observable, or the noncommutativity of the operators measured at distinct times.

\section{Measuring quasiprobabilities}\label{subsec:meas_QP}

In this section, we are going to present two approaches that allows the reconstruction of a QD: the first is based on  performing only projective measurements~\cite{johansen2007quantum,hernandez2022experimental}, while the second (based on interferometry or assisted by a detector) is aimed at  measuring directly the characteristic function of the QD under scrutiny. More than these two approaches have been formulated so far to achieve such a reconstruction~\cite{hofmann2010complete,lundeen2011direct,LundeenPRL2012,Buscemi2013direct,Buscemi_2014,thekkadath2016direct}; the reader can find more details in Ref.~\cite{LostaglioQuantum2023} where alternative methods have been surveyed and some of them extended. Moreover, it is also worth mentioning Ref.~\cite{wagner2023quantum} that investigates the use of quantum circuits for the measurement of weak values and KDQ distributions.

\subsection{Weak two-point measurement scheme}\label{sec:weak_TPM}

The real part of the KDQ distribution, defined in Sec.~\ref{sec:beyond_TPM} as the Margenau-Hill (MH) distribution, can be determined by resorting only to a scheme entirely based on projective measurements. We have already proved that a procedure directly using sequential projective measurements cannot carry out this task. Instead, the combination of projective measurement schemes accomplishes the task. This is indeed enabled by the weak two-point measurement (wTPM) scheme for the measurement of quantum time correlators~\cite{johansen2007quantum,LostaglioQuantum2023}. The main feature of the wTPM scheme is to cancel the measurement back-action, thus attaining the back-action-free limit and restoring a condition of no measurement invasiveness~\cite{RyuPRA2013}. 

As noticed in Ref.~\cite{hernandez2022experimental}, the wTPM scheme can be effectively seen as a \emph{probabilistic error cancellation technique}, a technique largely employed in quantum computing from sampling noisy circuits~\cite{cai2022quantum}.

Let us consider the MHQ $q_{\rm MHQ}(s_1,s_2) = {\rm Re}\left[\,{\rm Tr}[ \Pi_{s_2}^{H}(t_2)\Pi_{s_1}(t_1)\rho ] \right]$ and the wTPM probability:  
\begin{eqnarray}\label{eq:wTPM_prob}
   w(s_1,s_2) &\equiv& {\rm Tr}\Big[ \Pi_{s_2}^{H}(t_2)\Big( \Pi_{s_1}(t_1)\rho\,\Pi_{s_1}(t_1)+\nonumber\\             &+& \Pi_{s_1}^{\perp}(t_1)\rho\,\Pi_{s_1}^{\perp}(t_1) \Big) \Big],   
\end{eqnarray}
where $\Pi_{s_1}^{\perp}(t_1)$ has been defined in Eq.~\eqref{eq:P_s1_orthogonal}. The wTPM probability has a clear physical meaning and can be obtained via a proper measurement procedure. In fact, the transformation
\begin{equation}\label{eq:non_selective_meas}
    \rho \, \longrightarrow \, \Pi_{s_1}(t_1)\rho\,\Pi_{s_1}(t_1) + \Pi_{s_1}^{\perp}(t_1)\rho\,\Pi_{s_1}^{\perp}(t_1) 
\end{equation}
is associated to the events ``the outcome $o_{s_1}$ is recorded'' or ``the outcome $o_{s_1}$ is not recorded'', both at the initial time $t_1$. For this reason, being given by a binary measurement result, the transformation Eq.~\eqref{eq:non_selective_meas} is denoted as \emph{nonselective measurement}, and applies to a given projector of the quantum observable of interest---in this case, the projector $\Pi_{s_1}(t_1)$ of $\mathcal{O}_{1}(t_1)$.

We introduced the wTPM probability because one can infer the MHQ from $w(s_1,s_2)$. To see this, we just need to substitute Eq.~\eqref{eq:P_s1_orthogonal} in Eq.~\eqref{eq:wTPM_prob}, and write the explicit expression of $w(s_1,s_2)$ as a function of $\Pi_{s_1}(t_1)$; we get: \begin{equation}
    w(s_1,s_2) = 2p(s_1,s_2) + p_{s_2}(t_2) - 2q_{\rm MHQ}(s_1,s_2),
\end{equation}
with the result that
\begin{equation}\label{eq:infer_q_MHQ}
    q_{\rm MHQ}(s_1,s_2) = p(s_1,s_2) + \frac{1}{2}\Big( p_{s_2}(t_2) - w(s_1,s_2) \Big).
\end{equation}
Eq.~\eqref{eq:infer_q_MHQ} is the way the MHQ can be experimentally reconstructed via the wTPM scheme, as done e.g.~in Ref.~\cite{hernandez2022experimental} where a pictorial representation of the scheme is provided. In fact, the TPM joint probability $p(s_1,s_2)$ can be obtained via a procedure of sequential projective measurements, and $p_{s_2}(t_2)$ is the unperturbed single-time probability to measure one of the outcomes $o_{s_2}(t_2)$ of $\mathcal{O}_{2}(t_2)$ at the final time $t_2$. Finally, the wTPM probability $w(s_1,s_2)$ is returned via a procedure based on nonselective projective measurements, as already explained above.

Notice that the probability $p_{s_2}(t_2)$ also enters the so-called \emph{end-point measurement} (EPM) scheme~\cite{GherardiniPRA2021,HernandezGomez_entropy_coherence} that, by construction, singles out the presence of quantum coherence in the initial state $\rho$ by performing single measurements at the end of the quantum process under scrutiny. A discussion about the conceptual difference of the KDQ and the joint probabilities stemming from the EPM scheme can be found in Ref.~\cite{LostaglioQuantum2023}.

We conclude this subsection by observing that, for qubits, the expression of $w(s_1,s_2)$ simplifies. This is because
\begin{equation}
    \Pi_{s_1}(t_1)\rho\,\Pi_{s_1}(t_1) + \Pi_{s_1}^{\perp}(t_1)\rho\,\Pi_{s_1}^{\perp}(t_1) = \mathcal{D}_{1}[\rho]\,,
\end{equation}
where $\mathcal{D}_{1}[\rho]$ is the dephasing superoperator defined in Eq.~\eqref{eq:dephasing_operator}. As a result, the wTPM probability reduces to the marginal of the TPM joint probability $p(s_1,s_2)$ over the outcomes $s_1$ of the initial observable, i.e.,
\begin{equation}
    w(s_1,s_2) = {\rm Tr}\left[ \Pi_{s_2}^{H}(t_2)\mathcal{D}_{1}[\rho] \right] = \sum_{s_1}p(s_1,s_2),
\end{equation}
such that
\begin{eqnarray}
    q_{\rm MHQ}(s_1,s_2) &=& p(s_1,s_2) + \frac{1}{2}\Big( p_{s_2}(t_2) - \sum_{s_1}p(s_1,s_2) \Big)=\nonumber\\
    &=& p(s_1,s_2) + \frac{1}{2}{\rm Tr}\left[ \Pi_{s_2}^{H}(t_2)\chi \right],
\end{eqnarray}
where $\chi$, defined in Eq.~\eqref{eq:initial_quantum_coherence}, contains the quantum coherence in $\rho$.

\subsection{Interferometric scheme}\label{sec:interferometry}

Another approach for the inference of the KDQ distribution $P[\Delta o]$, which we consider in this tutorial, is an \emph{interferometric scheme}. This method consists in encoding on an auxiliary system $\mathcal{A}$ the real and imaginary parts of the characteristic function $\mathcal{G}(u)$ of $P[\Delta o]$ for a given quantum system $\mathcal{S}$. The use of an auxiliary system allows one to infer both the real and imaginary parts of the KDQ by implementing a unique scheme. As explained in Sec.~\ref{sec:weak_TPM}, if our aim is to reconstruct only the real part of a KDQ, we can resort to a  procedure that is only based on projective measurements.

The interferometric scheme we are going to present here is a simplified variant of the theoretical proposals discussed in Refs.~\cite{dorner2013extracting,mazzola2013measuring,mazzola2014detecting,DeChiara2018Chapter,LostaglioQuantum2023}, and has similarities with the experimental schemes employed in Refs.~\cite{BatalhaoPRL2014,CamatiPRL2016}. It has been recently realized in Ref.~\cite{hernandezArXiv2024Interfero} using an electron-nuclear spin system associated with a nitrogen-vacancy center in diamond.
However, all these interferometers lead to the same result, namely the direct measurement of the characteristic function $\mathcal{G}(u)$. Notably, the observed $\mathcal{G}(u)$ can belong to both a probability distribution stemming from a procedure of sequential projective measurements (thus, a TPM distribution), and a quasiprobability one.

When the auxiliary system $\mathcal{A}$ is taken as a qubit, the real and imaginary parts of $\mathcal{G}(u)$, can be extracted from the expectation values of two Pauli matrices with respect to the state of $\mathcal A$ at the end of the scheme. As it will be clearer later, in order to implement the interferometer, $u$ is taken as a \emph{real number} with the dimension of a time $t$. By collecting several values of the pairs $\left( {\rm Re}[\mathcal{G}], {\rm Im}[\mathcal{G}] \right)$ for different $u$, we can reconstruct the (quasi)probability distribution $P[\Delta o]$ by applying the inverse Fourier transform to $\mathcal{G}$. The Fourier transform is performed numerically, and hence is subject to finite-time and finite-resolution constraints; see for example Ref.~\cite{plonka2019numerical}.

Let us present the interferometric scheme for quantum systems subject to unitary dynamics by assuming $\mathcal A$ is a qubit. The extension to open quantum systems, i.e., nonunitary dynamics, is straightforward through the substitution of the unitary operator with a CPTP map $\Phi$, as long as the environment does not affect the auxiliary system. 
\begin{figure}[t]
\begin{center}
\includegraphics[width=\columnwidth]{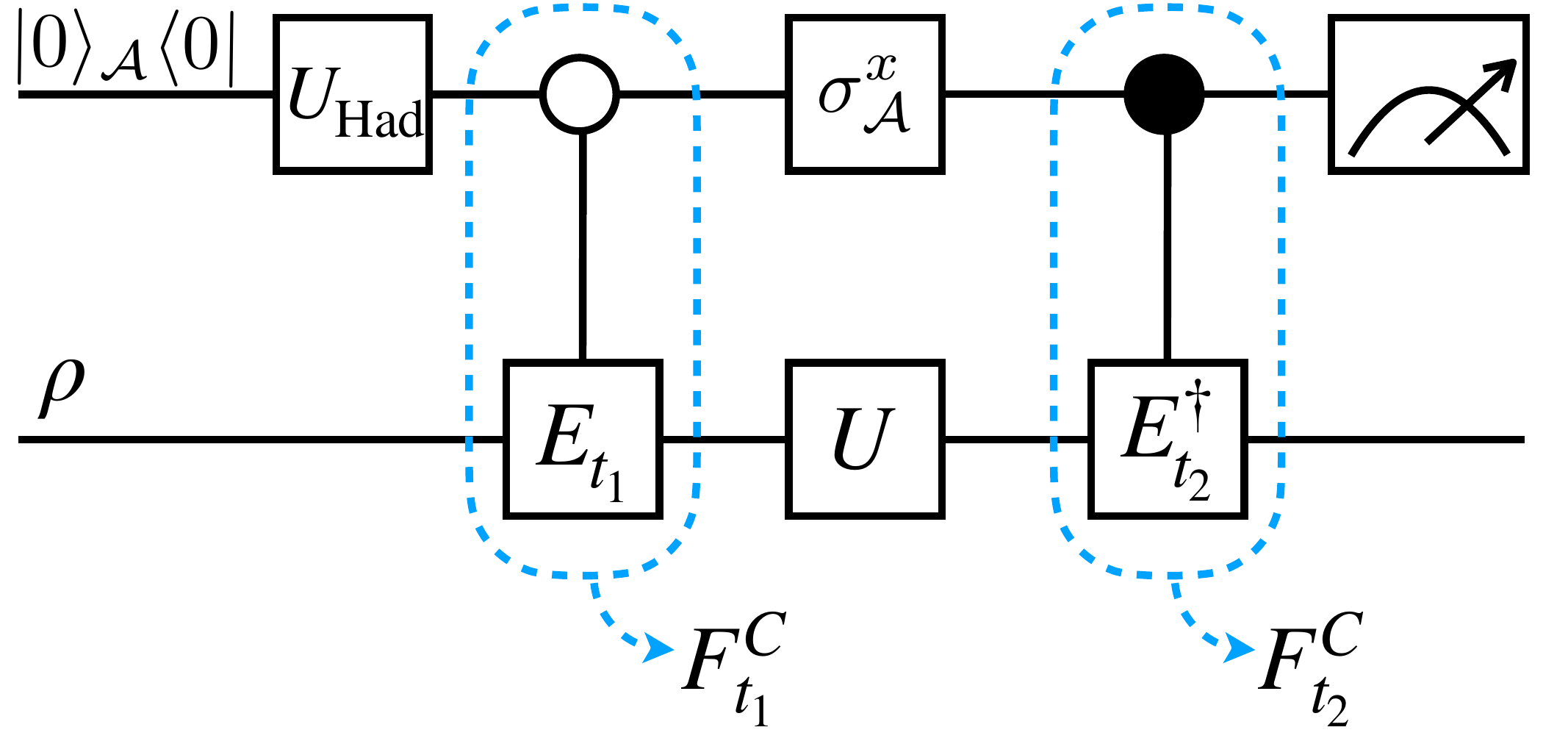}
\caption{
Pictorial representation of the interferometric scheme to directly access the characteristic function $\mathcal{G}(u)$ of a KDQ distribution $P[\Delta o]$. The scheme encodes the information on the real and imaginary parts of $\mathcal{G}(u)$ associated to the quantum system $\mathcal{S}$ of interest that is initialized in the generic density operator $\rho$. Such an encoding is operated on the auxiliary system $\mathcal{A}$ via the two conditional gates $F_{t_1}^{C}$ and $F_{t_2}^{C}$, which applies the operations $E_{t_1}$ and $E_{t_2}^{\dagger}$ on $\mathcal{S}$ whether $\mathcal{A}$ is in $|0\rangle_{\mathcal{A}}$ (white dot in the figure) or $|1\rangle_{\mathcal{A}}$ (black dot) respectively. The other gates involved are the Hadamard gate $U_{\rm Had}$, the system's evolution operator $U$ and the gate   $\sigma^x_{\cal A}$ for the auxiliary system. A detector-like box on $\cal A$ denotes its final measurement. 
}
\label{fig:interferometric_scheme}
\end{center}
\end{figure}

As pictorially represented in Fig.~\ref{fig:interferometric_scheme}, the working principle of the scheme is to initialize the auxiliary system $\mathcal{A}$ in the state $|0\rangle_{\mathcal{A}}\langle 0|$ where we denote with $\ket{i}_{\mathcal A}$ the eigenstates of the Pauli matrix $\sigma^z_{\cal A}$ for the auxiliary system $\mathcal A$ ($i=0,1$). We then perform a Ramsey interferometric scheme on $\mathcal{A}$~\cite{RamseyPR1950,foot2005atomic}. Between the application of both the Hadamard $U_{\rm Had}=2^{-1/2}(\sigma^x_{\cal A}+\sigma^z_{\cal A})$ and  $\sigma^x_{\cal A}$ gates, and the final projective measurement of $\mathcal{A}$ (with respect to the observables $\sigma^x_{\mathcal{A}}$, $\sigma^y_{\mathcal{A}}$), the auxiliary system interacts with the quantum system $\mathcal{S}$ via the conditional (C) gates
\begin{eqnarray}
    F_{t_1}^{C} &\equiv& E_{t_1} \otimes |0\rangle_{\mathcal{A}}\langle 0| + \mathbb{I}_{\mathcal{S}} \otimes |1\rangle_{\mathcal{A}}\langle 1|, \\
    F_{t_2}^{C} &\equiv& \mathbb{I}_{\mathcal{S}} \otimes |0\rangle_{\mathcal{A}}\langle 0| + E_{t_2}^{\dagger} \otimes |1\rangle_{\mathcal{A}}\langle 1|,
\end{eqnarray}
where 
\begin{eqnarray}
    E_{t_j} &\equiv& \exp\left( -i\,\mathcal{O}_{j}(t_j)u \right),\quad j=1,2.
\end{eqnarray}
The latter can be thought as unitary evolution operators corresponding to the effective Hamiltonian $\mathcal{O}_{j}(t_j)$ for a time $u$.
Moreover, between the two conditional gates $F_{t_1}^{C}$ and $F_{t_2}^{C}$, the quantum system $\mathcal{S}$ undergoes the actual unitary dynamics of the process ruled by the evolution operator $U$.

The result of implementing the interferometric scheme of Fig.~\ref{fig:interferometric_scheme} is that the reduced state $\rho_{\mathcal{A}}'$ of $\mathcal{A}$, before the final measurement of $\sigma^x_{\mathcal{A}}$ and $\sigma^y_{\mathcal{A}}$, is
\begin{eqnarray}
     \rho_{\mathcal{A}}' &=& \frac{1}{2}\mathbb{I}_{\mathcal{A}} + \frac{1}{4}\mathcal{G}(u)\left( \sigma^x_{\mathcal{A}} -i\,\sigma^y_{\mathcal{A}} \right) + \frac{1}{4}\mathcal{G}^{*}(u)\left( \sigma^x_{\mathcal{A}} + i\,\sigma^y_{\mathcal{A}} \right) = \nonumber \\
     &=& \frac{1}{2}\mathbb{I}_{\mathcal{A}} + \frac{1}{2}{\rm Re}\left[\mathcal{G}(u)\right]\sigma^x_{\mathcal{A}} + \frac{1}{2}{\rm Im}\left[\mathcal{G}(u)\right]\sigma^y_{\mathcal{A}}\,. 
\end{eqnarray}
In this way, we can obtain:
\begin{eqnarray}
    {\rm Re}\left[ \mathcal{G}(u) \right] &=& \langle \sigma^x_{\mathcal{A}}\rangle (u) = {\rm Tr}\left[ \rho_{\mathcal{A}}'\sigma^x_{\mathcal{A}} \right],\label{eq:ReG_interfer} \\
    {\rm Im}\left[ \mathcal{G}(u) \right] &=& \langle \sigma^y_{\mathcal{A}}\rangle (u) = {\rm Tr}\left[ \rho_{\mathcal{A}}'\sigma^y_{\mathcal{A}} \right].\label{eq:ImG_interfer}
\end{eqnarray}
In Sec.~\ref{sec:3examples}, we will illustrate how the interferometric scheme works in a simple qubit case. 

\subsection{Detector-assisted measurement of quasiprobabilities}
\label{sec:det_assisted_meas}

Here we provide a more general framework for the measurement of quasiprobabilities considering a quantum model for a detector coupled to the system of interest. This framework extends the TPM and the Ramsey scheme, by realizing that the observation of the change of an observable $\mathcal{O}(t)$ can be attained not through von Neumann projective measurements, but rather via a \emph{generalized measurement} or, more precisely, a \emph{positive operator-valued measure} (POVM). This was first introduced by Roncaglia et al.~in Ref.~\cite{RoncagliaPRL2014} to assess the thermodynamic work done on a quantum system (see Sec.~\ref{sec:QTD}), while a proposal for its measurement in cold atoms was reported immediately after in Ref.~\cite{DeChiaraNJP15}. Moreover, an experimental realization of the POVM to measure quantum work done on a Bose-Einstein condensate is in Ref.~\cite{Cerisola2017}. In a series of papers, Solinas and his collaborators formalized this approach establishing a profound connection between fluctuations of quantum observables, quasiprobabilities and the full counting statistics~\cite{SolinasPRE2015,Solinas2016probing,SolinasPRA2017, SolinasPRA2021,SolinasPRA2022}.

We will describe two possible schemes: one that provides access to the characteristic function of the distribution associated to $\Delta o$, and the other providing directly the (quasi)probability distribution~\cite{Solinas2016probing}.

Let us imagine a system coupled to a detector that is modelled as a quantum free particle moving in one dimension. The detector is described by the canonical position $X$ and momentum $P$ operators. We assume that the system-detector interaction Hamiltonian is
\begin{equation}
    H_{SD} = -b(t) \, P \otimes\, \mathcal{O}(t)\,,
\end{equation}
where the time-dependent coupling constant $b(t) = \kappa [\delta(t-t_2)-\delta(t-t_1)]$ is such that the detector is impulsively coupled to the system only at times $t_1$ and $t_2$ with strength $\kappa$. The operator $\mathcal{O}(t)$ is the observable to be measured and, without loss of generality, can be thought of being $\mathcal{O}_1(t_1)$ at time $t_1$ and $\mathcal{O}_2(t_2)$ at time $t_2$, as formalized in Sec.~\ref{section_quasiprobs}.

In the same spirit of what we discussed above, we consider the initial state of the system to be $\rho$ and that of the detector to be pure, i.e.,
\begin{equation}
    \ket{\phi_D} = \int_{-\infty}^\infty dp \;G(p) \ket p,
\end{equation}
where $\ket p$ are eigenstates of the momentum operator with eigenvalue $p$ and $G(p)$ is the initial momentum distribution of the detector.
Moreover, for simplicity, let us assume the system to evolve with the unitary operator $U$ between times $t_1$ and $t_2$. The extension to the case of a non unitary evolution described by a CPTP map is straightforward. The detector may also evolve during these times, but the effect of this free evolution can be very small or compensated, and  we will therefore ignore it~\cite{Solinas2016probing}.

The final state of the detector, after tracing out the state of the system, can be cast in the following two equivalent forms, with $\Delta o_{s_1,s_2} = o_{s_2}(t_2) - o_{s_1}(t_1)$ [see Eq.~\eqref{eq:def_Delta_o}]:
\begin{widetext}
\begin{eqnarray}
    \rho_D(t_2) &=&  \sum_{s_1,s_1',s_2}
                     \int_{-\infty}^\infty dp \int_{-\infty}^\infty dp' 
                     \mathfrak{q}(s_1,s_1',s_2)
                     G(p)G^*(p')
                     e^{i \kappa p  \Delta o_{s_1, s_2}}
                     e^{-i \kappa p' \Delta o_{s_1', s_2}}
                     \ketbra{p}{p'} = \label{eq:rhoDp}
                \\
                 &=& \sum_{s_1,s_1',s_2}
                     \int_{-\infty}^\infty dx \int_{-\infty}^\infty dx' 
                     \mathfrak{q}(s_1,s_1',s_2)
                     g(x-\kappa  \Delta o_{s_1, s_2})
                     g^*(x'-\kappa \Delta o_{s_1', s_2}) 
                     \ketbra{x}{x'},      \label{eq:rhoDx}      
\end{eqnarray}    
\end{widetext}
where we have used the expression $\mathfrak{q}$ of the NDQP, see Eq.~\eqref{eq:def_NDQP}. The last two equations are written in the detector momentum and position representations respectively, whereby we have introduced the detector position distribution $g(x)$ as the Fourier transform of $G(p)$.

Some considerations are in order: since the detector momentum $P$ is a conserved quantity (as it commutes with the interaction Hamiltonian), the sole effect of the evolution is to induce phase shifts in the momentum eigenstates $\ket p$. By measuring these phase shifts we access information about the observable change $\Delta o$. On the other hand, the evolution operator  $\exp{-i \kappa P \Delta o}$ associated with the system-detector interaction is effectively a translation or displacement operator of the detector position. Therefore, the quantities $\Delta o$ can be also observed by measuring the detector position distribution. In the following, we will describe two schemes that follow these two ideas, respectively.

In the first scheme, the detector is initially prepared in a superposition of two states with opposite momenta of magnitude $p_0$:
\begin{equation}
    \ket{\Phi_D} = A\left( \ket{p_0/2}+\ket{-p_0/2} \right),
\end{equation}
where $A$ is a real normalization constant~\footnote{If the states $\ket{\pm p_0/2}$ are the infinitely localised eigenstates of the detector momentum operator $P$, the state $\ket{\Phi_D}$ is not normalizable. However, we can assume that $\ket{\pm p_0/2}$ are normalized narrow wave packets with negligible overlap, in which case $A=2^{-1/2}$.}. This corresponds to a momentum distribution $G(p)=A[\delta(p-p_0/2)+\delta(p+p_0/2)]$. After the evolution, the state of the detector can be written as
\begin{eqnarray}
    \rho_D(t_2)= A^2 \Big( \ketbra{p_0/2}{p_0/2} + \ketbra{-p_0/2}{-p_0/2} + \nonumber \\
    + e^{i\phi}\ketbra{p_0/2}{-p_0/2} + e^{-i\phi}\ketbra{-p_0/2}{p_0/2} \Big),
\end{eqnarray}
where information about the dynamics is contained in the phase $\phi$ given that $P$ is a conserved quantity (see also considerations above).
If we now measure the phase $\phi$ accumulated during the whole evolution between the eigenstates $\ket{p_0/2}$ and $\ket{-p_0/2}$, we have access to a modified characteristic function:
\begin{eqnarray}\label{eq:exp_i_phi}
    \tilde{\mathcal G} &\equiv& e^{i\phi} = \sum_{s_1,s_1',s_2} \mathfrak{q}(s_1,s_1',s_2)
    \times \nonumber \\
    &\times& \exp{i \kappa p_0  \left[o_{s_2}(t_2)-\frac{o_{s_1}(t_1)+o_{s_1'}(t_1)}{2}\right]} \,,
\end{eqnarray}
which resembles the characteristic function $\mathcal{G}(u)$ defined in Eq.~\eqref{eq:characteristic}, but this time computed over the NDQP of Eq.~\eqref{eq:def_NDQP}. Hence, to all effects, Eq.~\eqref{eq:exp_i_phi} represents the characteristic function of a nondemolition quasiprobability distribution. As already outlined at the level of quasiprobabilities in Sec.~\ref{sec:simple_case_study}, $\tilde{\mathcal G}$ is a symmetric version of a KDQ characteristic function, where the symmetrization is done over the indexes $s_1$ and $s_1'$ labelling the outcomes of the initial observable $\mathcal{O}_{1}(t_1)$. It is interesting to notice that Eq.~\eqref{eq:exp_i_phi} can also be obtained by the trace of Eq.~(26) in Ref.~\cite{EspositoRMP2009}.
Thus, the inverse Fourier transform of $\tilde{\mathcal G}$ 
returns the NDQP distribution
\begin{eqnarray}
    && P_{\rm NDQP}[\Delta o] = \sum_{s_1,s_1',s_2} \mathfrak{q}(s_1,s_1',s_2) \times \nonumber \\
    && \times \delta\left\{\Delta o -  \left[o_{s_2}(t_2)-\frac{o_{s_1}(t_1)+o_{s_1'}(t_1)}{2}\right]\right\}
\end{eqnarray}
that is real (since $\mathfrak{q}(s_1,s_1',s_2)+\mathfrak{q}(s_1',s_1,s_2)$ is real) but can assume negative values due to the non commutativity of $\rho, \mathcal{O}(t_1)$ and $\mathcal{O}(t_2)$. When the initial state of the system $\rho$ does not have any coherence in the basis of eigenstates of $\mathcal O(t_1)$, then $\rho_{s_1 s_1'}=0$ for $s_1\neq s_1'$ and the inverse Fourier transform of $\tilde{\mathcal G} $ reduces to the TPM probability distribution, see Eq.~\eqref{eq:joint_TPM}.

In the second scheme, we analyze directly Eq.~\eqref{eq:rhoDx}, which leads to the position probability distribution for the detector:
\begin{eqnarray}\label{eq:Pxdet}
    && P(x) = \bra x \rho_D \ket x = \\
    &&  =\sum_{s_1,s_1',s_2} \mathfrak{q}(s_1,s_1',s_2)
         g(x-\kappa  \Delta o_{s_1, s_2})
         g^*(x-\kappa \Delta o_{s_1', s_2}) , \nonumber
\end{eqnarray}
that is real and never negative as it derives from the expectation value of a Hermitian and positive semi-definite density operator. If we assume an initially localized detector position, $g(x)=\delta(x)$, then 
$$\delta(x-\kappa\Delta o_{s_1, s_2})
 \delta(x-\kappa \Delta o_{s_1', s_2}) = \delta(x-\kappa  \Delta o_{s_1, s_2}) \delta_{s_1,s_1'},$$
and Eq.~\eqref{eq:Pxdet} reduces to
\begin{equation}
   P(x) =  \sum_{s_1,s_2} \rho_{s_1,s_1} p(s_1,s_2) \delta(x-\kappa\Delta o_{s_1,s_2}),
\end{equation}
that corresponds to the TPM probability distribution for $x=\kappa\Delta o$.
In contrast to the case in which $g(x)$ is delocalised, in this case there is a unique relation connecting $x$ and $\Delta o$ allowing perfect reconstruction of the TPM probability distribution for $\Delta o$ from the statistics of the detector's position $X$.

Even though $P(x)$ in Eq.~\eqref{eq:Pxdet} is real and positive semi-definite, effects due to initial quantum coherences can manifest themselves when the detector's initial wave function $g(x)$ is not localized and has a width comparable or larger than the typical changes $\kappa\Delta o_{s_1,s_1'}$. In fact, imagine that $\rho_{s_1,s_1'}\neq 0$, then in Eq.~\eqref{eq:Pxdet} the functions $g(x-\kappa  \Delta o_{s_1, s_2})$ and $g^*(x-\kappa \Delta o_{s_1', s_2})$ may have an overlap that results in a modification of the position probability distribution if compared to the one provided by the TPM scheme.

\subsection{Examples}\label{sec:3examples}

We conclude this section by providing examples of the quasiprobability distributions obtained using the schemes presented above. 

\subsubsection{Weak-TPM scheme}

Let us start with an application of the weak two-point measurement scheme to a three-level system (or a spin-$1$) that is initialized in a generic density operator $\rho$, and whose spin is consecutively measured along the orthogonal axes $z$ and $x$. In particular, we take
\begin{eqnarray*}
\mathcal{O}_{1}(t_1)&=&S_z=\begin{pmatrix}
    1 & 0 & 0 \\
    0 & 0 & 0 \\
    0 & 0 & -1 
\end{pmatrix}\\ 
\mathcal{O}_{2}(t_2)&=&S_x=\begin{pmatrix}
    0 & \frac{1}{\sqrt{2}} & 0 \\
    \frac{1}{\sqrt{2}} & 0 & \frac{1}{\sqrt{2}} \\
    0 & \frac{1}{\sqrt{2}} & 0 
\end{pmatrix}
\end{eqnarray*}
that share the same set of eigenvalues $o_{s_{1}}(t_1),o_{s_{2}}(t_2) = -1,0,1$, with eigenvectors 
\begin{equation*}
\left\{ |\phi_z^{(-1)}\rangle, |\phi_z^{(0)}\rangle, |\phi_z^{(1)}\rangle\right\} = \left\{ \begin{pmatrix} 0 \\
0 \\
1
\end{pmatrix},
\begin{pmatrix} 0 \\
1 \\
0
\end{pmatrix},
\begin{pmatrix} 1 \\
0 \\
0
\end{pmatrix}
\right\}
\end{equation*}
and 
\begin{equation*}
\Big\{ |\phi_x^{(-1)}\rangle, |\phi_x^{(0)}\rangle, |\phi_x^{(1)}\rangle \Big\} =
{\small 
\frac{1}{2}\left\{ 
\begin{pmatrix} 1 \\
-\sqrt{2} \\
1
\end{pmatrix},
\begin{pmatrix} -\sqrt{2} \\
0 \\
\sqrt{2}
\end{pmatrix},
\begin{pmatrix} 1 \\
\sqrt{2} \\
1
\end{pmatrix}
\right\}
}
\end{equation*}
respectively. As in the qubit example in Sec.~\ref{sec:simple_case_study}, no quantum dynamics occurs in between the projective measurements of $\mathcal{O}_{1}(t_1)$ and $\mathcal{O}_{2}(t_2)$ i.e. $U=\mathbb{I}$. We now write: the expression of the MHQ $q_{\rm MHQ}(s_1,s_2)$, the joint probability $p(s_1,s_2)$ of the TPM scheme, the unperturbed final-time probability $p_{s_2}(t_2)$ and the wTPM probability $w(s_1,s_2)$. We recall that $p(s_1,s_2), p_{s_2}(t_2), w(s_1,s_2)$ can be all experimentally measured via a procedure based on single or sequential measurements. Moreover, by linearly combining them together according to Eq.~\eqref{eq:infer_q_MHQ}, any MHQ can be fully reconstructed~\cite{hernandez2022experimental}. Thus, for the example under scrutiny, 
\begin{eqnarray}
    q_{\rm MHQ}(s_1,s_2) &=& {\rm Re}\left[\langle\phi_{x}^{(s_2)}|\phi_{z}^{(s_1)}\rangle \langle\phi_{z}^{(s_1)}|\,\rho\,|\phi_{x}^{(s_2)}\rangle \right ]\\
    p(s_1,s_2) &=& \left| \langle\phi_{x}^{(s_2)}|\phi_{z}^{(s_1)}\rangle \right|^{2} \langle\phi_{z}^{(s_1)}|\,\rho\,|\phi_{z}^{(s_1)}\rangle \\
    p_{s_2}(t_2) &=& \langle\phi_{x}^{(s_2)}|\,\rho\,|\phi_{x}^{(s_2)}\rangle\,,
\end{eqnarray}
and $w(s_1,s_2)$ is given by Eq.~\eqref{eq:wTPM_prob} with $\Pi_{s_2}^{H}(t_2)=|\phi_{x}^{(s_2)}\rangle\!\langle\phi_{x}^{(s_2)}|$ and $\Pi_{s_1}(t_1)=|\phi_{z}^{(s_1)}\rangle\!\langle\phi_{z}^{(s_1)}|$. In this example, $\Pi_{s_1}^{\perp}(t_1)=\mathbb{I}-\Pi_{s_1}(t_1)$ are projectors with rank-2 and describe the collapse of the spin-1 state onto a two-dimensional subspace. For completeness, the explicit expressions of $\Pi_{s_1}^{\perp}(t_1)$ with $o_{s_{1}}(t_1) = -1,0,1$ are:
\begin{eqnarray*}
    &\Pi_{-1}^{\perp}(t_1)=\begin{pmatrix} 
    1 & 0 & 0 \\
    0 & 1 & 0 \\
    0 & 0 & 0
    \end{pmatrix},& \\
    &\Pi_{0}^{\perp}(t_1)=\begin{pmatrix} 
    1 & 0 & 0 \\
    0 & 0 & 0 \\
    0 & 0 & 1
    \end{pmatrix}, \quad
    \Pi_{1}^{\perp}(t_1)=\begin{pmatrix} 
    0 & 0 & 0 \\
    0 & 1 & 0 \\
    0 & 0 & 1
    \end{pmatrix}.&
\end{eqnarray*}

Let us now take the initial density operator $\rho = \ketbra{\psi}{\psi}$ with 
\begin{equation}
    \ket{\psi}=\frac{1}{\sqrt{2}}\left( |\phi_z^{(-1)}\rangle - |\phi_z^{(0)}\rangle \right) = \frac{1}{\sqrt{2}}\begin{pmatrix}
        0 \\ -1 \\ 1
    \end{pmatrix}.
\end{equation}
In the following, we provide the analytical expressions of $q_{\rm MHQ}(s_1,s_2)$, $p(s_1,s_2)$, $p_{s_2}(t_2)$ and $w(s_1,s_2)$ for a single pair $(s_1,s_2)$: $s_1=-1$, $s_2=1$. This choice ensures that $q_{\rm MHQ}(-1,1) < 0$. In doing this, we will show a specific example of how Eq.~\eqref{eq:infer_q_MHQ} effectively works:
\begin{equation}\label{eq:weak_TPM_example}
    q_{\rm MHQ}(-1,1) = p(-1,1) + \frac{1}{2}\Big( p_{s_2}(1) - w(-1,1) \Big).
\end{equation}
From direct calculations, one can find that 
\begin{eqnarray*}
    &\langle\phi_{x}^{(1)}|\phi_{z}^{(-1)}\rangle = \frac{1}{2}, \quad 
    &\langle\phi_{z}^{(-1)}|\,\rho\,|\phi_{x}^{(1)}\rangle = \frac{1-\sqrt{2}}{4}, \\
    &\langle\phi_{z}^{(-1)}|\,\rho\,|\phi_{z}^{(-1)}\rangle = \frac{1}{2}, \quad 
    &\langle\phi_{x}^{(1)}|\,\rho\,|\phi_{x}^{(1)}\rangle = \frac{3-2\sqrt{2}}{8}\,.
\end{eqnarray*}
Therefore,
\begin{eqnarray}
    q_{\rm MHQ}(-1,1) &=& \frac{1-\sqrt{2}}{8} \approx -0.0518 < 0 \\
    p(-1,1) &=& \frac{1}{8} \\
    p_{s_2}(1) &=& \frac{3-2\sqrt{2}}{8} \approx 0.0214 \,.
\end{eqnarray}
Moreover, 
\begin{eqnarray}
    w(-1,1) &=& {\rm Tr}\left[ |\phi_{x}^{(1)}\rangle\!\langle\phi_{x}^{(1)}| \Big( \langle\phi_{z}^{(-1)}|\,\rho\,|\phi_{z}^{(-1)}\rangle|\phi_{z}^{(-1)}\rangle\!\langle\phi_{z}^{(-1)}| \right. \nonumber \\
    &+& \left.\Pi_{-1}^{\perp}(t_1)\,\rho\,\Pi_{-1}^{\perp}(t_1) \Big) \right] = \frac{3}{8} 
\end{eqnarray}
that validates Eq.~\eqref{eq:weak_TPM_example}. 

\subsubsection{Interferometric scheme}\label{sec:example_interferometry}

We consider the Ramsey interferometric scheme applied to a spin-$1/2$ particle that is sequentially measured along the $z$ and $x$ axis, as in the example reported in Sec.~\ref{sec:simple_case_study}. We recall that, in the case the initial density operator $\rho$ does not commute with $\mathcal{O}_1(t_1)=\sigma^z$, any procedure based on sequential projective measurements is invasive and unavoidably cancels out the quantum coherence contained in $\rho$, with respect to the eigenbasis of $\mathcal{O}_1(t_1)$. As a result, the statistics of the outcome pairs $(z_k(t_1),x_j(t_2))$ is distorted. The unperturbed statistics of the outcome pairs is provided by the KDQ distribution that, as shown in Sec.~\ref{sec:simple_case_study}, can exhibit negative and imaginary quasiprobabilities.

The interferometric scheme finds application to such a case study by making the substitution $E_{t_1} = \exp(-i\,\sigma^{z}u)$, $U = \mathbb{I}$, and $E_{t_2}=\exp(-i\,\sigma^{x}u)$. In this way, by measuring the expectation values $\langle\sigma^x_{\mathcal{A}}\rangle (u)$ and $\langle\sigma^y_{\mathcal{A}}\rangle (u)$, we can recover the real and imaginary parts of the KDQ characteristic function providing the statistics of outcome pairs $(z_k(t_1),x_j(t_2))$. In this case, the characteristic function reads as [see also Eq.~\eqref{eq:characteristic}]
\begin{equation}\label{eq:G_Stern-Gerlach}
    \mathcal{G}(u) = {\rm Tr}\left[ e^{-i\,\sigma^{z}u}\rho\,e^{i\,\sigma^{x}u} \right].
\end{equation}
Using the initial state defined in Eq.~\eqref{eq:qubitstateexample}, we obtain: 
\begin{eqnarray}\label{eq:Guqubit}
    {\mathcal G}(u)&=&\cos^2(u) + 2i\sin(u)\times\nonumber \\
    &\times&\left( \cos(u)\Re\left[\rho_{0,1}\right] +\sin(u)\Im\left[\rho_{0,1}\right]\right),
\end{eqnarray}
such that the analytical expressions of the expectation values for the auxiliary system are
\begin{eqnarray}
    \langle\sigma^x_{\mathcal{A}}\rangle &=& \cos^2(u),
    \\
    \langle\sigma^y_{\mathcal{A}}\rangle &=& 2\sin(u)\left( \cos(u)\Re\left[\rho_{0,1}\right] +\sin(u)\Im\left[\rho_{0,1}\right]\right).\nonumber\\
\end{eqnarray}
Taking the inverse Fourier transform, we get the following QD for $\Delta o$:
\begin{eqnarray}
    P[\Delta o]
    &=& \frac{1+2i\Im\left[ \rho_{0,1} \right] }{2} \delta(\Delta o)+
    \nonumber\\
    &+&  \frac {1-2\rho_{0,1}}{4}  \delta(\Delta o-2)+
    \label{eq:Pqubit} \\
    &+&\frac {1+2\rho_{0,1}^*}{4}\delta(\Delta o+2)\,, 
    \nonumber
\end{eqnarray} 
which may be complex depending on the
form of $\rho_{0,1}$.

It is instructive to point out that, by defining the effective Hamiltonian operators $\mathcal{H}_{1} \equiv \omega\sigma^z$ and $\mathcal{H}_{2} \equiv \omega\sigma^x$, then Eq.~\eqref{eq:G_Stern-Gerlach} takes the more general expression
\begin{equation}\label{eq:LE_Stern-Gerlach}
    \mathcal{G}\left( \omega t \right) = {\rm Tr}\left[ e^{i\mathcal{H}_{2}t}\,e^{-i\mathcal{H}_{1}t}\rho \right]
\end{equation}
with $u = \omega t$. From this, we observe that the characteristic function of the KDQ distribution can be identically equal to the so-called Loschmidt echo~\cite{PeresPRA84,JalabertPRL2001}. Hence, thanks to the link with the Loschmidt echo, $\mathcal{H}_{1}$ and $\mathcal{H}_{2}$ can be interpreted as the Hamiltonian operators governing, respectively, the \emph{forward} and \emph{backward} evolution of a perturbed quantum system~\footnote{In Eq.~\eqref{eq:LE_Stern-Gerlach}, the perturbation is introduced in considering two different Hamiltonian operators $\mathcal{H}$ for the forward and backward dynamics that return the Loschmidt echo.}, and $t$ as the time instant at which the reversal operation takes place.

In Sec.~\ref{sec:Link_with_LE} we will show that the connection between the characteristic function of a KDQ distribution and Loschmidt echos does not hold only in specific examples, but it is valid in general for any quantum system. In this respect, condensed-matter physics and quasiprobabilities are deeply related.  

\subsubsection{Detector-assisted scheme} 

Let us consider the modified characteristic function $\tilde {\mathcal G}$ in Eq.~\eqref{eq:exp_i_phi}. By choosing the initial state qubit to be Eq.~\eqref{eq:qubitstateexample} as in \ref{sec:example_interferometry}, we find:
\begin{equation}
    \tilde {\mathcal G} = \frac 12 \Big[ 1 + \cos( 2\kappa p_0 ) + 4i \Re\left[ \rho_{0,1}\sin( \kappa p_0 ) \right] \Big]\,,
\end{equation}
which clearly depends on the presence of the off-diagonal element $\rho_{0,1}$ of the system's density operator $\rho$. 

Consequently, taking the Fourier transform, we obtain the NDQP:
\begin{eqnarray}
    P_{\rm NDQP}[\Delta o]
    &=& \frac 12 \delta(\Delta o)+\frac 14 \Big( \delta(\Delta o-2)+\delta(\Delta o+2) \Big) + \nonumber \\
    &+& \Re\left[ \rho_{0,1} \right] \Big( \delta(\Delta o+1)-\delta(\Delta o-1) \Big),
    \label{eq:NDQPexample}
\end{eqnarray} 
which contains peaks at $\Delta o=\pm 1$ that are absent in the KDQ (or for an incoherent initial state) and that can be negative depending on the sign of $\Re\left[ \rho_{0,1} \right]$. Another difference with the KDQ extracted from the Ramsey scheme, Eq.~\eqref{eq:Pqubit}, is that Eq.~\eqref{eq:NDQPexample} is strictly real.

Let us now consider the signal observed in the position representation of the detector assuming for concreteness the detector's wave function $g(x)=(2\pi\sigma^2)^{1/4}\exp[-x^2/(4\sigma^2)]$, albeit any localized function would be suitable. From Eq.~\eqref{eq:Pxdet}, we obtain $P(x) = P_{\rm inc}(x)+P_{\rm coh}(x)$, where the incoherent and coherent parts of the distributions read as
\begin{eqnarray}
    P_{\rm inc}(x) &=& \frac{ e^{-\frac{x^2}{2\sigma^2}}
    + \frac{1}{2} e^{-\frac{(x-2\kappa)^2}{2\sigma^2}}
    + \frac{1}{2} e^{-\frac{(x+2\kappa)^2}{2\sigma^2}} }{2(2\pi\sigma^2)^{1/2}}\,,\\
    P_{\rm coh}(x) &=& \frac{\Re\left[ \rho_{0,1} \right] e^{-\frac{\kappa^2}{2\sigma^2}} }{(2\pi\sigma^2)^{1/2}}   
    \left( e^{-\frac{(x-\kappa)^2}{2\sigma^2}} - e^{-\frac{(x+\kappa)^2}{2\sigma^2}}  \right).\nonumber \\
\end{eqnarray}
The probability density $P_{\rm inc}(x)$ would always appear in the expression of $P(x)$, even in the absence of initial coherence. It represents a coarse-grained version of the TPM probability distribution; see Eq.~\eqref{eq:prob_distribution_TPM} for the general definition. On the other hand, $P_{\rm coh}(x)$ is  proportional to $\Re\left[ \rho_{0,1} \right]$ and can be negative (although $P(x)$ can never be negative).

\begin{figure}[t]
\begin{center}
\includegraphics[width=0.95\columnwidth]{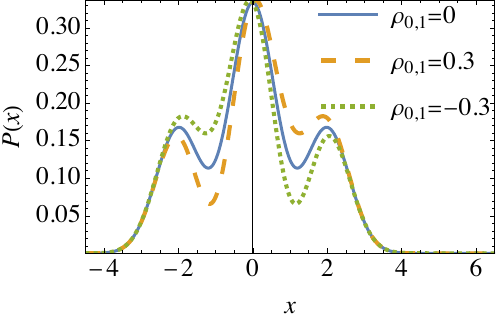}
\caption{Distribution function $P(x)$ observed by an imperfect detector for $\rho_{0,1}=0, \rho_{0,1}=0.3$ and $\rho_{0,1}=-0.3$. We have chosen units in which $\kappa=1$ and $\sigma=0.6$. 
}
\label{fig:Px}
\end{center}
\end{figure}

We show $P(x)$ in Fig.~\ref{fig:Px} and compare the cases when $\rho_{0,1}=0$ and $\rho_{0,1}\neq 0$. The latter case brings an asymmetry to the function $P(x)$ and, for sufficiently large $\sigma$, a slight shift in the position of the peaks and deeper dips in between peaks. This effect increases for larger $\rho_{0,1}$ and $\kappa$ and is due to interference between probability amplitudes for the different measurement outcomes. 

\section{Quantum thermodynamics}
\label{sec:QTD}

In Sec.~\ref{section_quasiprobs}, we have argued that, in the general case of $\rho$ and $\mathcal{O}_{1}(t_1)$ arbitrary noncommuting operators, the first measurement of the TPM scheme is invasive. Specifically, the TPM scheme does not allow one to recover the unperturbed single-time probability $p(s_2)$ by marginalizing the joint probability $p(s_1,s_2)$ of the TPM scheme over the outcomes $s_1$ of the first measurement. As a result, applying the TPM scheme breaks the no-signaling in time condition, failing to capture noncommutativity in the statistics of the measurement outcomes taken at times $t_1$ and $t_2$~\cite{lostaglio2018quantum,beyer2021joint}. This evidence is proven to be true for arbitrary quantum observables $\mathcal{O}_{1}(t_1)$ and $\mathcal{O}_{2}(t_2)$. Consequently, the same considerations shall hold in a thermodynamic context, where the measured observables are Hamiltonian operators.

Kirkwood-Dirac quasiprobabilities can be employed to investigate {\it nonclassical energetic processes}, where here `nonclassical' indicates the presence of negative and imaginary values in the quasiprobability distribution of the thermodynamic quantity of interest, for instance, work, heat or entropy. In the current literature, Margenau-Hill quasiprobability distributions~\cite{allahverdyan2014nonequilibrium,halliwell2016leggett,lostaglio2018quantum,levy2020quasiprobability,hernandez2022experimental,PeiPRE2023}, the real parts of Kirkwood-Dirac ones, have been already discussed and employed to characterize nonclassical work distributions~\cite{diaz2020quantum,hernandez2022experimental}, as well as the statistics of anomalous heat exchanges due to quantum correlations~\cite{levy2020quasiprobability}.

In this section, we discuss the KDQ approach to characterize the statistics of internal energy fluctuations in a generic quantum system, close or open. Then, we will focus on the ways the presence of nonclassicality is responsible for \emph{anomalous energy exchanges}~\cite{maffei2022anomalous} (see below for a proper definition) that can be identified, e.g., in the average and variance of work and heat distributions.

As we are going to show in Sec.~\ref{sec:extractable_work}, negative probabilities in a KDQ distribution of work find an interpretation as nonclassical energy transitions that make use of quantum coherence to transform absorbed energy in extractable work. 
In this regard, we will show how KDQ can take into account genuinely quantum features in energy-change fluctuations, and outline thermodynamic advantages. For example, this becomes evident when noting that, without quantum coherence, stochastic work processes can generate a lower amount of extractable work and thus be less performing in operating a quantum device.

Before proceeding, it is worth stressing that possible thermodynamic advantages enabled by quantum features are inherent to any quantum system subject to a thermodynamic transformation that does not spoil the state of the system with respect to the expected output. For example, a measurement procedure based on projective measurements, such as the TPM scheme that strongly perturbs the system's state, cannot be considered as a thermodynamic transformation preserving quantum features, in the same way as any decoherence process. Approaches using quasiprobabilities, instead, have been thought to characterize the quantum statistics originating from measuring the energy of a given quantum system at multiple times. Of course, to attain this characterization experimentally and possibly in real time, one has to resort to a measurement procedure that is the least invasive possible. The use of a detector weakly interacting with the measured system (see Sec.~\ref{sec:det_assisted_meas}) could represent a solution.

Overall, acting directly on a quantum system while a given thermodynamic transformation is active leads to the spoil of functionalities entailed by quantum coherence or correlations. However, a proper quasiprobability distribution can retain this kind of information, and its knowledge could assist to build up a minimally invasive measurement protocol that is able to return results beyond average values while still characterizing the statistics of energy records.

\subsection{Quantum internal energy distribution}

From a stochastic thermodynamic point of view, any internal energy difference of a quantum transformation is a \emph{stochastic process}. This holds even for isolated systems, since energy-change fluctuations---that however average to zero---are induced by the measurement apparatus. We recall that the latter irreversibly perturbs the measured (thermodynamic) system in any procedure sequential projective measurements  that are directly performed on the system.

In the following, we are going to introduce the concepts of \emph{stochastic internal energy} and \emph{stochastic quantum work} in a generic quantum scenario with arbitrary density operators and time-dependent thermodynamic transformations.
Let us identify the time-dependent quantum observables $\mathcal{O}_{1}(t_1)$ and $\mathcal{O}_{2}(t_2)$ with the Hamiltonian operators $\mathcal{H}(t_1)$ and $\mathcal{H}(t_2)$ at the initial and final times of the transformation under scrutiny.
The Hamiltonian operators admit the spectral decomposition: 
\begin{eqnarray}
\mathcal{H}(t_1)&=&\sum_{i}E_i(t_1)\Pi_{i}(t_1)\label{eq:spectrum_Ht1} \\
\mathcal{H}(t_2)&=&\sum_{f}E_f(t_2)\Pi_{f}(t_2)\,,\label{eq:spectrum_Ht2}    
\end{eqnarray}
where $i,f$ denote the indexes on the initial and final energies, respectively.
From Eqs.~\eqref{eq:spectrum_Ht1}-\eqref{eq:spectrum_Ht2}, the definition of the stochastic internal energy $\Delta\mathcal{U}_{if}$ follows. $\Delta\mathcal{U}_{if}$ is defined within the time interval $[t_1,t_2]$, and it is given by the differences $\Delta\mathcal{U}_{if} \equiv E_{f} - E_{i}$. Notice that $\Delta \mathcal{U}_{if}$ depends only on the eigenvalues of the Hamiltonian $\mathcal{H}$ at the initial and final times of the thermodynamic transformation described by the CPTP map $\Phi$ which the system is subject to, and not directly on $\Phi$ itself. On the other hand, what is dependent on $\Phi$ is the probability distribution ruling the occurrence of any value of $\Delta\mathcal{U}_{if}$. In order to describe such a distribution, we introduce the Kirkwood-Dirac quasiprobability $q_{if} \equiv q(E_i,E_f)$ defined as
\begin{equation}
    \label{eq:qif}
    q_{if} = {\rm Tr}\left[ \Pi_{f}^{H}(t_2)\Pi_{i}(t_1)\rho \right],
\end{equation}
with $\rho$ denoting the initial density operator, such that the QD $P[\cdot]$ of $\Delta\mathcal{U}_{if}$ is 
\begin{equation}
\label{eq:PDeltaU}
    P\left[ \Delta\mathcal{U}\right] = \sum_{i,f}q_{if}\delta\left( \Delta\mathcal{U} - \Delta\mathcal{U}_{if}\right).
\end{equation}
Notice that, here, we have adopted a simplified notation for the KDQ $q_{if}$ using as subscript the labels for the initial and final energies, respectively.

Following what we previously stated in Sec.~\ref{sec:beyond_TPM} about the properties of a KDQ, all the information about the statistics of the stochastic internal energy $\Delta\mathcal{U}_{if}$ is also contained in the characteristic function 
\begin{equation}
\mathcal{G}_\mathcal{U}(u) = \int_{-\infty}^{+\infty}P[\Delta\mathcal{U}_{if}]e^{iu\Delta\mathcal{U}}d\Delta\mathcal{U},  
\end{equation}
obtained by the Fourier transform of $P[\Delta\mathcal{U}_{if}]$. As such, the KDQ distribution of the internal energy variation can be directly evaluated by means of the interferometric scheme discussed in Sec.~\ref{sec:interferometry}. As in the general case, the characteristic function $\mathcal{G}_\mathcal{U}(u)$---as well as each KDQ $q_{if}$---is formally a quantum correlation function that takes the form
\begin{eqnarray}
    \mathcal{G}_\mathcal{U}(u) &=& \sum_{i,f}q_{if}\,e^{iu(E_f(t_2)-E_i(t_1))} =\nonumber \\
    &=& {\rm Tr}\left[ e^{-iu\mathcal{H}(t_1)}\rho\,\Phi^{\dagger}\left[e^{iu\mathcal{H}(t_2)}\right]\right].
\end{eqnarray}
For the case of time-dependent unitary dynamics (possibly leading to work fluctuations), 
\begin{equation}
\Phi^{\dagger}[e^{iu\mathcal{H}(t_2)}] = U^{\dagger}e^{iu\mathcal{H}(t_2)}U = e^{iu\mathcal{H}^{H}(t_2)},    
\end{equation}
where $\mathcal{H}^{H}(t_2)=U^{\dagger}\mathcal{H}(t_2)U$ is the evolution of the system's Hamiltonian at the final time of the work protocol in the Heisenberg picture.

It is worth stressing that, when $[\rho,\Pi_{i}(t_1)]\neq 0$ or $[\Pi_{i}(t_1),\Pi_{f}^{H}(t_2)]\neq 0$, the corresponding KDQ $q_{if}$ can be a complex number, with possibly a negative real part. We recall that we have denoted this circumstance as being nonclassical. Even in this quantum thermodynamics case, the nonclassicality of the internal energy distribution $P[\Delta\mathcal{U}_{if}]$, in the time interval $[t_1,t_2]$, can be measured via the functional $\aleph$ computed over the KDQ $q_{if}$.

\subsection{Quantum work \& KDQ-correction to Jarzynski equality}
\label{subsec:quantum_work}

In any closed quantum system that is driven by a time-dependent Hamiltonian $\mathcal{H}$ in the time interval $[t_1,t_2]$, the internal energy difference corresponds to the exerted work $W$. This means that $\Delta\mathcal{U} = W$, being the dissipated heat equal to zero in such a case.

If the initial state of the system is a an equilibrium thermal state at inverse temperature $\beta$:
\begin{equation}
\rho=\frac{e^{-\beta \mathcal H (t_1)}}{Z},
\end{equation}
where $Z(t_k) \equiv {\rm Tr}[e^{-\beta\mathcal{H}(t_k)}]$ is the system's partition function, then the TPM probability distribution of work fulfils the celebrated
%
%
Jarzynski equality (JE)~\cite{JarzynskiPRL1997,JarzynskiPRX2015}
%
%
\begin{equation}\label{eq:JE}
\langle e^{-\beta W} \rangle_{\rm TPM} = e^{-\beta\Delta F}.
\end{equation}
Eq.~\eqref{eq:JE} relates a fluctuating physical quantity (the work $W$) measured for an out-of-equilibrium system in a given time during the work protocol, and the equilibrium free-energy difference 
\begin{equation}\label{eq:free-energy}
    \Delta F \equiv -\beta^{-1}\ln\left( \frac{ Z(t_2) }{ Z(t_1) }\right).
\end{equation}

The equilibrium free-energy difference is achieved asymptotically by the driven quantum system under the assumptions that, once the work protocol is over, (i) the Hamiltonian of the system is assumed constant and equal to $\mathcal{H}(t_2)$; (ii) the system is put in contact with a thermal bath at inverse temperature $\beta$. Moreover, in Eqs.~\eqref{eq:JE}-\eqref{eq:free-energy}, it is implicitly assumed that the quantum system is connected to the thermal bath at inverse temperature $\beta$ also before the work protocol is applied.

As surveyed in Ref.~\cite{RasteginJSTAT2013,AlbashPRE2013,RasteginPRE2014, Gherardini2022energy}, the symmetries allowing for the JE in Eq.~\eqref{eq:JE} to hold are generally maintained as long as (I) the initial density operator is thermal at inverse temperature $\beta$, namely $\rho = \rho_{\rm th}(t_1) \equiv e^{-\beta\mathcal{H}(t_1)}/Z(t_1)$; (II) the dynamics of the quantum system is \emph{unital}, meaning that the identity $\mathbb{I}$ is a fixed point of the quantum map to which the system is subject: $\Phi[\mathbb{I}]=\mathbb{I}$. Notice that unitary dynamics are a subgroup of such more general family of maps. Therefore, a requirement for the validity of the JE is that $\rho$ is a thermal state, i.e., both (a) $[\rho,\mathcal{H}(t_1)]=0$, and (b) the diagonal elements of $\rho$ (with respect to the eigenbasis of $\mathcal{H}(t_1)$) follow a Boltzmann distribution.

As a result, the JE in Eq.~\eqref{eq:JE} can be obtained by applying the TPM scheme, which returns the following characteristic function for any given work distribution:
\begin{eqnarray}
    \mathcal{G}^{\rm TPM}_{W}(u) &=& \sum_{i,f}p_{if}\,e^{iu(E_f(t_2)-E_i(t_1))} =\nonumber \\
    &=& {\rm Tr}\left[ U^{\dagger} e^{iu\mathcal{H}(t_2)} U e^{-iu\mathcal{H}(t_1)} \rho_{\rm th}(t_1) \right]
\end{eqnarray}
with $p_{if} = {\rm Tr}[ U^{\dagger}\Pi_{f}(t_2)U\Pi_{i}(t_1)\rho_{\rm th}(t_1)\Pi_{i}(t_1) ]$ the joint probabilities of the TPM scheme, and $u$ complex number. Hence, by  setting $u=i\beta$, we get
\begin{equation}\label{eq:derivation_JE_TPM}
    \mathcal{G}^{\rm TPM}_{W}(i\beta) \equiv \langle e^{-\beta W}\rangle_{\rm TPM} = \frac{ Z(t_2) }{ Z(t_1) } = e^{-\beta\Delta F}\,.
\end{equation}
Moreover, if one applies the Jensen inequality on both sides of Eq.~\eqref{eq:derivation_JE_TPM}, we directly get the inequality 
\begin{equation}
    \langle W\rangle_{\rm TPM} \geq \Delta F,
\end{equation}
that is one of the formulation of the second law of thermodynamics in relation with the Clausius theorem.

Let us now start connecting these results with the discussion undertaken in the previous sections. In this regard, we already know from Sec.~\ref{section_quasiprobs} that, if the density operator $\rho$ at the beginning of the work protocol does not contain quantum coherence $\chi$ with respect to the eigenbasis of $\mathcal{H}(t_1)$ ($\rho=\mathcal{D}_1[\rho]$), then the first energy measurement of the TPM scheme is not invasive. Hence, in such a case, $\langle W\rangle_{\rm TPM} - \Delta F$ denotes, without any ambiguities, the \emph{dissipated work} that is the amount of work that cannot be converted in extracted work.

However, the JE breaks down if $\rho$ is not a thermal state. The failure of the JE also occurs when $\rho$ is thermal but the dynamics of the quantum system is nonunital, possibly leading to heat dissipation~\cite{CampisiPRL2009,SagawaPRL2010,CampisiNJP2015,HernandezGomezPRR2020}. As a consequence, one ends up with an expression similar to the JE that exhibits a correction that is not a state function and depends on both the dynamical map to which the quantum system is subject and its initial state. We stress that, in the general case, a correction is present also in the case no quantum coherence $\chi$ is contained in $\rho$, i.e., $\rho=\mathcal{D}_1[\rho]$. Accordingly, by setting $\rho=\mathcal{D}_1[\rho]$, the characteristic function of the work distribution provided by the TPM scheme reads as~\cite{MorikuniJSP11,KafriPRA2012,AlbashPRE13,GardasSciRep2018,HernandezGomez2021nonthermal,SoretPRA2022} 
\begin{equation}
\langle e^{-\beta W}\rangle_{\rm TPM} = e^{-\beta\Delta F}\gamma\,,
\end{equation}
where 
\begin{equation}
\gamma \equiv {\rm Tr}\left[ \left( \rho_{\rm th}(t_1) \right)^{-1}\mathcal{D}_1[\rho]\,\Phi^{\dagger}\left[ \rho_{\rm th}(t_2) \right] \right]
\end{equation}
with $\rho_{\rm th}(t_2) \equiv e^{-\beta\mathcal{H}(t_2)}/Z(t_2)$. Being expressed as a function of the dynamical transformation applied to the system, the efficacy $\gamma \geq 0$ depends on the time $t_2$, making $\gamma$ generally a time-dependent quantity. 
Of course, $\gamma=1$ $\forall t_1, t_2$ if $\rho=\rho_{\rm th}(t_1)$ and $\Phi$ is a unital map. We recall that the difference $\mathcal{D}_1[\rho] - \rho_{\rm th}(t_1)$ is commonly known as \emph{athermality} as it quantifies the nonthermal contributions in the diagonal of $\rho$ with respect to the eigenbasis of $\mathcal{H}(t_1)$. The athermality can be a significant thermodynamic resource if the quantum system undergoes dynamics with feedback~\cite{CampisiNJP2017,GiachettiCondMatt2020,HernandezGomez2021nonthermal}.

If we now abandon the use of the TPM scheme and we consider in the more general framework of a quasiprobability distribution, how is the average exponentiated work $\langle e^{-\beta W}\rangle$ modified when $[\rho,\mathcal{H}(t_1)] \neq 0$, namely when quantum coherences are present in the initial density operator $\rho$? It is indeed clear that, if $\rho$ is not thermal, the JE in Eq.~\eqref{eq:JE} is no longer valid and a further correction has to be included to attain a modified JE expression. Notice that a different correction has to be considered for all the protocols that go beyond the TPM scheme~\cite{MicadeiPRL2020,SonePRL2020,GherardiniPRA2021,MicadeiPRL2021}. 

The use of KDQ to describe quantum work fluctuations leads to the relation
\begin{equation}
\langle e^{-\beta W}\rangle_{\rm KDQ} = \mathcal{G}_{W}(i\beta) = e^{-\beta\Delta F}\Gamma\,, 
\end{equation}
where
\begin{equation}
\Gamma \equiv {\rm Tr}\left[ \left( \rho_{\rm th}(t_1) \right)^{-1}\rho\,\Phi^{\dagger}\left[ \rho_{\rm th}(t_2) \right] \right]
\end{equation}
is the KDQ-correction to the JE that holds for any CPTP map $\Phi$. In conformity with the results shown in Sec.~\ref{section_quasiprobs}, $\Gamma=\gamma$ when $\rho=\mathcal{D}_1[\rho]$, i.e., under the commutative condition $[\rho,\mathcal{H}(t_1)]=0$. Similar to the efficacy $\gamma$, the KDQ-correction $\Gamma$ to the JE is not a state function, and therefore depends on the specific thermodynamic transformation that is performed on the system. However, in contrast to the TPM result, $\Gamma$ is in general a complex number, whose real part can take both negative and positive values. Consequently, as a possible application, if one measured ${\rm Re}\left[\Gamma\right] < 0$ or ${\rm Im}\left[\Gamma\right] \neq 0$, it would imply the presence of nonclassicality, since the nonpositivity functional $\aleph$ in Eq.~\eqref{eq:def_aleph} would necessarily be greater than zero. 
%
%
The thermodynamic meaning of the KDQ-correction $\Gamma$ is still lacking, and further investigations are thus needed. 

\subsection{nonclassical work exerted by qubits: a case study}\label{subsec:qubit_example_work}

In this section, we discuss a simple example to analyze the KDQ distribution of work done on a single qubit that is driven by a work protocol described by a unitary operator $U$. Assuming the system does not interact with any external bath, the internal energy change can be fully identified as work. Albeit simple, this model can be solved analytically and finds experimental applications in nuclear magnetic resonance (NMR) spin systems~\cite{MicadeiPRL2021} and nitrogen-vacancy (NV) centers~\cite{HernandezGomez_entropy_coherence,hernandez2022experimental} (point defects in the diamond lattice), where experiments of quantum thermodynamics beyond TPM have been recently performed. 

Let us assume the Hamiltonian of the qubit to be
\begin{equation}
{\mathcal H}(t)=\frac 12\Big[ \Omega \left( \cos\delta t \, \sigma^x + \sin\delta t \, \sigma^y \right) + \delta\sigma^z \Big],    
\end{equation}
corresponding to a spin-1/2 particle subject to an effective magnetic field rotating around the $z$-axis. In the rotating frame, described by the unitary operator $U_{\rm rot}=e^{i \delta \sigma^z t/2}$, the effective Hamiltonian governing the dynamics of the qubit becomes time-independent and reads
\begin{equation}
    \tilde {\mathcal H} = U_{\rm rot} {\mathcal H} U_{\rm rot}^\dagger +i \, \dot U_{\rm rot} U_{\rm rot}^\dagger = \frac{\Omega}{2} \sigma^x \,,
\end{equation}
so that the system's evolution operator (in the original frame) is
\begin{equation}\label{eq:example_qubit-unitary}
    U=e^{-i \delta \sigma^z t/2}e^{-i \Omega \sigma^x t/2}.
\end{equation}

To find the statistics of work done between times $t_1=0$ and $t_2=t$, we use the spectral decomposition of the time-dependent Hamiltonian, i.e.,
\begin{eqnarray}
    {\mathcal H}(t) &=& \sum_{\alpha=\pm} E_\alpha \Pi_\alpha(t)\,,\\
    E_\pm &=& \pm\frac 12 \Delta\,,\label{eq:qubit_energies}\\
    \Pi_\pm(t) &=& \frac{\mathbb{I}}{2}\pm\frac{\Omega(\sigma^x\cos\delta t+\sigma^y\sin\delta t)+\delta\sigma^z}{2\Delta}\,,\label{eq:qubit_projectors}
\end{eqnarray}
where we have defined a generalized Rabi frequency $\Delta \equiv \sqrt{\delta^2+\Omega^2}$.
Moreover, we assume the system to have quantum coherence in the eigenbasis of ${\mathcal H}(0)$, so that 
\begin{equation}\label{eq:initial_state_qubit}
    \rho=\begin{pmatrix}
        p & c \\ 
        c & 1-p
    \end{pmatrix},
\end{equation}
where $0\le p\le 1$ corresponds to the populations of the initial eigenstates, and $c$ is the quantum coherence that we have chosen to be real for simplicity. Using the definition of the quasiprobability distribution for work, see Eq.~\eqref{eq:PDeltaU}, we find:
\begin{eqnarray}
    P[W] &=& \left( q_{--}+q_{++} \right)\delta(W)+
    \nonumber\\
    &+&q_{+-}\delta(W+\Delta)
      +q_{-+}\delta(W-\Delta)\,,
\end{eqnarray}
where the KDQ $q$ are [see Eq.~\eqref{eq:qif}]:
\begin{widetext}
\begin{eqnarray}
q_{--}&=&\frac{p(\delta^2 + 2\Omega^2) -c\delta\Omega+\delta(p\delta+c\Omega)\cos\Omega t +i c\delta\Delta \sin\Omega t}{2\Delta^2}\,,\label{eq:qubit_qmm}
\\
q_{-+}&=&\delta\sin\frac{\Omega t}{2}\left[-\frac{ic\cos(\Omega t/2)}{\Delta}+\frac{(p\delta+c\Omega)\sin(\Omega t/2)}{\Delta^2} \right]\,,
\\
q_{+-}&=&\frac{\delta[(1-p)\delta-c\Omega](1-\cos\Omega t)-ic\delta\Delta\sin\Omega t}{2\Delta^2}\,,
\\
q_{++}&=&\frac{(1-p)(\delta^2 + 2\Omega^2)+c\delta\Omega+\delta[(1-p)\delta-c\Omega]\cos\Omega t +i c\delta\Delta \sin\Omega t}{2\Delta^2}\,.\label{eq:qubit_qpp}
\end{eqnarray}    
\end{widetext}

\begin{figure}
    \centering
    \includegraphics[width=0.95\columnwidth]{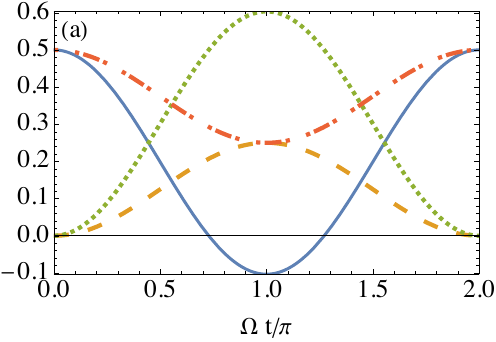}
    \includegraphics[width=0.95\columnwidth]{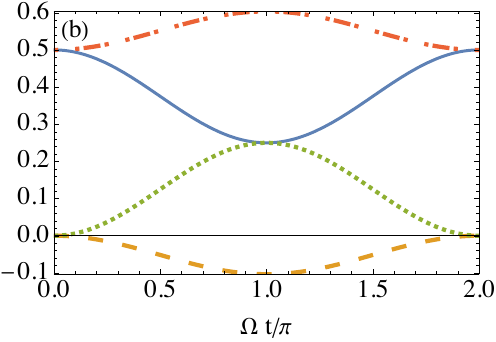}
    \caption{
    Real parts of the quasiprobabilities $q_{if}$ defined in Eqs.~\eqref{eq:Reqmm}-\eqref{eq:Reqpp} (with $p=c=1/2$) as a function of time $t=t_2$ for $\Omega=(\sqrt 2-1)\delta$ in panel (a) and $\Omega=(\sqrt 2+1)\delta$ in panel (b). $\Re[q_{--}]$ (blue solid line), $\Re[q_{+-}]$ (orange dashed line), $\Re[q_{-+}]$ (green dotted line), $\Re[q_{++}]$ (red dot-dashed line).
    } 
    \label{fig:Req}
\end{figure}

First, we notice that the imaginary parts of $q_{if}$ are always proportional to the coherence $c$. 
Second, the real parts of $q_{if}$ may become negative. To see this, we specify, for simplicity, initial conditions and take the maximum possible coherence: $p=c=1/2$. In this case, the real parts $\Re[q_{if}]$ become:
\begin{eqnarray}
   \Re[q_{--}] &=& \frac{\delta ^2-\delta  \Omega +2 \Omega^2+\delta  (\delta +\Omega ) \cos\Omega t}{4 \Delta^2},
   \label{eq:Reqmm}
   \\
   \Re[q_{-+}] &=&\frac{\delta  (\delta +\Omega ) (1-\cos\Omega t)}{4 \Delta^2},
   \\
   \Re[q_{+-}] &=&\frac{\delta  (\delta -\Omega ) (1-\cos\Omega t)}{4 \Delta^2},
   \\
   \Re[q_{++}] &=& \frac{\delta^2+\delta \Omega + 2 \Omega^2 +\delta (\delta -\Omega) \cos\Omega t}{4 \Delta^2}.
   \label{eq:Reqpp}
\end{eqnarray}
It is possible to see that the minimum value of $\Re[q_{--}]$ is $(1-\sqrt 2)/4 < 0$ that is obtained for $\Omega = (\sqrt 2 - 1)\delta$. Similarly, the minimum value of $\Re[q_{+-}]$ is also $(1 - \sqrt 2)/4$ attained for $\Omega = (\sqrt 2 + 1)\delta$.
The time-dependence of the quasiprobabilities $q_{if}$ is shown in Fig.~\ref{fig:Req} for the two cases $\Omega=(\sqrt 2\pm 1)\delta$. It is interesting to see that only one of the $\Re[q_{if}]$ may become negative for each value of $\Omega$. Moreover, choosing $c$ complex may lead to another of the $\Re[q_{if}]$ to become negative, but the minimum value is still $(1-\sqrt 2)/4$. 

\subsection{Enhancement of extractable work}\label{sec:extractable_work}

In this section, we are going to explain the meaning of \emph{nonclassical work extraction} and \emph{anomalous energy exchange} or \emph{variation}. 

In any system that is subject to a work protocol, the extractable work is defined as the amount of energy that is left over, with respect to the energy of the system
at the beginning of the transformation. Accordingly, if a protocol admits non zero extractable work, then the average energy at the end of the work protocol, $\langle\mathcal{H}(t_2)\rangle = {\rm Tr}[U\rho \, U^{\dagger}\mathcal{H}(t_2)]$, is smaller than the average energy at the beginning, $\langle\mathcal{H}(t_1)\rangle = {\rm Tr}[\rho\mathcal{H}(t_1)]$, so that the extra energy amount can be used by a work reservoir~\cite{MyersAVS2022} or stored in a battery~\cite{CampaioliArXiv2023}. Hence, the requirement for work extraction is that
\begin{equation}
\langle W\rangle \equiv \langle\mathcal{H}(t_2)\rangle - \langle\mathcal{H}(t_1)\rangle < 0 \, .
\end{equation}

Recently, it has been discussed whether the negativity of the terms composing a quasiprobability work distribution may correspond to an enhancement of work extraction, and whether this circumstance can be witnessed by violating an inequality that is instead valid under the commutative conditions $[\rho,\mathcal{H}(t_1)]=0$ and $[\mathcal{H}(t_1),\mathcal{H}(t_2)]=0$, i.e., when $\aleph=0$. The answer to both these questions is positive~\cite{hernandez2022experimental}.

In order to see this, at the level of energy transitions, let us consider the fact that an \emph{excitation} process $\Delta\mathcal{U}_{if} \equiv E_f - E_i > 0$ (indexes $i$, $f$ label the initial and final energies, respectively) occurring in a quantum process with negative quasiprobability $q_{if}$ (not neccessarily KDQ) is equivalent to a \emph{de-excitation} process $\Delta\mathcal{U}_{if} < 0$ in a classical work transformation with probability $|q_{if}|$.

During an excitation (stochastic) process, the system absorbs energy and uses this energy to carry out a transition between the energy levels. 
On the other hand, any de-excitation process that is operated by a thermodynamic transformation contributes to increase the amount of the extractable work. Therefore, the presence of negative quasiprobabilities can be effectively exploited as a resource to enhance work extraction, beyond what any classical stochastic process can achieve. Such enhancement is deemed as `nonclassical', and the internal energy variations $\Delta\mathcal{U}_{if}$ associated to negative probabilities, enabling it, are called `anomalous'. Thus, anomalous energy variations denote energy exchanges that are inherently quantum mechanical, and heralded by nonpositivity.

Let us see how the enhancement of work extraction occurs. 
If one uses a work protocol (such as the TPM scheme) returning positive joint probabilities $p_{if}$ of the work distribution, the work extraction is maximized if we minimize (with sign) $\langle W\rangle_{\rm clas} = \sum_{i,f}p_{if}\Delta\mathcal{U}_{if}$, where the subscript `clas' here specifically stands for `classical' in the sense of positive joint probabilities. Without specifying anything about the thermodynamic transformation, the necessary condition to achieve the largest extractable work is to set
\begin{equation}
    \begin{cases}
    p_{if} = 0 \quad \text{for} \quad \Delta\mathcal{U}_{if} > 0, \\
    p_{if} > 0 \quad \text{for} \quad \Delta\mathcal{U}_{if} \leq 0, \\
    \end{cases}
\end{equation}
such that 
\begin{equation}\label{eq:classical_work_extraction}
\mathcal{W}_{\rm clas}
=-\langle W\rangle_{\rm clas}=\sum_{E_i\geq E_f}p_{if}\left(E_i-E_f\right) \ge 0
\end{equation}
leads to extractable work (in absolute value).

If instead the statistics of the internal energy variations $\Delta\mathcal{U}_{if}$ are described by some quasiprobabilities $q_{if}$ (for example when $[\rho,\mathcal{H}(t_1)] \neq 0$, as shown in Sec.~\ref{section_quasiprobs}), then the extractable work can be enhanced beyond what is obtained by a work protocol returning positive joint probabilities. For such a purpose, one can set
\begin{equation}\label{conditions_enhanced_WE}
    \begin{cases}
    {\rm Re}\left[q_{if}\right] < 0 \quad \text{for} \quad \Delta\mathcal{U}_{if} > 0, \\
    {\rm Re}\left[q_{if}\right] > 0 \quad \text{for} \quad \Delta\mathcal{U}_{if} \leq 0 \,. \\
    \end{cases}
\end{equation}
In this way, the magnitude of the extractable work
\begin{equation}\label{eq:quantum_work_extraction}
\mathcal{W} = -\langle W\rangle = \sum_{i,f}q_{if}(E_i - E_f) \geq 0
\end{equation}
can be effectively maximized in order to satisfy the inequality
\begin{equation}\label{eq:WE_beyond_TPM}
\mathcal{W}^{\rm max} \ge \mathcal{W}_{\rm clas}^{\rm max}.
\end{equation}
To achieve this, any excitation process $\Delta\mathcal{U}_{if} > 0$ has to be associated to a negative ${\rm Re}[q_{if}]$, while any de-excitation process $\Delta\mathcal{U}_{if} < 0$ should occur with a positive quasiprobability. 
It is worth observing that, in the case a KDQ distribution of work is taken into account, the imaginary parts of KDQ do not play an effective role for the task of work extraction, since they do not affect the average work.

What really matters to get enhanced work extraction is to ensure nonclassical behaviours in the time-distribution of negativity, namely the distribution over time of the quasiprobabilities with negative real part. In fact, it is not necessarily crucial for the nonpositivity functional $\aleph$ to take a large value, but that a significant negativity is associated to a positive `anomalous' energy variations $\Delta\mathcal{U}_{if}>0$. At the same time, work extraction is enhanced when negative values of $\Delta\mathcal{U}_{if}$ occur with the largest possible positive quasiprobability $q_{if}$. The interplay of all these conditions is model-dependent and depends on the specific parameters that rule the dynamics of the work process. Therefore, it is evident that attaining enhanced work extraction stems from an optimization routine that makes Eqs.~\eqref{conditions_enhanced_WE} valid in a given time interval of the work protocol.

In Ref.~\cite{hernandez2022experimental}, the electronic spin of an NV center in bulk diamond at room temperature was considered as the system to which a work protocol would be applied. Work extraction was observed, and its maximum values were associated to negative 
%
%
MHQ fulfilling Eq.~\eqref{eq:WE_beyond_TPM}. The work extraction enhancement observed in Ref.~\cite{hernandez2022experimental} originates from a sub-optimal solution for the optimization of work extraction against the time duration of the work protocol. In fact, due to the experimental constraints, only one internal energy change $\Delta\mathcal{U}$, corresponding to the largest possible value, was associated with a negative MHQ. At the same time, the smaller internal energy variation $-\Delta\mathcal{U}$ occurred with positive quasiprobability, with all other MHQ being negligible.

\subsubsection{Enhanced extractable work from violating a classical inequality}

We are going to show that fulfilling Eq.~\eqref{eq:WE_beyond_TPM} implies the violation of an inequality for work extraction that holds as long as the commutativity condition $[\rho,\mathcal{H}(t_1)]=0$ is obeyed; as in~\cite{hernandez2022experimental}, we consider MHQ. The violation of such an inequality cannot occur in any experiment realizing a work protocol yielding positive work joint probabilities, as the TPM scheme.

Let us thus consider that the projectors $\Pi_i$ and $\Pi_f$ of the Hamiltonian at the initial and final times $t_1$ and $t_2$ of the work protocol are rank-1 operators. This means that $\Pi_i=|E_i(t_1)\rangle\!\langle E_i(t_1)|$ and $\Pi_f=|E_{f}(t_2)\rangle\!\langle E_{f}(t_2)|$. Moreover, we assume, for simplicity,  the initial density operator $\rho= |\psi\rangle\!\langle\psi|$ to be pure.  Under these assumptions, the MHQ takes the form
\begin{equation}
    {\rm Re}\left[q_{if}\right] = {\rm Re}\Big[ \langle\psi|E_i\rangle \langle E_i|U^\dag|E_f\rangle \langle E_f|U|\psi\rangle \Big]\,.
\end{equation}
Interestingly, all the terms $\langle\psi|E_i\rangle$, $\langle E_i|U^\dag|E_f\rangle$, $\langle E_f|U|\psi\rangle$ are complex numbers whose real parts are linked with a standard probability amplitude, either defined at a single time or measurable by means of the TPM scheme. In particular, for the probability $p_i$ to measure the initial energy of the system, one has
\begin{equation}\label{eq:initial_prob_pure}
    p_i \equiv | \langle E_i|\psi\rangle |^2 \quad \Longrightarrow \quad \langle\psi|E_i\rangle = e^{-i\phi_i}\sqrt{p_i}\,,
\end{equation}
where $\phi_i$ is a phase factor. Then, in the same spirit, we can write
\begin{equation}\label{eq:cond_prob_TPM}
    p_{f|i} \equiv | \langle E_f|U^\dag|E_i\rangle |^2 \quad \Longrightarrow \quad \langle E_i|U^\dag|E_f\rangle = e^{-i\varphi_{if}}\sqrt{ p_{f|i} }
\end{equation}
and
\begin{equation}\label{eq:final_prob_END}
    p_{f} \equiv | \langle\psi|U|E_f\rangle |^2 \quad \Longrightarrow \quad \langle E_f|U|\psi\rangle = e^{-i\theta_f}\sqrt{p_f} \,,
\end{equation}
where $\varphi_{if},\theta_f$ are the corresponding phase factors. 

In Eq.~\eqref{eq:cond_prob_TPM}, $p_{f|i}$ is the conditional probability (associated to the TPM scheme) of measuring the energy $E_f$ at time $t_2$ conditioned to have measured $E_i$ at time $t_1$. Instead, in Eq.~\eqref{eq:final_prob_END}, $p_f$ is the probability to measure the energy $E_f$ at the end of the work protocol, by initializing the system in $\rho=|\psi\rangle\!\langle\psi|$. Notice that, by construction, the probability $p_f$ encodes information on the quantum coherence that is initially present in $\rho$; for this reason, $p_f$ is a key element of the EPM scheme~\cite{GherardiniPRA2021,HernandezGomez_entropy_coherence,Gianani2022diagnostics}. 

Overall, combining Eqs.~\eqref{eq:initial_prob_pure}-\eqref{eq:final_prob_END} we arrive at
\begin{equation}\label{eq_MHQ_WE}
    {\rm Re}\left[q_{if}\right] = {\rm Re}\Big[ \Lambda_{if}\sqrt{p_{if}\,p_f} \Big],
\end{equation}
where, by definition, $p_{if}=p_{f|i}p_{i}$ is the joint probability returned by the TPM scheme, and $\Lambda_{if} \equiv \cos( \phi_i + \varphi_{if} + \theta_f )$ that is named {\it activity}~\cite{hernandez2022experimental}. The latter brings information on the quantum interference fringes among the eigenbasis of $\rho$, $\Pi_i(t_1)$ and $\Pi_f^H(t_2)$. It is indeed the activity $\Lambda_{if}$ that is responsible for the negativity of ${\rm Re}\left[q_{if}\right]$, such that ${\rm Re}\left[q_{if}\right] < 0$ {\it if and only if} $\Lambda_{if} < 0$.

If we substitute Eq.~\eqref{eq_MHQ_WE} into the expression of the work extraction in \eqref{eq:quantum_work_extraction}, we find
\begin{equation}\label{eq:bound_work_extraction}
   \mathcal{W}_{\rm clas} \leq
   \sum_{E_i \geq E_f}\left(E_i - E_f\right)\sqrt{p_{if}\,p_f}
\end{equation}
whenever $\Lambda_{if}\geq 0$ $\forall\,i,f$. The inequality in Eq.~\eqref{eq:bound_work_extraction} gives an upper bound, dependent on the work protocol, to the amount of extractable work when $\aleph = 0$, and applies also to initial quantum mixed states. Hence, a violation of this bound, as experimentally tested in Ref.~\cite{hernandez2022experimental}, is a witness of the presence of negativity, as well as of nonclassical work extraction.

\begin{figure}
    \centering
    \includegraphics[width=0.95\columnwidth]{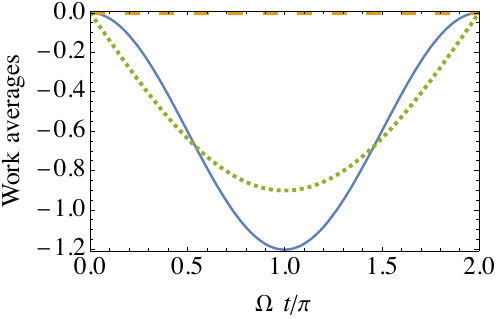}
    \caption{Average work in units of $\Omega$ for the spin-1/2 model described in Sec.~\ref{subsec:qubit_example_work} as a function of the final protocol time $t_2=t$. We consider the parameters $p=1/2$, $c=-1/2$ and $\Omega=(1+\sqrt 2)\delta$. In particular, we plot the average $\langle W\rangle$ of the KDQ distribution of work (blue solid line), the average work $\langle W\rangle_{\rm TPM}$ returned by the TPM scheme (orange dashed line)---equal to zero for any time---and the classical bound from Eq.~\eqref{eq:bound_work_extraction} (green dotted line) taken with opposite sign.} 
    \label{fig:avwork}
\end{figure}

In Fig.~\ref{fig:avwork}, we show an example of the enhancement of work extraction aided by negativity using the work protocol introduced in \ref{subsec:qubit_example_work}, applied to a driven qubit. In particular, we plot the average work of the TPM and MHQ probability distributions using the energies and Hamiltonian projectors in Eqs.~\eqref{eq:qubit_energies}-\eqref{eq:qubit_projectors}, as well as the quasiprobabilities in Eqs.~\eqref{eq:qubit_qmm}-\eqref{eq:qubit_qpp}, with $p=1/2$, $c=-1/2$ ($c=0$ to get the work statistics of the TPM scheme), and $\delta=\Omega/(\sqrt{2}+1)$.

Interestingly, if the initial state of the qubit [see Eq.~\eqref{eq:initial_state_qubit}] is fully mixed ($c=0$), then the average work is zero for any value of the final time $t_2$, see Fig.~\ref{fig:avwork}. On the other hand, turning on the quantum coherence in $\rho$ and making use of quasiprobability to attain the work distribution $P[W]$, the energy injected by the driving field is transformed into extractable work, beyond the classical bound [right-hand-side of Eq.~\eqref{eq:bound_work_extraction}] in the interval $(\Omega t)/\pi \in [0.6,1.4]$ approximately. 
In this case study with a qubit the classical bound amounts to 
\begin{eqnarray*}
    &(E_{+}-E_{-})\sqrt{ p_{+-} p_{-} } =&\\ 
    &\displaystyle{ = \delta \left( \frac{ (1-p)( 1 - \cos(\Omega t))}{2} \, {\rm Tr}\left[U\rho \,U^{\dagger}\Pi_{-} \right] \right)^{1/2} }&
\end{eqnarray*}
with $U$ given by Eq.~\eqref{eq:example_qubit-unitary}.

\subsection{Work variance in the KDQ setting}

In the previous section, we have shown how using KDQ to describe the work fluctuations makes the average work $\langle W\rangle=\sum_{i,f}q_{if}\left(E_f - E_i\right)$ equal to the corresponding value that is unperturbed by the measurement disturbance. Even though the KDQ $q_{if}$ are complex numbers, the average work $\langle W\rangle$ is always a real number, with a clear interpretation with classical physics, as shown above.

In the following, we analyze how the fact that $q_{if}$ are complex numbers affects the second moment of the KDQ distribution of work, $P[W]$, i.e., the work variance $(\Delta W)^2$. As noticed in Ref.~\cite{FuscoPRX2014}, the variance of work is in general not equal to the variance of the operator $\mathcal{H}^{H}(t_2)-\mathcal{H}(t_1)$ calculated for the initial state $\rho$.  Instead, this is formally defined by
\begin{eqnarray}\label{eq:work_variance}
    (\Delta W)^2 &=& \sum_{i,f}q_{if}(E_f-E_i)^2 - \Big( \sum_{i,f}q_{if}(E_f-E_i)\Big)^2 =\nonumber \\
    &=& \langle W^2\rangle - \langle W\rangle^2 \,, 
\end{eqnarray}
where, as before, all the averages $\langle\cdot\rangle$ are performed with respect to $\rho$, and the second statistical moment $\langle W^2\rangle$ reads as
\begin{eqnarray}\label{eq:second_moment}
    &\langle W^2\rangle = \sum_{i,f}q_{if}\left( E_i^2 + E_f^2 - 2E_i E_f\right) =& \nonumber\\
    &=\langle\mathcal{H}^2(t_1)\rangle + \langle\mathcal{H}^{H}(t_2)^2\rangle - 2{\rm Tr}\left[ \mathcal{H}^{H}(t_2) \mathcal{H}(t_1) \rho \right].&\nonumber\\ 
\end{eqnarray}

The quantity ${\rm Tr}\left[ \mathcal{H}^{H}(t_2) \mathcal{H}(t_1) \rho \right]$ in the right-hand-side of Eq.~\eqref{eq:second_moment} is a two-time quantum correlation function for the Hamiltonian $\mathcal H$ and is generally complex, making $\langle W^2\rangle$ also complex. This means that the imaginary part of $\langle W^2\rangle$ is equal to the imaginary part of ${\rm Tr}[ \mathcal{H}^{H}(t_2) \mathcal{H}(t_1) \rho ]$, whose meaning lies in the presence of phase correlations in the scalar products of the eigenvectors of $\rho$ and $\mathcal{H}$ at the times $t_1, t_2$. For this reason, the quantum correlation function for $\mathcal{H}$ preserves information about the quantum coherence contained in $\rho$, and this feature is transferred to the work variance $(\Delta W)^2$. In this regard, we are going to show that the imaginary part of $(\Delta W)^2$, ${\rm Im}\left[ (\Delta W)^2 \right]$, is directly linked with the noncommutativity between $\rho$ and $\mathcal{H}$. From this, following the time-energy Schr\"{o}dinger-Robertson uncertainty relation~\cite{RobertsonPR1929,SchrodingerPmK1930,schrodinger1999heisenberg}, ${\rm Im}\left[ (\Delta W)^2 \right]$ is bounded by the product of the uncertainties of $\mathcal{H}(t_1)$ and $\mathcal{H}^H(t_2)$ respectively.

A first expression of the work variance is obtained by combining Eqs.~\eqref{eq:work_variance}-\eqref{eq:second_moment}, so that:
\begin{eqnarray}\label{eq:work_variance_2}
    &(\Delta W)^2 = \left( \Delta\mathcal{H}(t_1) \right)^2 + \left( \Delta\mathcal{H}^{H}(t_2) \right)^2 + & \nonumber \\
    & -2{\rm Tr}\Big[ \Big( \, \mathcal{H}^{H}(t_2) - \langle\mathcal{H}^{H}(t_2)\rangle \, \Big) \Big( \, \mathcal{H}(t_1) - \langle\mathcal{H}(t_1)\rangle \, \Big) \rho \Big],&\nonumber \\ 
\end{eqnarray}
where 
\begin{equation}
    \Big( \Delta\mathcal{H}(t_1) \Big)^2 = {\rm Tr}\Big[ \rho \, \Big( \, \mathcal{H}(t_1) -  \langle\mathcal{H}(t_1)\rangle \, \Big)^2 \Big]\in\mathbb{R}
\end{equation}
and
\begin{equation}
    \Big( \Delta\mathcal{H}^{H}(t_2) \Big)^2 = {\rm Tr}\Big[ \rho \, \Big( \, \mathcal{H}^{H}(t_2) -  \langle\mathcal{H}^{H}(t_2)\rangle \, \Big)^2 \Big]\in\mathbb{R}\,.
\end{equation}

The last term in the right-hand-side of \eqref{eq:work_variance_2} identifies the way Hamiltonian operators at distinct times correlate in a quantum work protocol. The real and imaginary parts of ${\rm Tr}\left[ \big( \, \mathcal{H}^{H}(t_2) - \langle\mathcal{H}^{H}(t_2)\rangle \, \big) \big( \, \mathcal{H}(t_1) - \langle\mathcal{H}(t_1)\rangle \, \big) \rho \right]$ are equal respectively to~\footnote{For any density operator $\rho$ and two quantum observables $\mathcal{O}_1$ and $\mathcal{O}_2$, one can split the trace into real and imaginary parts, such that ${\rm Tr}\left[ \rho \, \mathcal{O}_1 \mathcal{O}_2 \right] = \frac{1}{2}{\rm Tr}\left[ \rho \left\{ \mathcal{O}_1, \mathcal{O}_2 \right\} \right] - i \frac{1}{2}{\rm Tr}\left[ i \rho \left[ \mathcal{O}_1, \mathcal{O}_2 \right] \right]$.}:
\begin{eqnarray}\label{eq:quantum_covariance}
    && \frac{1}{2}{\rm Tr}\left[ \rho \Big\{ \big( \mathcal{H}^{H}(t_2) - \langle\mathcal{H}^{H}(t_2)\rangle \big), \big( \mathcal{H}(t_1) - \langle\mathcal{H}(t_1)\rangle \big) \Big\} \right]\nonumber\\
    && \equiv {\rm Cov}\Big( \mathcal{H}(t_1), \mathcal{H}^{H}(t_2) \Big) \in \mathbb{R}
\end{eqnarray}
and
\begin{eqnarray}\label{eq:trace_commutator}
    && - \frac{1}{2}{\rm Tr}\left[ i \rho \Big[ \big( \mathcal{H}^{H}(t_2) - \langle\mathcal{H}^{H}(t_2)\rangle \big), \big( \mathcal{H}(t_1) - \langle\mathcal{H}(t_1)\rangle \big) \Big] \right]\nonumber\\
    && = - \frac{1}{2}{\rm Tr}\Big[ i \rho \, \left[ \mathcal{H}^H(t_2),\mathcal{H}(t_1)\right] \Big] \in \mathbb{R}.
\end{eqnarray}

Eq.~\eqref{eq:quantum_covariance} defines the {\it quantum covariance} of $\mathcal{H}(t_1)$ and $\mathcal{H}^{H}(t_2)$. Instead, in Eq.~\eqref{eq:trace_commutator}$, {\rm Tr}[ \rho \, [ \mathcal{H}^H(t_2), \mathcal{H}(t_1) ] ]$ is a purely imaginary number and, by definition, is the expectation value of the commutator $[\mathcal{H}^H(t_2), \mathcal{H}(t_1)]$ with respect to the initial density operator $\rho$.

\begin{figure}
    \centering
    \includegraphics[width=0.95\columnwidth]{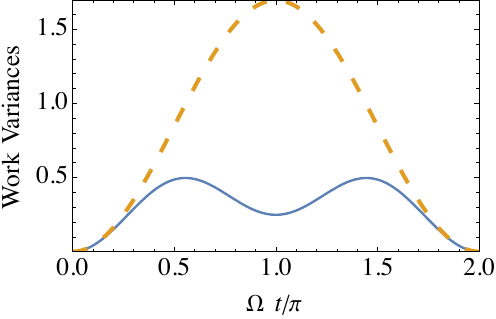}
    \caption{Plot of the work variances of the KDQ (real part, blue solid line) and TPM (orange dashed line) distributions, respectively. The parameters are the same as in Fig.~\ref{fig:avwork}.} 
    \label{fig:varwork}
\end{figure}

This derivation demonstrates that the work variance has both a real and an imaginary part. The real part has a clear correspondence with the thermodynamics of classical systems, as
\begin{eqnarray}\label{real_part_variance}
    {\rm Re}\left[ (\Delta W)^2 \right] &=& \left( \Delta\mathcal{H}(t_1) \right)^2 + \left( \Delta\mathcal{H}^{H}(t_2) \right)^2 + \nonumber \\
    &-& 2\,{\rm Cov}\Big( \mathcal{H}(t_1), \mathcal{H}^{H}(t_2) \Big).
\end{eqnarray}
In addition, the fact that the commutator $[\rho,\mathcal{H}(t_1)] \neq 0$ or $[ \mathcal{H}(t_1), \mathcal{H}^H(t_2) ]\neq 0$ may lead to a {\it decreased} work variance, namely to ${\rm Re}\left[(\Delta W)^2\right] \leq (\Delta W_{\rm TPM})^2$. We show this for the driven qubit of Sec.~\ref{subsec:qubit_example_work} and report the results in Fig.~\ref{fig:varwork} where we assume the same values used for Fig.~\ref{fig:avwork}. Interestingly, apart from $\Omega t/\pi = 0,2$ where both the work average and variances are zero, the real part of the KDQ work variance ${\rm Re}\left[(\Delta W)^2\right]$ has a local minimum at $\Omega t/\pi = 1$ that is the time instant with maximum negativity. 

On the other hand, the imaginary part of the work variance is
\begin{equation}\label{imag_part_variance}
    {\rm Im}\left[(\Delta W)^2\right] = {\rm Tr}\Big[ i \rho \left[ \mathcal{H}^H(t_2), \mathcal{H}(t_1) \right] \Big] 
\end{equation}
that quantifies the  noncommutativity of $\mathcal{H}(t_1)$ and $\mathcal{H}^H(t_2)$. The magnitude of ${\rm Im}\left[(\Delta W)^2\right]$ can be bounded from above by making use of the time-energy Schr\"{o}dinger-Robertson uncertainty relation~\cite{RobertsonPR1929,SchrodingerPmK1930,schrodinger1999heisenberg}. In fact, the latter states that, for any quantum observables $\mathcal{O}_1$, $\mathcal{O}_2$ and density operator $\rho$, 
\begin{equation}\label{eq:sr_uncertainty}
    \left|
    \frac{\langle \, [\mathcal{O}_1, \mathcal{O}_2] \, \rangle}{2i}
    \right|
    \leq
    \Delta\mathcal{O}_1 \, \Delta\mathcal{O}_2 \,,
\end{equation}
where $\langle \, [\mathcal{O}_1, \mathcal{O}_2] \, \rangle = {\rm Tr}\big[ \rho \, [\mathcal{O}_1, \mathcal{O}_2] \big]$ and 
\begin{equation}
\Delta\mathcal{O} = \sqrt{\langle\mathcal{O}^2 \rangle -\langle\mathcal{O}\rangle^2}      
\end{equation}
with $\langle\mathcal{O}^k\rangle = {\rm Tr}[\rho \, \mathcal{O}^k]$. Therefore, by applying the inequality in Eq.~\eqref{eq:sr_uncertainty} to our setting, we obtain
\begin{equation}
    \Big| {\rm Im}\left[(\Delta W)^2\right] \Big| \leq 2 \, \Delta\mathcal{H}^H(t_2) \,  \Delta\mathcal{H}(t_1) \,. 
\end{equation}

\subsection{Heat fluctuations in the quantum regime}

In this section, we no longer deal with work distributions, and we will focus on heat fluctuations. For this purpose, we consider the paradigmatic model that consists in placing into contact a cold and a hot quantum system, which globally undergo a unitary quantum dynamics. Depending on the initial quantum states of the cold and hot systems, different results as well as thermodynamic interpretations can be drawn. In this context, a first relevant result is in Ref.~\cite{JarzynskiPRL2004} and goes under the name of {\it Jarzynski-W\'ojcik exchange fluctuation theorem}. In Ref.~\cite{JarzynskiPRL2004}, two quantum systems $B_c$ and $B_h$ with finite Hilbert-space dimension are prepared in two equilibrium thermal states at different temperatures $\beta_c$ and $\beta_h$ with $\beta_c>\beta_h$. Then, they are made weakly interacting with one another for a given time interval. Under this assumption, one gets that
\begin{equation}\label{eq:JWE}
\langle e^{-\Delta\beta \, Q}\rangle = 1 \,,  
\end{equation}
where $Q$ is the stochastic heat exchanged by the two bodies, and $\Delta\beta=\beta_c-\beta_h$ denotes the difference of the inverse temperatures of the initial thermal states for the two bodies.

If the initial global state of the two systems is a product state then the average $\langle\cdot\rangle$ in Eq.~\eqref{eq:JWE} can be performed also with respect to the TPM distribution of the exchanged heat. To find this, it is sufficient to measure the sum of the local Hamiltonian operators of the two bodies, i.e., $\mathcal H=\mathcal{H}_{B_c} + \mathcal{H}_{B_h}$ (a time-independent Hermitian operator). Furthermore, throughout this section, we also implicitly assume the energy-preserving condition for the unitary operator $U$ that describes the quantum dynamics of the two bodies: 
\begin{equation}\label{eq:energy-preserving}
[\mathcal{H}\, , \, U]=0\,.
\end{equation}
Equation~\eqref{eq:energy-preserving} physically entails that, at any time $t$, the average energy variation in a body is minus the corresponding average energy variation in the other body. Such symmetry allows one to study fluctuations of exchanged energy between the two bodies by just measuring one of them.

In the literature, it has been considered also the cases of an initial quantum state that is locally thermal as in Ref.~\cite{JarzynskiPRL2004}, or classically correlated~\cite{JevticPRE2015}. This kind of correlation makes nonthermal the diagonal of the initial density operator $\rho$ for the two bodies taken individually, but does not add off-diagonal elements in $\rho$ with respect to $\mathcal{H}$. As shown in Ref.~\cite{JevticPRE2015}, a generalized exchange fluctuation relation, extending Eq.~\eqref{eq:JWE}, can be still obtained, as we will discuss next. 

Let us now introduce the spectral decomposition of the local Hamiltonians $\mathcal{H}_{B_c}$ and $\mathcal{H}_{B_h}$ for each of the two bodies:
\begin{eqnarray}
    \mathcal{H}_{B_k} &=& \sum_{\ell_k} E_{\ell_k}\Pi_{\ell_k}
\end{eqnarray}
with $k=c,h$ and $\ell=i,f$. This implies that the projectors of the total Hamiltonian $\mathcal{H}$  are $\Pi_{i_{c}i_{h}} = \Pi_{i_c} \otimes \Pi_{i_h}$.

For the initial state $\rho$, we require that the reduced states of the each body is in equilibrium at inverse temperature $\beta_k$:
\begin{eqnarray}
    \rho_{{\rm th},B_c} &=& {\rm Tr}_h \left[ \rho \right] = \frac{ e^{-\beta_{c}\mathcal{H}_{B_c}} }{ Z_c } \label{eq:reduced_thermal_c}
    \\
    \rho_{{\rm th},B_h} &=& {\rm Tr}_c \left[ \rho \right] = \frac{ e^{-\beta_{h}\mathcal{H}_{B_h}} }{ Z_h }
    \label{eq:reduced_thermal_h}   
\end{eqnarray}
where $Z_k \equiv {\rm Tr}\left[ e^{-\beta_{k}\mathcal{H}_{B_k}} \right]$ are the local partition functions. We hence have: $\rho=\rho_{{\rm th},B_c} \otimes \rho_{{\rm th},B_h}$. While the reduced states are diagonal in the eigenbasis of $\mathcal{H}_{B_k}$, in general the global state $\rho$ may contain off-diagonal elements, with respect to the local energy eigenbasis, that may be the signature of the presence of quantum correlations.

We are now in the position to define the average heat flow that, due to the energy-preserving condition, can be inferred from the energy change of either the cold or the hot body. Without loss of generality, we choose to measure it through the cold system as in Ref.~\cite{levy2020quasiprobability}. The average heat flow at the final time $t_2$ of the thermodynamic transformation is 
\begin{eqnarray}\label{eq:average_heat}
    \langle Q\rangle &\equiv& {\rm Tr}\left[ (\rho-\rho')\mathcal{H}_{B_c} \right],
\end{eqnarray}
where $\rho' = U \rho U^{\dagger}$ denotes the evolved density operator of the two bodies. 

According to Eq.~\eqref{eq:average_heat}, $\langle Q\rangle \leq 0$ denotes heat flowing on average from the hot to the cold body, as naturally requested by the second law of thermodynamics with the intervention of no external drive. On the other hand, by resorting to additional resources, it can also occur that $\langle Q\rangle > 0$ meaning that on average heat flows from the cold body to the hot one, as in a refrigerator. Summarising,
\begin{eqnarray*}
    &&\langle Q\rangle \leq 0 \quad \Longrightarrow \quad \text{hot-to-cold heat flow}, \\
    &&\langle Q\rangle > 0 \quad \Longrightarrow \quad \text{cold-to-hot heat backflow}.
\end{eqnarray*}
Moreover, if the amount of exchanged heat from the cold to the hot body exceeds in magnitude the value of $(\Delta\beta)^{-1}\ln d$, with $d$ the Hilbert-space's dimension of each body, then the heat backflows are called {\it strong}. Notably, the observation of strong backflows indicates that the actual quantum state of the bipartite system, on which heat fluctuations are evaluated, is entangled~\cite{JenningsPRE2010}.

Let us introduce the quasiprobabilities $q_{i_{c}i_{h}f_{c}f_{h}}$ associated to the energy variations $\Delta E_{i_{c}i_{h}f_{c}f_{h}}$, eigenvalues of the total Hamiltonian $\mathcal H$ of the two bodies. Using again the definition of the QD in Eq.~\eqref{eq:PDeltaU}, one has that  
\begin{equation}\label{eq:KDQ_heat}
    q_{i_{c}i_{h}f_{c}f_{h}} = p_{i_{c}i_{h}f_{c}f_{h}} + {\rm Tr}\left[ \Pi_{f_{c}f_{h}}^H \Pi_{i_{c}i_{h}} \chi \right],
\end{equation}
where $\Pi_{f_{c}f_{h}}^H = U^{\dagger}\Pi_{f_{c}f_{h}} U$ and
\begin{equation}\label{eq:joint_TPM_heat}
    p_{i_{c}i_{h}f_{c}f_{h}} = {\rm Tr}\left[ \Pi_{f_{c}f_{h}}^H \Pi_{i_{c}i_{h}} \mathcal{D}_1[\rho] \right]
\end{equation}
is the corresponding joint probabilities returned by the TPM scheme. As before, in Eq.~\eqref{eq:joint_TPM_heat} we have employed the dephasing operator $\mathcal{D}_1[\rho]=\sum_{i_ci_h} \Pi_{i_ci_h}\rho\,\Pi_{i_ci_h}$. In Eqs.~\eqref{eq:KDQ_heat}-\eqref{eq:joint_TPM_heat} the initial quantum state $\rho$ of the two bodies is linearly decomposed as $\rho = \mathcal{D}_1[\rho] + \chi$ [see Eq.~\eqref{eq:dephasing_operator}], where both the diagonal and off-diagonal parts of $\rho$ are considered with respect to the eigenbasis of $\mathcal{H}$.

Based on this framework, we describe an exchange fluctuation relation that is also valid in the noncommutative regime of $[\rho, \mathcal{H}] \neq 0$, due to the presence of quantum correlations or entanglement in the initial state. For this purpose, let us introduce the {\it stochastic mutual information} $I$ with elements
\begin{equation}
    I_{j_{c}j_{h}} \equiv \ln\left( \frac{ {\rm Tr}\left[ \Pi_{j_{c}j_{h}}\rho \right] }{ {\rm Tr}\left[ \Pi_{j_{c}}\rho_{{\rm th}, B_c} \right]{\rm Tr}\left[ \Pi_{j_{h}}\rho_{{\rm th}, B_h} \right] } \right)
\end{equation}
where $j=i,f$, such that $\Delta I \equiv I_{f_{c}f_{h}} - I_{i_{c}i_{h}}$. We also recall that the energy variation in the cold body is $Q \equiv E_{i_c} - E_{f_c}=E_{f_h} - E_{i_h}$, assuming the energy-preserving condition for $U$ and a resonant interactions between the bodies. Hence, we find~\cite{levy2020quasiprobability}
\begin{equation}\label{eq:QXFT}
    \langle e^{ \Delta I + \Delta\beta Q }\rangle = 1 + \Upsilon ,
\end{equation}
where the average $\langle\cdot\rangle$ in Eq.~\eqref{eq:QXFT} is made with respect to the quasiprobabilities $q_{i_{c}i_{h}f_{c}f_{h}}$, and
\begin{equation}
    \Upsilon \equiv \sum_{i_c,i_h,f_c,f_h} \frac{ {\rm Tr}\left[ \Pi_{f_{c}f_{h}}^H \Pi_{i_{c}i_{h}} \chi \right] {\rm Tr}\left[ \Pi_{f_{c}f_{h}} \rho \right] }{ {\rm Tr}\left[ \Pi_{i_{c}i_{h}} \rho \right] } \,.
\end{equation}

The correction to the exchange fluctuation theorem, $\Upsilon$, is equal to zero if $[\rho, \mathcal{H}] = 0$, which is equivalent to $\rho = \mathcal{D}_1[\rho]$ and $\chi=0$. In this case the application of the TPM scheme suffices. The exchange fluctuation relation of Eq.~\eqref{eq:QXFT} reduces to the Jarzynski-W\'ojcik identity in Eq.~\eqref{eq:JWE} in the case $\rho = \rho_{{\rm th}, B_c} \otimes \rho_{{\rm th}, B_h}$, whereby $\Delta I = 0$. 
Instead, if the diagonal elements of $\rho$ are not thermally distributed---due to classical correlations in the $\mathcal{H}$ eigenbasis---and $\chi=0$, then one recovers the exchange fluctuation relation in Ref.~\cite{JevticPRE2015}, i.e.,
\begin{equation}\label{eq:XFT_Jevtic}
    \langle e^{ \Delta I + \Delta\beta Q }\rangle = 1 \,.
\end{equation}

We conclude this theoretical analysis about heat fluctuations in the quantum regime, by providing the thermodynamic interpretation of the fluctuation profiles associated to the quasiprobability distribution of heat exchanges.  
Previously, we have seen that the presence of quantum correlations in the initial state can enhance the amount of heat backflows, such that $\langle Q\rangle \geq 0$ according to the used convention. The explanation of this phenomenon lies in the possibility to associate negative quasiprobabilities ${\rm Re}\left[ q_{i_{c}i_{h}f_{c}f_{h}} \right]$ to positive heat exchanges $Q = E_{i_c} - E_{f_c} > 0$ corresponding to heat flowing from the cold body to the hot one. Such a process, which needs an external energy source for its activation, is triggered by quantum correlations. 

Notice that, in order for quantum correlations to be effectively considered as a resource for heat backflows, it is required that negative heat exchanges $Q \leq 0$ (i.e., energy fluxes from the hot to the cold bodies) occur with positive quasiprobabilities ${\rm Re}\left[ q_{i_{c}i_{h}f_{c}f_{h}} \right]$, similarly to what happens to work extraction in Sec.~\ref{sec:extractable_work}. When the enhancement induced by quantum correlations in $\rho$ allows for strong cold-hot heat backflows, then one can state that the corresponding energy exchange process takes nonclassical traits.

\subsubsection{Example: two-qubit system}

We now apply the theoretical framework introduced above to a pair of interacting qubits, at inverse temperatures $\beta_c$ and $\beta_h$, respectively, and local Hamiltonians $\mathcal{H}_{B_k}=\Omega \sigma^{z}_{k}, k=c,f$. The two qubits are initialized in a global state $\rho$ containing off-diagonal elements with respect to $\mathcal H= \mathcal{H}_{B_c} + \mathcal{H}_{B_h}$.

As proven in \cite{levy2020quasiprobability}, a general form of the initial state for a two-qubit system fulfilling the requirements \eqref{eq:reduced_thermal_c}-\eqref{eq:reduced_thermal_h} is
\begin{equation}\label{eq:initial_coherent_state_heat}
    \rho = \begin{pmatrix}
    p & 0 & 0 & 0 \\
    0 & \alpha_{c}^{-1}-p & \eta \, e^{i\xi} & 0 \\
    0 & \eta \, e^{-i\xi} & \alpha_{h}^{-1}-p & 0 \\
    0 & 0 & 0 & \frac{\alpha_{c}\alpha_{h}-\alpha_{c}-\alpha_{h}}{\alpha_{c}\alpha_{h}}+p
    \end{pmatrix}
\end{equation}
where $p \in [0,1]$ is a population term, $\alpha_{k} \equiv 1 + e^{-\beta_{k}}$ ($k=c,h$), $\xi \in [0,2\pi]$, and $|\eta| \leq \sqrt{(\alpha_{c}^{-1}-p)(\alpha_{h}^{-1}-p)}$ such that $\rho \geq 0$. In Ref.~\cite{levy2020quasiprobability}, it is also shown that the energy-preserving condition [Eq.~\eqref{eq:energy-preserving}] is responsible to set a minimal form for the unitary operator $U$:
\begin{equation}\label{eq:U_heat_2qubits}
    U = \begin{pmatrix}
    1 & 0 & 0 & 0 \\
    0 & \cos(\theta) & -\sin(\theta) & 0 \\
    0 & \sin(\theta) & \cos(\theta) & 0 \\
    0 & 0 & 0 & 1
    \end{pmatrix}
\end{equation}
with $\theta\in [0,2\pi]$, equivalent to a partial swap transformation.

\begin{figure}
    \centering
\includegraphics[width=0.95\columnwidth]{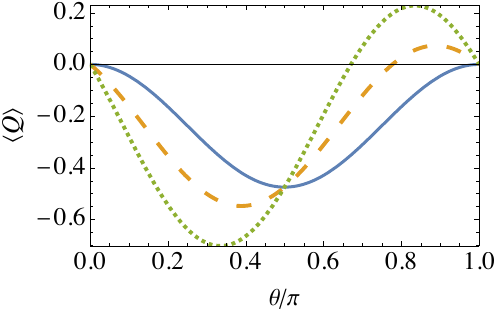}
    \caption{Average heat exchanged $\langle Q\rangle$, see Eq.~\eqref{eq:average_heat_2qubits}, against $\theta$ for different values of the coherence strength $\eta=0$ (solid line), $\eta=0.2$ (dashed line), $\eta=0.4$ (dotted line). Note that $\langle Q\rangle = \langle Q\rangle_{\rm TPM}$ for $\eta=0$. Other parameters: $p=0,\; \xi=0,\; \beta_c=10,\; \beta_h=0.1$.}
    \label{fig:Qav}
\end{figure}

Under these assumptions, the analytical expression of the heat exchange \eqref{eq:average_heat} between the two bodies reads as~\cite{levy2020quasiprobability}
\begin{equation}\label{eq:average_heat_2qubits}
    \langle Q\rangle = - \eta\cos(\xi)\sin(2\theta) + \langle Q\rangle_{\rm TPM}\,,
\end{equation}
where 
\begin{equation}\label{eq:average_heat_2qubits_TPM}
\langle Q\rangle_{\rm TPM} = \sin^{2}(\theta)\left( \frac{1}{ 1 + e^{\beta_c} } - \frac{1}{1 + e^{\beta_h}} \right)
\end{equation}
is the average heat flow obtained by applying the TPM scheme, or by setting $\eta  = 0$ (no quantum coherence) in Eq.~\eqref{eq:average_heat_2qubits}. For any value of $\theta$, $\beta_c$ and $\beta_h$, it can be easily found that $\langle Q\rangle_{\rm TPM} \leq 0$, which means that no cold-to-hot heat backflow is possible. This is evident in Fig.~\ref{fig:Qav} where we plot Eq.~\eqref{eq:average_heat_2qubits} against $\theta$ for $\eta=0$ and for $\eta\neq 0$ for fixed $p,\xi,\beta_c,\beta_h$. From the figure it can be observed that, for some values of $\theta$, $\langle Q\rangle > 0$ (cold-to-hot heat backflows) is possible when $\eta\neq 0$. The parameter $\eta$ also affects the magnitude of the heat exchanged between the cold and hot systems.

\section{Quantum many-body systems}

Quasiprobability distributions play an instrumental role in the understanding of quantum many-body systems. In this section, we will present their applications in the context of quantum scrambling and out-of-time-ordered correlators (Sec.~\ref{sec:OTOC}), the Loschmidt echo (Sec.~\ref{sec:Link_with_LE}) and a technique for efficiently calculating the KDQ distribution of free fermion systems (Sec.~\ref{sec:quadratic}).

\subsection{Quantum scrambling and out-of-time-ordered correlators (OTOCs)}
\label{sec:OTOC}

Out-of-time-ordered correlators (OTOCs) allow the study of quantum information scrambling, a phenomenon in which localized quantum information rapidly spreads across multiple degrees of freedom in many-body systems~\cite{Larkin1969,SwingleNatPhys2018}. OTOCs have recently found extensive applications in diverse fields, such as condensed-matter physics, quantum chaos, holography, and the study of black holes. Their versatility has thus pushed the development of numerous experimental proposals aimed at measuring OTOCs~\cite{SwinglePRA2016,YoshidaPRX2019,VermerschPRX2019,DagPRA2019,SundarNJP2022,WangPRR2022}, with some experiments already realized~\cite{LiPRX2017, GarttnerNatPhys2017, LandsmanNature2019,NiePRL2020,BraumullerNatPhys2022,LiScience2023}.

In this section, we review the basic definition of OTOCs and show how they relate to quasiprobabilities. This connection has been recently noted in various works~\cite{HalpernPRA2017,HalpernPRA2018,dressel2018strengthening,AlonsoPRL2019}.

Following Ref.~\cite{HalpernPRA2017}, let us consider a system that is initially prepared at time $t=0$ in the pure state $\ket{\psi}$ and initially perturbed by a unitary operator $V$, acting locally on a part of the system. For instance, if the system is made of qubits, we may consider the spin flip $V=\sigma^x_{j}$ that acts on qubit $j$. The system is then evolved for a time interval $t$ following a time evolution described by the unitary operator $U$. 
At the end of the dynamics, the system is perturbed by the application of another unitary operator $Y$~\footnote{Although in the OTOC literature, this is normally denoted as $W$, we choose a different symbol to avoid confusion with the symbol for the work.}, acting locally on another disjoint part of the system. Finally, the system is evolved backward in time through the operator $U^\dagger$. At the end of this protocol, the state of the system is: $\ket{\psi'}=U^\dagger Y U V \ket\psi$. Now, suppose that we perform an alternative protocol in which the perturbation $V$ is applied, not after the initial preparation of $\rho$, but after the backward evolution. This results in the state $\ket{\psi''}=VU^\dagger Y U \ket\psi$. The overlap between these two states equals the OTOC
\begin{equation}\label{eq:OTOCF}
    F(t) \equiv \braket{\psi''}{\psi'} = \left\langle Y^\dagger_t V^\dagger Y_t V \right\rangle,
\end{equation}
where we have defined $Y_t = U^\dagger Y U$ that denotes the perturbation operator $Y$ evolved in the Heisenberg picture.
The definition in Eq.~\eqref{eq:OTOCF} can be extended to initial mixed states $\rho$ and from now on we will assume: $\langle \cdot \rangle = {\rm Tr}[\rho\;\cdot]$, so that $F(t)={\rm Tr}\left[ \rho \, Y^\dagger_t V^\dagger Y_t V \right]$.

While the OTOC is in general a complex number, one can consider a real quantity by introducing the OTO commutator:
\begin{equation}
C(t) \equiv \frac 12 \left\langle [Y_t,V]^\dagger [Y_t,V] \right\rangle.
\end{equation}
Its interpretation is the following. Initially, at time $t=0$ ($U=\mathbb{I}$), the operators $Y_t = Y$ and $V$ commute as they have support on spatially separated parts of the system: $[Y,V]=0$. However, as time progresses, the effects of the perturbation $Y_t$ may reach the support region of $V$ and their commutator might become nonzero: $[Y_t,V]\neq 0$. The quantity $C(t)$ measures the magnitude of this commutator.
 If both operators $V$ and $Y$ are unitary, the OTO commutator is related to the OTOC by the relation
\begin{eqnarray}
C(t)&=&\frac 12 \left\langle (V^\dagger Y_t^\dagger -   Y_t^\dagger  V^\dagger)(Y_tV-VY_t) \right\rangle =
\nonumber \\
&=&1 - \frac 12 \left\langle   Y_t^\dagger  V^\dagger Y_tV+   V^\dagger Y_t^\dagger VY_t   \right\rangle =
\nonumber \\
&=& 1-\Re[F(t)]\,.
\end{eqnarray}
This shows that the growth of the commutator $C(t)$ is associated with a decay of the real part of $F(t)$.

Next, we are going to prove that an OTOC is equal to the characteristic function of a KDQ. To this end, let us follow Ref.~\cite{CampisiPRE2017thermodynamics}, which introduces the {\it wing-flap protocol} and express the unitary operator $V$ in exponential form as
%
%
$V(u)=e^{i u \mathcal O}$, with $\mathcal O$ a Hermitian operator and $u$ a real scalar. The spectral decomposition of the observable $\mathcal O$ reads: $\mathcal O = \sum_m o_m \Pi_m$, in terms of its real eigenvalues $o_m$ and the corresponding orthogonal projectors $\Pi_m$. Using these definitions, $V$ can be expressed as
\begin{equation}
V(u)=\sum_m e^{iu o_m}\Pi_m \,.
\end{equation}
The wing-flap protocol consists of the following steps:
\begin{enumerate}
    \item Prepare the system in the state $\rho$.
    \item Measure $\mathcal O$.
    \item Evolve the system forward in time with $U$.
    \item Apply the perturbation $Y$.
    \item Evolve the system backward in time with $U^\dagger$.
    \item Measure $\mathcal O$.
\end{enumerate}

Recalling the definition in Eq.~\eqref{eq:def_KDQ}, we can define the KDQ 
\begin{equation}\label{eq:KDQ_OTOC}
q_{nm}(t)=\left\langle Y^\dagger_t \Pi_m  Y_t \Pi_n \right\rangle,
\end{equation}
for the change $\Delta o_{nm}=o_m-o_n$ in the eigenvalues of $\mathcal{O}$ when two measurements of $\mathcal O$ are performed at times $t_1=0$ and $t_2=t$, respectively (steps 2 and 6 of the wing-flap protocol). In Eq.~\eqref{eq:KDQ_OTOC}, the operator $Y_t$ plays the role of the complete evolution operator $Y_t = U^\dagger Y U$ that combines the steps 3-5 of the protocol between the two measurements of $\mathcal{O}$. Notice that, in order to denote a quasiprobability, we continue using the simplified notation adopted in Sec.~\ref{sec:QTD}, whereby the subscript in $q_{nm}$ contains the indexes labelling the measurement outcomes at the initial and final times of a two-time procedure.

As a result, the quasiprobability distribution to observe a change $\Delta o(t)$ at time $t$ is given by
\begin{equation}
P[\Delta o,t] = \sum_{n,m} q_{nm}(t) \delta(\Delta o(t) - \Delta o_{nm}),
\end{equation}
and the characteristic function of $P[\Delta o,t]$ is its Fourier transform (see Sec.~\ref{sec:distr_and_char-func_KDQ}), such that
\begin{eqnarray}
\label{eq:G_KDQ}
\mathcal{G}(-u,t) &=& \int_{-\infty}^\infty P[\Delta o,t]e^{-iu\Delta o}d\Delta o 
= \nonumber \\
&=&\sum_{n,m}q_{nm}(t)e^{-iu\Delta o_{nm}}=
\nonumber \\
&=&\sum_{n,m}\left\langle Y_t^{\dagger}\Pi_m Y_t\Pi_n\right \rangle e^{-iu(o_m-o_n)} = \nonumber \\
&=& \left\langle Y_{t}^{\dagger}V^\dagger(u) Y_{t} V(u)\right\rangle = F(t) 
\end{eqnarray}
that can thus be expressed as an OTOC.

In general, whenever $[\rho, \mathcal O]\neq 0$, both the KDQ $P[\Delta o,t]$ and its characteristic function $\mathcal{G}(u,t)$ are complex numbers. 
When $[\rho, \mathcal O]=0$ (as in Ref.~\cite{CampisiPRE2017thermodynamics}), the KDQ is real and positive, as explained in Sec.~\ref{section_quasiprobs}.

Similar to Secs.~\ref{section_quasiprobs}-\ref{sec:QTD}, we can define the corresponding MHQ distribution that can be associated with an OTOC. Such a distribution is the real part of the corresponding KDQ distribution, that is $r_{nm} = \Re\left[ q_{nm}(t) \right]$, and its characteristic function reads as
\begin{equation}
\mathcal{G}_{\rm MHQ}(-u,t)=\frac{\mathcal G(-u,t)+\mathcal G^*(u,t)}{2}\,.
\end{equation}
Interestingly, using the equality
\begin{eqnarray}
\mathcal{G}(u,t) &=& \left\langle Y_t^{\dagger} V^{\dagger}(-u)Y_t V(-u)\right \rangle= 
\nonumber \\
&=&\left\langle Y^{\dagger}_t V(u) Y_t V^{\dagger}(u)\right\rangle,
\end{eqnarray}
we obtain
\begin{eqnarray}
&&
\mathcal{G}_{\rm MHQ}(-u,t)
=\nonumber \\
&&=\frac 12 \left[ \left\langle Y_{t}^{\dagger}V^\dagger(u) Y_{t} V(u)\right\rangle
+\left\langle V(u) Y^{\dagger}_t V^{\dagger}(u) Y_t\right\rangle \right]=
\nonumber
\\
&&
%
%
={\rm Tr}\left[ \left\{ \rho , V(u) \right\} Y^{\dagger}_t V^{\dagger}(u) Y_t \right].
\end{eqnarray}
%

\begin{figure*}[t]
\begin{center}
\includegraphics[width=0.7\columnwidth]{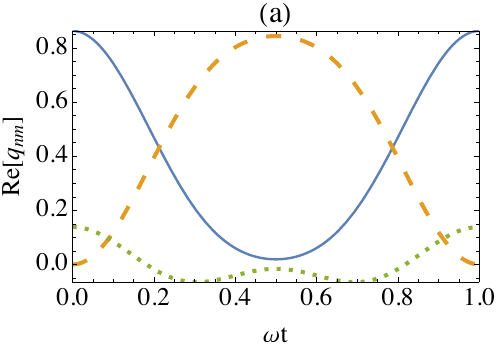}
\includegraphics[width=0.75\columnwidth]{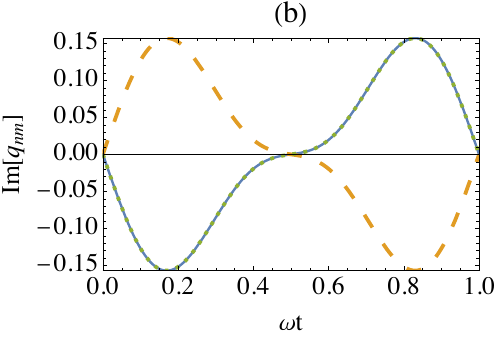}
\\
\includegraphics[width=0.75\columnwidth]{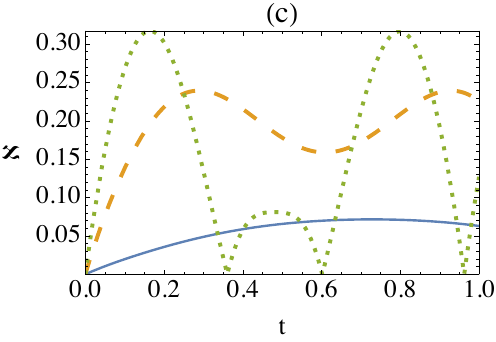}
\includegraphics[width=0.7\columnwidth]{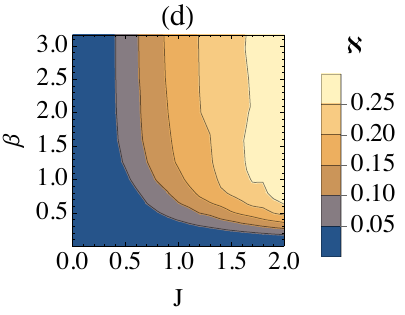}
\caption{
We plot the real and imaginary parts of $q_{nm}$ in panels (a) and (b) respectively, as a function of $\omega t$, for $J=2$ (lines, blue solid: $nm=11$, orange dashed: $nm=10$, green dotted: $nm=00$). The nonpositivity functional $\aleph$ [see Eq.~\eqref{eq:def_aleph}] is shown in panel (c), as a function of time for different values of the coupling $J$ (lines, blue solid: $J=0.5$, orange dashed: $J=1.5$, green dotted: $J=2.5$). (d) Contour plot of the the minimum value of $\aleph$ in the time interval $0\le t \le 20$ as a function of $\beta$ and $J$. Other parameters: $u=\pi/2, B_1=1, B_2=1.1, \beta=10$.
}
\label{fig:OTOC1}
\end{center}
\end{figure*}

Let us now consider a practical example and let 
\begin{equation}
\mathcal{H} = B_1 \sigma^z_1+B_2\sigma^z_2 + J \sigma^x_1\sigma^x_2
\end{equation}
be the Hamiltonian of two qubits initially in the thermal state
\begin{equation}
\rho = \frac{e^{-\beta \mathcal{H}}}{{\rm Tr}\left[e^{-\beta \mathcal{H}}\right]}
\end{equation}
with inverse temperature $\beta$. Then, we choose $Y=\sigma^z_1$ and $\mathcal O=\sigma^z_2$. Notice that neither $Y$ nor $\cal O$ commute with either $\mathcal{H}$ or $\rho$. Therefore, this is an ideal setting to test the possible presence of nonpositivity in the KDQ \eqref{eq:KDQ_OTOC}.
For this purpose, we write the measurement observable as 
$\mathcal O = \ketbra{0}{0} - \ketbra{1}{1}$, from which $\Pi_0 = \ketbra{0}{0}$ and $\Pi_1 = \ketbra{1}{1}$ with the corresponding eigenvalues $o_0=1$ and $o_1=-1$. We stress that the projectors $\Pi_0, \Pi_1$ of $\mathcal O$ act locally  on qubit 2. One can see that the evolution operator $U = \exp(-i\mathcal{H}t)$ is periodic with the frequency $\omega=2\sqrt{(B_1+B_2)^2+J^2}$.

In Fig.~\ref{fig:OTOC1}(a)-(b), we show the real and imaginary parts of $q_{nm}$. The quasiprobability $q_{00}$ is the only one whose real part becomes negative. Since the probabilities need to fulfil the normalization condition, it means that $q_{11}$ is amplified due to the presence of quantum coherences in the initial state $\rho$, compared to a case in which the initial state commutes with the observable $\mathcal O$.

In Fig.~\ref{fig:OTOC1}(c), the nonpositivity functional $\aleph$ is plotted against the time for different values of $J$. In Fig.~\ref{fig:OTOC1}(d), we show a contour plot of the minimum value of $\aleph$ in the time interval $0\le t \le 20$ as a function of $\beta$ and $J$. First, as expected, we confirm that larger values of the interaction strength $J$ always lead to a stronger nonpositivity. Second, the nonclassicality reduces as the temperature increases. 

\subsection{Link with the Loschmidt echo}\label{sec:Link_with_LE}

Another interesting connection of quasiprobability distributions with the irreversibility of many-body systems arises in the context of quantum chaos and decoherence. Consider a system of many particles, classical or quantum, that evolves in time for a period $t$ according to a time-independent Hamiltonian $\mathcal{H}_0$. If we were to invert all momenta and run the evolution backward we should be able to recover the initial state. However, little imperfections in the inverted evolution or decoherence induced by an external environment may cause some deviations. For an initial pure state $\ket\psi$, one can define the Loschmidt echo (LE) $\mathcal{L}(t) = |\mathcal G(t)|^2$ as the absolute square value of the complex amplitude
\begin{equation}\label{eq:LEGt}
    \mathcal G(t) = \bra{\psi} e^{i\mathcal{H}_0 t} e^{-i \mathcal{H}_\delta t}\ket{\psi},
\end{equation}
whose generalization to mixed states and non unitary evolutions is straightforward. 

Mathematically, $\mathcal{L}(t)$ represents the fidelity, in terms of the overlap, between the initial state $\ket{\psi}$ evolved with the unperturbed Hamiltonian $\mathcal{H}_0$ and the state $\ket{\psi}$ evolved with the perturbed Hamiltonian $\mathcal{H}_\delta$. Peres transferred the LE idea in the quantum domain~\cite{PeresPRA84}, while Ref.~\cite{CucchiettiPRL2003} used the LE to analyze the decoherence of a many-body spin system and the relation to chaos. For the quantum version of systems with a
classically chaotic Hamiltonian (for instance a particle moving in a driven double well, see Ref.~\cite{CucchiettiPRL2003}) the rate at which the information about the initial state is destroyed by the environment, within a range of couplings to the environment, is set by the classical maximal Lyapunov exponent. Under these assumptions, the LE decays exponentially in time with the Lyapunov exponent, thus revealing the underlying classical chaotic behaviour; see also Ref.~\cite{JalabertPRL2001}.

Moreover, the LE was beneficial to uncover a new type of phase transition occurring in time. In this regard, if we assume that the initial state $|\psi\rangle$ is the ground state of $\mathcal{H}_0$ with zero energy, then the LE amplitude \eqref{eq:LEGt} reduces to
\begin{equation}
    \mathcal{G}(t) = \bra{\psi} e^{-i\mathcal{H}_\delta t}\ket{\psi},
\end{equation}
which looks like the partition function of the Hamiltonian $\mathcal{H}_\delta$ but with an imaginary inverse temperature $it$. Since classical phase transitions arise because of {\it singularities} in the partition function, Heyl and coworkers discovered dynamical quantum phase transitions as those that give rise to singularities in the LE at specific instants of time, see Refs.~\cite{HeylPRL2013,HeylRPP2018}.

The LE is also strongly connected with the statistics of work as mentioned in Sec.~\ref{sec:QTD} and described in detail in Ref.~\cite{SilvaPRL2008}. In fact, Eq.~\eqref{eq:LEGt} can be interpreted as the characteristic function of the work done on a quantum system initially in the state $\ket\psi$, whose Hamiltonian is subject to a {\it quench dynamics} that instantaneously changes $\mathcal{H}_0$ to $\mathcal{H}_\delta$.

Let us now formalize the connection between the LE and the KDQ. First, we write the spectral decomposition of the two Hamiltonian operators: $\mathcal{H}_0=\sum_n E_n \Pi_n$ and  $\mathcal{H}_\delta=\sum_m E^{(\delta)}_m \Pi^{(\delta)}_m$. With these assumptions, let us write an expression for the LE for a generic mixed initial state:
\begin{eqnarray}
    \label{eq:LEgen}
    {\mathcal G}(t) &=& {\rm Tr}\left[ \rho \, e^{i \mathcal{H}_0 t} e^{-i \mathcal{H}_\delta t} \right]=\nonumber
    \\
    &=& \sum_{n,m} e^{-i (E^{(\delta)}_m -E_n)t} q_{nm}\,,
\end{eqnarray}
where we have introduced the KDQ $q_{nm}$ of the random variable $W = E^{(\delta)}_m - E_n$, defined as
\begin{equation}
    q_{nm} = {\rm Tr}\left[\rho\,\Pi_n  \Pi^{(\delta)}_m \right].
\end{equation}
Thus, the inverse Fourier transform of ${\mathcal G}(t)$ with respect to time $t$, i.e.,
\begin{eqnarray}
    P[W] &=& \int^{\infty}_{-\infty} {\mathcal G}(t) e^{i W t}dt =\nonumber\\
    &=&
    \sum_{n,m} \delta(W-E^{(\delta)}_m+E_n) q_{nm},
\end{eqnarray}
can be interpreted as the quasiprobability distribution for the work $W=E^{(\delta)}_m-E_n$ done on the quantum system, which initially is in the state $\rho$ and whose Hamiltonian is suddenly changed from $\mathcal{H}_0$ to $\mathcal{H}_\delta$. It is worth noting that in contrast to the case of the characteristic function of work obtained using the TPM scheme~\cite{SilvaPRL2008}, in the general case the initial state $\rho$ may not commute with any of the two Hamiltonian operators $\mathcal{H}_0$ and $\mathcal{H}_\delta$. This fact, as we have seen in other examples earlier, may give rise to a distribution $P[W]$ with nonpositive values.

\begin{figure}[t]
\begin{center}
\includegraphics[width=0.95\columnwidth]{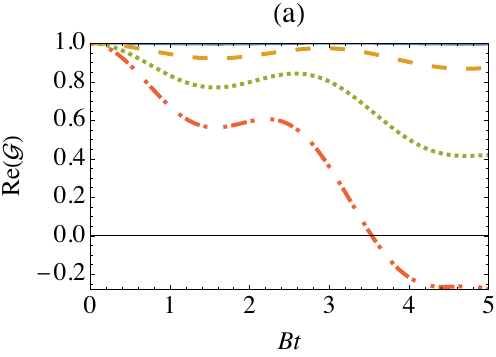}
\includegraphics[width=0.95\columnwidth]{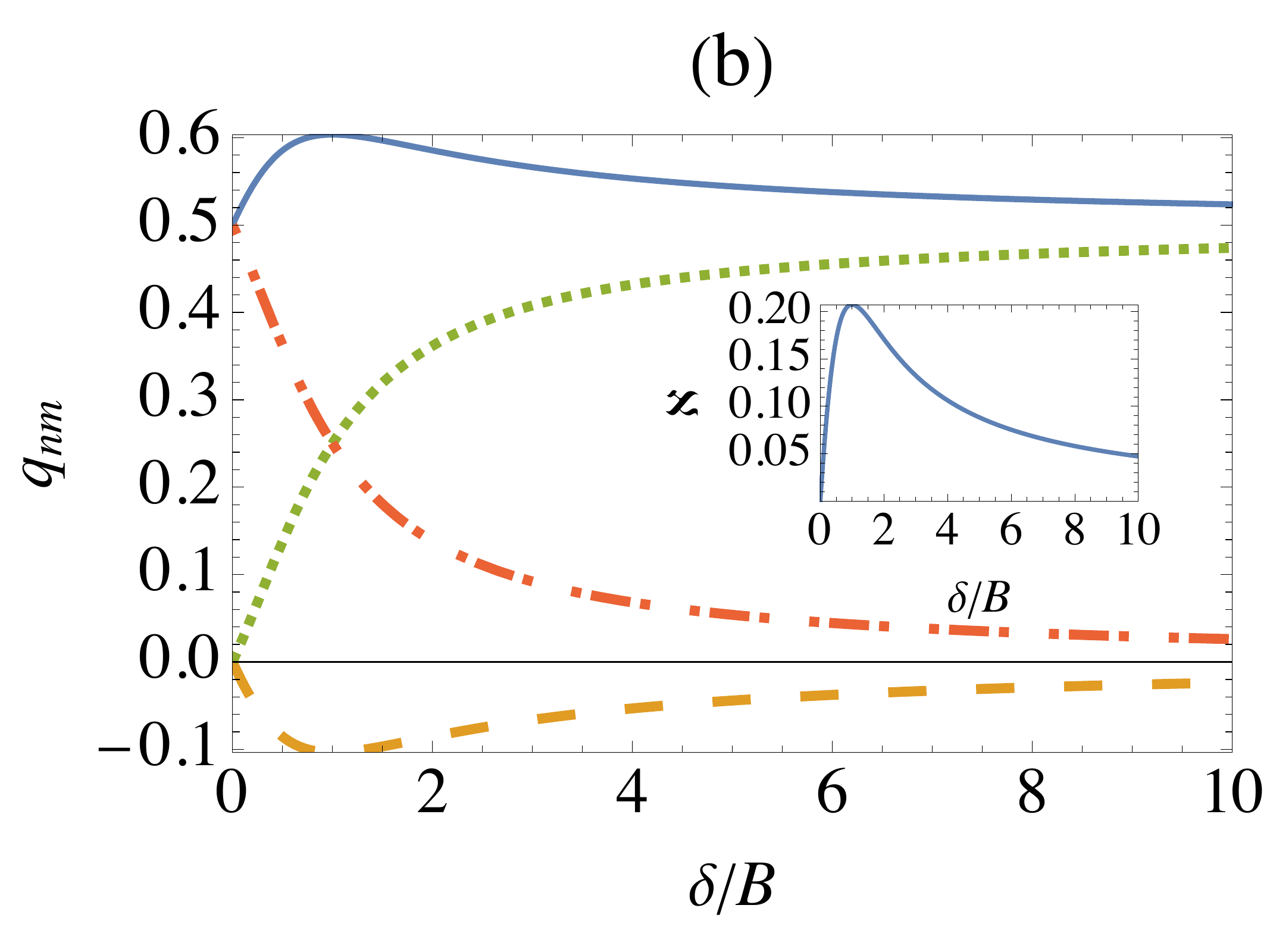}
\caption{(a) Real part of LE $\mathcal G(t)$ as a function of time for different values of $\delta$: 0.1 (blue solid line), 0.4 (orange dashed line) 0.7 (green dotted line) 1.0 (red dot-dashed line). (b) Quasiprobabilities $q_{nm}$ associated with the LE of Eq.~\eqref{eq:LEqubit} as a function of $\delta/B$: $q_{00}$ (blue solid line), $q_{01}$ (orange dashed line), $q_{10}$ (green dotted line), $q_{11}$ (red dot-dashed line). Inset: nonpositivity functional $\aleph$ [see Eq.~\eqref{eq:def_aleph}] as a function of $\delta/B$.}
\label{fig:LE}
\end{center}
\end{figure}

We now proceed to illustrate these concepts with a simple example. Let us consider a qubit in the pure initial state
\begin{equation}
\ket{\psi} =\frac{\ket 0+\ket 1}{\sqrt 2}
\end{equation}
and let us choose the Hamiltonian operators $\mathcal{H}_0$ and $\mathcal{H}_\delta$ as
\begin{eqnarray}
    \mathcal{H}_0 &=& B\sigma^z,\\
    \mathcal{H}_\delta &=& \mathcal{H}_0 + \delta\sigma^x.
\end{eqnarray}
The eigenstates of $\mathcal{H}_0$ are simply $\ket 0$ and $\ket 1$, with eigenvalues $\pm B$ respectively, and we define the projectors on these states as $\Pi_i=\ketbra{i}{i}$ with $i=0,1$.
Similarly, for $\mathcal{H}_\delta$, the eigenstates are
\begin{eqnarray}
    \ket{0_\delta} &=& \cos(\theta)\ket 0 + \sin(\theta)\ket 1, \\
    \ket{1_\delta}&=& -\sin(\theta)\ket 0 + \cos(\theta)\ket 1
\end{eqnarray}
with eigenvalues $\pm B_\delta$, where $B_\delta \equiv \sqrt{B^2+\delta^2}$ and the mixing angle $\theta$ defined by
\begin{equation}
    \tan(\theta) = \frac{\delta}{\delta^2+2B(B+B_\delta)}\,.
\end{equation}
Hence, for the LE, we get:
\begin{eqnarray}\label{eq:LEqubit}
     \mathcal G(t) &=& \cos(Bt)\cos(B_\delta t) + \nonumber \\
     &+& \frac{B \sin(Bt) -i\delta \cos(Bt)}{B_\delta}\sin(B_\delta t),
\end{eqnarray}
whose real part is plotted in Fig.~\ref{fig:LE}(a). 

One can observe that for very small $\delta$ the two evolutions associated with $\mathcal{H}_0$ and $\mathcal{H}_\delta$ are very similar and $\mathcal G(t)$ remains close to $1$. However, when $\delta$ increases, the perturbed Hamiltonian $\mathcal{H}_\delta$ induces a diverging trajectory for the initial state $|\psi\rangle$. As the system is small, large revivals of the LE are possible for longer times, but the short time response is symptomatic of what would happen for a much larger system.

Moreover, for the KDQ, we obtain:
\begin{equation}
    q_{nm} = \frac {1}{4 B_\delta}\left[B_\delta + (-1)^m (\delta + (-1)^n B)\right],
\end{equation}
where $n,m=0,1$.
After a close inspection, since $B_\delta<B+\delta$, it is evident that $q_{01}<0$ that corresponds to the transition between the highest and the lowest energy eigenstates of $\mathcal{H}$ and $\mathcal{H}_\delta$, respectively. Instead, all the other quasiprobabilities are strictly positive. They are plotted in Fig.~\ref{fig:LE}(b). In the inset of the figure, we plot the nonpositivity functional $\aleph$, which is always nonzero (since one quasiprobability $q_{01}$ is always negative) and peaks around $\delta\sim B$.

\subsection{Quantum work in quadratic fermionic systems}
\label{sec:quadratic}

In this section, we explain how to calculate the KDQ for systems of free fermions described by Hamiltonians that are quadratic in fermionic creation and annihilation operators. This is relevant not only for actual fermionic systems, for instance ultracold atoms in optical lattices, but also for systems that can be mapped onto free fermion models, for instance Ising and XY spin chains.

Quasiprobability distributions of work in quadratic fermionic systems have been recently calculated in a few references; see for instance Refs.~\cite{XuPRA2018,FrancicaPRE2023,SantiniPRB2023}. Inspired by the approach taken in Ref.~\cite{XuPRA2018}, here we showcase the calculation of the KDQ for an Ising spin chain with $N$ spin-$1/2$ particles, described by the Hamiltonian
\begin{equation}
\label{eq:HIsing}
    \mathcal{H}(\lambda) = -\sum_{j=1}^N \left(\lambda \sigma^z_j+\sigma_j^x \sigma^x_{j+1} \right),
\end{equation}
where periodic boundary conditions are assumed: $\sigma^\alpha_{N+1}\equiv\sigma^\alpha_{1}$, $\alpha=x,y,z$. The parameter $\lambda$ is an effective transverse magnetic field. The critical value $\lambda_c=1$ separates the ferromagnetic phase, existing for $\lambda<\lambda_c$, from the paramagnetic phase occurring for $\lambda >\lambda_c$. Within the ferromagnetic phase, in the thermodynamic limit, the ground state is doubly degenerate with a macroscopic magnetization along $x$, while in the paramagnetic phase the ground state is non degenerate and exhibits an induced magnetization along $z$.

To diagonalize the Ising Hamiltonian in Eq.~\eqref{eq:HIsing}, following Refs.~\cite{LiebAP1961,Pfeuty1970}, we first employ the {\it Jordan-Wigner transformation} that expresses the fermionic annihilation operators
\begin{eqnarray}
    a_i = \left(\prod_{j=1}^{i-1} \sigma^z_j\right) \sigma^-_i 
\end{eqnarray}
in terms of the spin ladder operators $\sigma^-_i \equiv \frac 12(\sigma^x_i-i\sigma^y_i)$.

Then, we transpose the problem to the quasimomentum space by defining the fermionic operators
\begin{equation}
    c_k = \frac{1}{\sqrt N}\sum_{j=1}^N e^{-i k j} a_j \,,
\end{equation}
where the possible values of the quasimomenta are $k=2\pi m/N$ with $m=-N/2+1,\dots, N/2$ (assuming for simplicity $N$ even). Let us thus define the fermionic operators $\gamma_k$ that are obtained by applying the following Bogoliubov rotation to the operators $c_k$:
\begin{equation}
    \gamma_k = \cos\left( \frac{\theta_k}{2} \right) c_k -i\sin\left( \frac{\theta_k}{2} \right) c_{-k}^\dagger\,.
\end{equation}
The fermionic operators $\gamma_k$ depend on the angles $\theta_k$, implicitly given by
\begin{equation}
    e^{i\theta_k} = \frac{\lambda-e^{-ik}}{\sqrt{\sin^2 k+(\lambda-\cos k)^2}}\,,
\end{equation}
and satisfy the canonical anticommutation relations
\begin{equation}
    \{\gamma_k,\gamma_{k'}^\dagger\} =\delta_{kk'}, \quad \{\gamma_k,\gamma_{k'}\}=0 \,.
\end{equation}
In terms of the operators $\gamma_k$, the Hamiltonian Eq.~\eqref{eq:HIsing} reduces to a diagonal form:
\begin{equation}
    \mathcal{H}(\lambda)=\sum_k \epsilon_k(\lambda) \left( \gamma_k^\dagger \gamma_k-\frac 12 \right),
\end{equation}
where 
\begin{equation}
    \epsilon_k(\lambda)=2\sqrt{\sin^2 k+(\lambda-\cos k)^2}
\end{equation}
are the single-particle eigenenergies.

In what follows, we consider a sudden change of the Hamiltonian $\mathcal{H}(\lambda)$ in which the magnetic field $\lambda$ is changed instantaneously from the initial value $\lambda_0$ to the final value $\lambda_1$. To calculate the  quasiprobabilities defined in Eq.~\eqref{eq:qif}, we need to express the fermionic operators $\gamma_k^{(1)}$, which diagonalize $\mathcal{H}(\lambda_1)$ with eigenenergies $\epsilon_k^{(1)}=\epsilon_k(\lambda_1)$, in terms of the fermionic operators $\gamma_k^{(0)}$ diagonalizing $\mathcal{H}(\lambda_0)$ with eigenenergies $\epsilon_k^{(0)} = \epsilon_k(\lambda_0)$. This is possible thanks to the linear Bogoliubov transformation~\cite{LiebAP1961}:
\begin{eqnarray}
    \gamma_k^{(1)} &=& \gamma^{(0)}_k \cos\left( \frac{\Delta_k}{2} \right) + \gamma_{-k}^{(0)\dagger}\sin\left( \frac{\Delta_k}{2} \right),
\end{eqnarray}
where $\Delta_k \equiv \theta_k^{(1)} - \theta_k^{(0)}$ denotes the difference of Bogoliubov angles $\theta_k^{(j)}$ corresponding to $\lambda_j$, with $j=0,1$. 
Let us now define the vacuum states $\ket{0_k}$ and $\ket{\tilde 0_k}$ that are such that $\gamma_k^{(0)}\ket{0_k} = 0$ and $\gamma_k^{(1)}\ket{\tilde 0_k} = 0$. The vacua of the two Hamiltonian operators $\mathcal{H}(\lambda_0)$ and $\mathcal{H}(\lambda_1)$ are related by the relation
\begin{equation}
    \label{eq:vacua}
    \ket{0_k0_{-k}} = \left( \cos\left(\frac{\Delta_k}{2}\right) +
       \sin\left(\frac{\Delta_k}{2}\right)\gamma_k^{(1)\dagger}\gamma_{-k}^{(1)\dagger}\right )\ket{\tilde 0_k \tilde 0_{-k}}.
\end{equation}
Here, the need for a quasiprobability approach arises whenever one chooses an initial state $\rho$ that has coherences in the eigenbasis of $\mathcal{H}(\lambda_0)$. In our example, we choose the following state:
\begin{equation}\label{eq:initial_state_fermions}
    \rho = p\ketbra{\Psi_G}{\Psi_G}+(1-p)\rho_G(\lambda_0),
\end{equation}
with $0\le p \le 1$. In Eq.~\eqref{eq:initial_state_fermions}, $\rho_G(\lambda_0)$ is the Gibbs equilibrium thermal state that corresponds to the initial Hamiltonian, i.e.,
\begin{eqnarray}\label{eq:Gibbs_thermal_state}
    & \displaystyle{ \rho_G(\lambda_0) = \frac{e^{-\beta \mathcal{H}(\lambda_0)}}{Z(\lambda_0)}=} & \\    
    & \displaystyle{ =\frac{1}{Z(\lambda_0)} \bigotimes_k\left[e^{-\beta\epsilon_k^{(0)}/2} \ketbra{1_k}{1_k}+e^{\beta\epsilon_k^{(0)}/2}\ketbra{0_k}{0_k}
    \right] }. & \nonumber 
\end{eqnarray}
In Eq.~\eqref{eq:Gibbs_thermal_state}, $\beta$ is the inverse temperature, and $Z(\lambda_0)=\prod_k Z_k(\lambda_0)$ is the total partition function, with $Z_k(\lambda_0) = 2{\rm cosh}(\beta\epsilon_k^{(0)}/2)$ denoting the partition function for each quasimomentum. Moreover, in Eq.~\eqref{eq:initial_state_fermions}, we have  introduced the coherent Gibbs state $\ket{\Psi_G}$:
\begin{equation}
    \label{eq:Gibbscoherent}
    \ket{\Psi_G}=\bigotimes_k \frac{1}{\sqrt{Z_k(\lambda_0)}}\left( e^{-\beta\epsilon_k^{(0)}/4} \ket{1_k} 
    +e^{\beta\epsilon_k^{(0)}/4} \ket{0_k} \right),  
\end{equation}
which has the same energy distribution of $\rho_G(\lambda_0)$ but is a pure state. Crucially, $\ketbra{\Psi_G}{\Psi_G}$ contains coherent terms in the initial energy eigenbasis, e.g., $\ketbra{0_k}{1_k}$.
This means that the initial state $\rho$ is a mixture of the Gibbs equilibrium state $\rho_G(\lambda_0)$, which is diagonal in the eigenbasis of the initial Hamiltonian, and of $\ketbra{\Psi_G}{\Psi_G}$ that exhibits non diagonal quantum coherence.

Let us now calculate the KDQ distribution of the work done by suddenly change the Hamiltonian from $\mathcal{H}(\lambda_0)$ to $\mathcal{H}(\lambda_1)$. Since $\epsilon_k=\epsilon_{-k}$, we can rewrite the initial state as:
\begin{eqnarray}
    \rho &=& \frac{1}{Z(\lambda_0)} \bigotimes_{k>0}\Big[
    e^{-\beta\epsilon_k^{(0)}}\ketbra{1_k1_{-k}}{1_k1_{-k}}     
    + \ketbra{1_k 0_{-k}}{1_k 0_{-k}} + \nonumber\\
    &+& \ketbra{0_k 1_{-k}}{0_k 1_{-k}} 
    +e^{\beta\epsilon_k^{(0)}}\ketbra{0_k 0_{-k}}{0_k 0_{-k}}     
    \Big] + \nonumber \\
    &+&\frac{p}{Z(\lambda_0)}\bigotimes_{k>0}\Big[ \ketbra{1_k1_{-k}}{0_k0_{-k}} +\ketbra{1_k0_{-k}}{0_k1_{-k}} + h.c.\Big]\nonumber
\end{eqnarray}
From Eq.~\eqref{eq:vacua}, we see that the eigenstates of $\mathcal{H}(\lambda_0)$ with quasimomenta $\pm k$ are transformed into the superposition of eigenstates of the $\mathcal{H}(\lambda_1)$ with the same pair of quasimomenta. Therefore, we can compute the work done for all possible transitions from $\ket{m_k,n_{-k}}$ to $\ket{m'_k,n'_{-k}}$ that correspond to the work instances $W_{mn,m'n'}(k)$. The only transitions that have nonzero quasiprobabilities 
\begin{equation}
    q_{mn,m'n'}(k) = \braket{m'_k n'_{-k}}{m_k,n_{-k}} \bra{m_k,n_{-k}} \rho \ket{m'_k n'_{-k}}
\end{equation}
are:
\begin{widetext}
    \begin{tabular}{l@{\hskip 1cm}l@{\hskip 1cm}l}
    $\ket{0_k,0_{-k}} \to \ket{0'_k,0'_{-k}} $
    &
    $W_{00,00}(k)=-\epsilon_k^{(1)}+\epsilon_k^{(0)}$
    &
    $q_{00,00}(k)= \dfrac{e^{\beta \epsilon_k^{(0)}}}{Z_k(\lambda_0)^2}\cos^2 \left( \dfrac{\Delta_k}{2} \right) -\dfrac{p\sin\left(\Delta_k\right)}{2Z_k(\lambda_0)^2}$
    \\
    $\ket{0_k,0_{-k}} \to \ket{1'_k,1'_{-k}}$
    &
    $W_{00,11}(k)=\epsilon_k^{(1)}+\epsilon_k^{(0)}$
    &
    $q_{00,11}(k)= \dfrac{e^{\beta \epsilon_k^{(0)}}}{Z_k(\lambda_0)^2}\sin^2\left(\dfrac{\Delta_k}{2}\right)+\dfrac{p\sin\left(\Delta_k\right)}{2Z_k(\lambda_0)^2}$
    \\
    $\ket{0_k,1_{-k}} \to \ket{0'_k,1'_{-k}}$
    &
    $W_{01,01}(k)=0$
    &
    $q_{01,01}(k)=\dfrac{1}{Z_k(\lambda_0)^2}$
    \\
    $\ket{1_k,0_{-k}} \to \ket{1'_k,0'_{-k}}$
    &
    $W_{10,10}(k)=0$
    &
    $q_{10,10}(k)=\dfrac{1}{Z_k(\lambda_0)^2}$
    \\
    $\ket{1_k,1_{-k}} \to \ket{0'_k,0'_{-k}}$
    &
    $W_{11,00}(k)=-\epsilon_k^{(1)}-\epsilon_k^{(0)}$ 
    &
    $q_{11,00}(k)= \dfrac{e^{-\beta \epsilon_k^{(0)}}}{Z_k(\lambda_0)^2}\sin^2\left(\dfrac{\Delta_k}{2}\right)-\dfrac{p\sin\left(\Delta_k\right)}{2Z_k(\lambda_0)^2}$
    \\
    $\ket{1_k,1_{-k}} \to \ket{1'_k,1'_{-k}}$
    &
    $W_{11,11}(k)=\epsilon_k^{(1)}-\epsilon_k^{(0)}$
    &
    $q_{11,11}(k)= \dfrac{e^{-\beta \epsilon_k^{(0)}}}{Z_k(\lambda_0)^2}\cos^2\left(\dfrac{\Delta_k}{2}\right)+\dfrac{p\sin\left(\Delta_k\right)}{2Z_k(\lambda_0)^2}$
    \end{tabular}
\end{widetext}
where, for each process, we have included the value of the stochastic work instances and the associated quasiprobability. As expected, the quasiprobabilities with non vanishing work also depends on the mixture parameter $p$, which weighs the contribution of the initial state $\rho$ containing quantum coherence in the eigenbasis of $\mathcal{H}(\lambda_0)$.

\begin{figure}[t!]
    \centering
    \includegraphics[width=0.875\columnwidth]{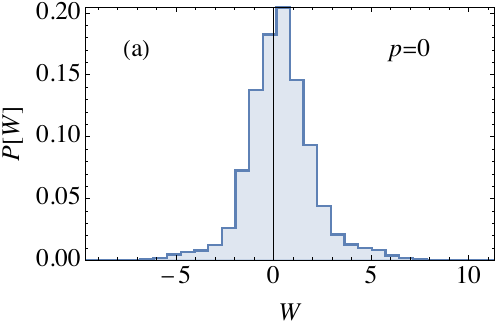}
    \includegraphics[width=0.875\columnwidth]{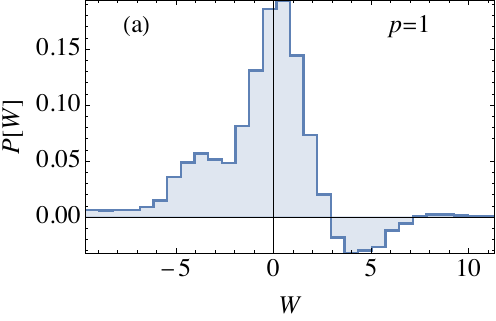}    
    \caption{KDQ distribution of work for the Ising model undergoing a sudden change of the Hamiltonian $\mathcal{H}(\lambda)$. Specifically, the Ising spin chain with $N=12$ spins is quenched from $\lambda_0=0$ to $\lambda_1=0.5$. Panel (a), $p=0$. Panel (b), $p=1$. Other parameters: $\beta=0.1$.}
    \label{fig:quadraticPW}
\end{figure}
\begin{figure}[t!]
    \centering
    \includegraphics[width=0.95\columnwidth]{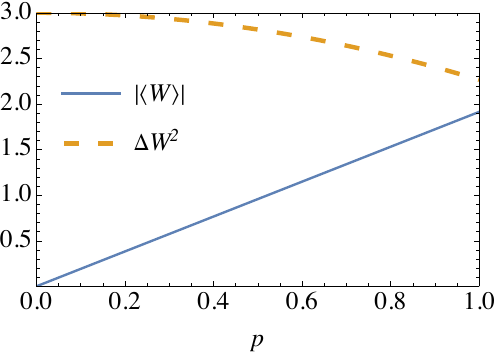}
    \caption{Average work (absolute value) and work variance for the quantum Ising model as a function of the weight $p$ introducing quantum coherence in $\rho$. The other parameters are the same as in Fig.~\ref{fig:quadraticPW}.}
    \label{fig:MomentsIsing}
\end{figure}
From the results above we can finally write the KDQ distribution of work by summing over all the quasimomenta $k>0$:
\begin{eqnarray}
    P[W] &=& \sum_{\textrm{All combinations}} \,\, \prod_{k>0} q_{mn,m'n'}(k)\times \nonumber \\
    &\times&\delta\left(W-\sum_{k>0}W_{mn,m'n'}(k)\right).
\end{eqnarray}

A coarse-grained histogram of $P[W]$ is shown in Fig.~\ref{fig:quadraticPW}. For $p=0$ the distribution is always nonnegative, while for $p=1$ some negative parts appear and are associated with positive values of $W$. As a consequence $P[W<0]$, leading to work extraction, tends to be enhanced, so that $\langle W\rangle<0$, as explained in Sec.~\ref{sec:extractable_work} and shown explicitly in Fig.~\ref{fig:MomentsIsing}. Notice that, in order for $P[W]$ to exhibit negativity, the temperature entering $\rho_G(\lambda_0)$ must be high enough for the two-body processes $00 \leftrightarrow 11$ to be significant. In contrast, if the initial state is close to the ground state these processes are suppressed and $P[W]$ is non negative everywhere.
Initial state coherence also leads to a reduction of the work fluctuations as measured by its variance, see Fig.~\ref{fig:MomentsIsing}.

\section{Discussion}

Quasiprobabilities have been quite elusive quantities so far, due to the difficulty for their experimental inference. As stressed in Sec.~\ref{section_quasiprobs}, procedures based on sequential projective measurements cannot reconstruct the quasiprobability distribution of a physical quantity that is defined over two times, as well as its corresponding statistical moments. 

Recently, however, we have witnessed a resurgence of quasiprobabilities, thanks to their direct link with two-point quantum correlation functions of the form $\langle\mathcal{O}_{1}(t_1)\mathcal{O}_{2}(t_2)\rangle$, with $\mathcal{O}_{1}(t_1)$, $\mathcal{O}_{2}(t_2)$ quantum observables, and the average $\langle\cdot\rangle$ performed with respect to a generic density operator $\rho$. 
Quantum correlation functions are a powerful tool to describe phase changes in quantum statistical mechanics. Hence, the possibility to express them in terms of quasiprobabilities opens the door for building a microscopic, nonequilibrium description of phenomena that naturally includes genuinely quantum resources as quantum coherence and correlations.

Beyond theoretical arguments, quasiprobabilities may turn out to be pivotal also for revealing advantages in quantum technology applications. In this tutorial we have seen several examples in quantum thermodynamic applications, particularly in experiments, including work extraction~\cite{hernandez2022experimental}. This is also relevant for the energetic assessment of quantum computation, where the energy exchange of a qubit with its environment can be continuously monitored through weak measurements~\cite{maffei2022anomalous}.

The contextual nature of quantum systems facilitates the emergence of anomalous weak values of the energy exchanges due to quantum coherence which manifest themselves through negative quasiprobability distributions. In the context of metrological applications, negative values of quasiprobability distributions may enhance parameter estimation increasing the precision of quantum sensing protocols~\cite{Lostaglio2020certifying,lupu2021negative,ArvidssonShukurReview2024}.

We conclude the tutorial by mentioning some possible future perspectives of the topics treated here. By now, it should be apparent how quasiprobabilities have connections with all main theoretical and experimental aspects of quantum theory. Here, we explored the direct link with  quantum measurement theory, fluctuation theorems, work and heat in quantum systems led by genuinely quantum resources, and the scrambling of information in many-body systems. Thus, further investigations on the following subjects could be considered: 
\begin{itemize}
\item[(i)] 
To determine how the thermodynamic entropy production in a nonequilibrium quantum process is expressed in terms of a quasiprobability distribution. A starting point could be taking Ref.~\cite{UpadhyayaArXiv2023}, where the entropy production as a quantifier of irreversibility is extended to a regime with noncommuting conserved quantities.
Afterwards, one might investigate the link with quantum information theory and feedback mechanism naturally including the so-called Maxwell's demon~\cite{GooldJPAMT2016}. In this regard, a quasiprobability formulation of quantum trajectories~\cite{HorowitzNJP2013} could be taken into account.
\item[(ii)]
The extension of two-time quasiprobability distribution to access multitime statistics in open quantum systems (see for instance Refs.~\cite{KrummPRA2017,Hofer2017quasiprobability} for similar attempts for other QDs). This could help investigating non-Markovianity arising because of memory effects in the environment. 
\item[(iii)]
To define to what extent the quasiprobability distribution underlying an OTOC can be a proper quantum sensing toolbox. In fact, given a quantum many-body system, different perturbations may scramble differently the state of the global system~\cite{LiScience2023}, and the corresponding quasiprobability distribution could give access to this information, measurable by means of an interferometric procedure.
\end{itemize}

We hope that the curious and interested reader can find new, fascinating ideas from this tutorial, and can develop some of the perspectives listed here, by opening in turn further open problems.

\section*{Data Availability Statement}

The codes implementing the computations for the figures of the tutorial are available as Supplemental Material~\footnote{See the SupplementalMaterial at \url{http://link.aps.org/supplemental/10.1103/PRXQuantum.5.030201} for Mathematica
notebooks for the main calculations of the paper.}. 

\section*{Acknowledgements}
We acknowledge enlightening discussions with Kenza Hammam, Paolo Solinas and Nicole Yunger-Halpern.
S.G. acknowledges financial support from the project PRIN 2022 Quantum Reservoir Computing (QuReCo), the PNRR MUR project PE0000023-NQSTI financed by the European Union--Next Generation EU, the MISTI Global Seed Funds MIT-FVG Collaboration Grant ``Revealing and exploiting quantumness via quasiprobabilities: from quantum thermodynamics to quantum sensing'', and the Royal Society Project IES\textbackslash R3\textbackslash 223086 ``Dissipation-based quantum inference for out-of-equilibrium quantum many-body systems''. G.D.C. acknowledges support from the UK EPSRC through Grant No. EP/S02994X/1.

\bibliography{biblio}

\end{document}